\documentclass{article}
\usepackage[final,dvips]{graphicx}
\usepackage{lscape}
\def\Journal#1#2#3#4{{#1} {\bf #2}, #3 (#4)}

\def\NCA{\em Nuovo Cimento}

\def\NIMA{{\em Nucl. Instrum. Methods} A}
\def\NPB{{\em Nucl. Phys.} B}
\def\NPA{{\em Nucl. Phys.} A}
\def\PLB{{\em Phys. Lett.}  B}
\def\PRL{\em Phys. Rev. Lett.}
\def\SJN{\em Sov. J. Nucl. Phys.}
\def\JETP{\em Sov. Phys. JETP}
\def\PRD{{\em Phys. Rev.} D}
\def\PRB{{\em Phys. Rev.} B}
\def\ZPC{{\em Z. Phys.} C}
\def\PRC{{\em Phys. Rev.} C}
\def\JPG{{\em J. Phys.} G}
\def\APP{{\em Acta Phys. Polon.} B}
\def\SJN{\em Sov. J. Nucl. Phys.}
\def\JETPL{\em JETP Lett.}
\def\EJC{{\em Eur. Phys. J.} C}
\def\EJA{{\em Eur. Phys. J.} A}
\def\CJP{\em Chin. J. Phys.}
\def\ANN{\em Ann. Rev. Nucl. Part. Sci.}
\def\PR{\em Phys. Rep.}

\def\IJE{{\em Int. J. Mod. Phys.} E}
\def\IJA{{\em Int J. Mod. Phys. } A}
\def\PTP{\em Prog. Theor. Phys.}
\def\APNY{\em Ann. Phys. (N. Y.)}
\def\MPLA{{\em Mod. Phys. Lett.} A}
\def\PPNP{\em Prog. Part. Nucl. Phys.}
\def\SPD{\em Sov. Phys. Doklady}
\def\FP{\em Fortschr. Phys.}
\def\YF{\em Yad. Fiz.}

\def\be{\begin{equation}}
\def\ee{\end{equation}}
\def\bea{\begin{eqnarray}}
\def\eea{\end{eqnarray}}
\def\gtorder{\mathrel{\raise.3ex\hbox{$>$}\mkern-14mu
             \lower0.6ex\hbox{$\sim$}}}
\def\ltorder{\mathrel{\raise.3ex\hbox{$<$}\mkern-14mu
             \lower0.6ex\hbox{$\sim$}}}

\begin{document}

\title{THE SPIN STRUCTURE OF THE NUCLEON}
\author{B. W. FILIPPONE$^*$ and XIANGDONG JI$^{**}$}
\maketitle
\centerline{$^*$W. K. Kellogg Laboratory, California Institute of 
Technology,} 
\centerline{Pasadena, CA 91125}
\centerline{$^{**}$Department of Physics, University of Maryland,}
\centerline{College Park, MD 20742}

\begin{abstract}
We present an overview of recent experimental and theoretical 
advances in our understanding of the spin structure of protons and
neutrons. 
\end{abstract}

\tableofcontents

\section{Introduction \label{sec:intro}}
Attempting to understand the origin of the 
intrinsic spin of the proton and neutron has been an active area of both 
experimental and
theoretical research for the past twenty years. With the confirmation 
that the proton and neutron were not elementary particles, physicists
were challenged with the task of explaining the nucleon's spin in terms
of its constituents. In a simple constituent picture  
one can decompose the nucleon's spin as
\begin{equation}
J_z^N = S_z^q + L_z^q + S_z^g + L_z^g = {1 \over 2} \ .
\label{eq:spin}
\end{equation}
where $S_z$ and $L_z$ represent the intrinsic and orbital angular momentum 
respectively
for quarks and gluons. A simple non-relativistic quark model 
(as described below) gives directly $S_z^q = {1 \over 2}$ and all the other 
components $ = 0$.

Because the structure of the nucleon is governed by 
the strong interaction, the components of the nucleon's spin 
must in principle be calculable from the fundamental theory: 
Quantum Chromodynamics (QCD). However since the spin is a 
low energy property, direct calculations with
non-perturbative QCD are only possible at present with 
primitive lattice simulations. The fact that the nucleon spin 
composition can be measured directly from experiments 
has created an important frontier in hadron physics 
phenomenology and has had crucial impact on our 
basic knowledge of the internal structure of the nucleon. 

This paper summarizes the status of our experimental and theoretical
understanding of the nucleon's spin structure. We begin with 
a simplified discussion of nucleon spin structure and how it can
be accessed through polarized Deep Inelastic Scattering (DIS). This is
followed by a theoretical overview of spin structure in terms of 
QCD. The experimental program is then reviewed where we discuss the 
vastly different techniques being applied in order to limit possible 
systematic errors in the measurements. We then address the variety of 
spin distributions associated with the nucleon: the total quark 
helicity distribution $\Delta\Sigma$ extracted from inclusive scattering,
the individual quark helicity distributions (flavor separation) determined by 
semi-inclusive scattering, and the gluon helicity distribution accessed by a 
variety of probes. We also discuss some additional distributions that 
have have recently been discussed theoretically but are only just being
accessed experimentally: the transversity distribution and the off-forward
distributions. Lastly we review a few topics closely related
to the spin structure of the nucleon. 

A number of reviews of nucleon spin structure have been published. 
Following the pioneering review of the field by Hughes and Kuti~\cite{hk83}
which set the stage for the very rapid development over the
last fifteen years, a number of reviews have summarized the recent
developments \cite{bof93,ael95,gr99,hv99,che00}. Also Ref.~\cite{rhicspin} 
presents a detailed review of the potential contribution of the 
Relativistic Heavy Ion Collider (RHIC) to field of 
nucleon spin structure. 

\subsection{A Simple Model for Proton Spin \label{sec:simple}}

A simple non-relativistic wave function for the proton comprising only the
valence up and down quarks can be written as 
\begin{equation}
|p\uparrow\rangle = {1 \over \sqrt 6}(2|u\uparrow u\uparrow d\downarrow
\rangle - |u\uparrow u\downarrow d\uparrow\rangle - 
|u\downarrow u\uparrow d\uparrow\rangle.
\label{eq:wavefut}
\end{equation}
where we have suppressed the color indices and permutations 
for simplicity but enforced the normalization.
Here the up and down quarks give all of the proton's spin. The 
contribution of
the $u$ and $d$ quarks to the proton's spin can be determined by the use 
of the following matrix element and projection operator:
\begin{eqnarray}
u^{\uparrow} = \langle p\uparrow | \hat{\mathcal{O}}_{u\uparrow} | p\uparrow
\rangle \\
\hat{\mathcal{O}}_{u\uparrow} = 
{1 \over 4}(1 + \hat\tau_3)(1 + \hat\sigma_3) \ .
\label{eq:operator}
\end{eqnarray}
where the matrix element gives the number of up quarks 
polarized along the direction of the proton's polarization. 
With the above matrix element and a similar one for the down quarks, the
quark spin contributions can be defined as
\begin{eqnarray}
\Delta u = u^{\uparrow} - u^{\downarrow} = 
{1 \over 2} (\langle\sigma_3\rangle + \langle\sigma_3 \tau_3\rangle) = {4 \over 3} \ , \\
\Delta d = d^{\uparrow} - d^{\downarrow} = 
{1 \over 2} (\langle\sigma_3\rangle - \langle\sigma_3 \tau_3\rangle) = -{1 \over 3} \ .
\label{eq:spindist}
\end{eqnarray}
Thus the fraction of the proton's spin carried by quarks in this simple model
is 
\begin{equation}
\Delta \Sigma \equiv \Delta u + \Delta d + \Delta s = \langle \sigma_3\rangle = 2 J_z^N = 1 \ ,
\label{eq:totspin}
\end{equation}
and all of the spin is carried by the quarks. Note however that this simple
model overestimates another property of the nucleon, namely the axial-vector
weak coupling constant $g_A$. In fact this model gives 
\begin{equation}
{g_A \over g_V} = \langle \sigma_3 \tau_3\rangle = \Delta u - \Delta d = {5 \over 3} \ , 
\label{eq:axial}
\end{equation}
compared to the experimentally measured value of 
${g_A / g_V} = 1.267 \pm 0.004$. The difference between the
simple non-relativistic model and the data is often attributed to 
relativistic effects. This ``quenching'' factor of $\sim$ 0.75 
can be applied to
the spin carried by quarks to give the following ``relativistic'' 
constituent quark model predictions:
\begin{eqnarray}
\Delta \Sigma \approx& 0.75 \nonumber \ , \\
\Delta u \approx& 1.0 \nonumber \ ,\\
\Delta d \approx& -0.25 \nonumber \ , \\
\Delta s \approx& 0 \ .
\label{eq:relativ}
\end{eqnarray}

\subsection{Lepton Scattering as a Probe of Spin Structure \label{sec:lepton}}

Deep-inelastic scattering (DIS) with charged lepton beams has been the
key tool for probing the structure of the nucleon.
With {\it polarized} beams and targets the ${\it spin}$
structure of the nucleon becomes accessible. Information from
neutral lepton scattering (neutrinos) is complementary to that from 
charged leptons but is generally of lower statistical quality. 
\begin{figure}
\hspace{1in}\includegraphics*[angle=270,width=3.5in]{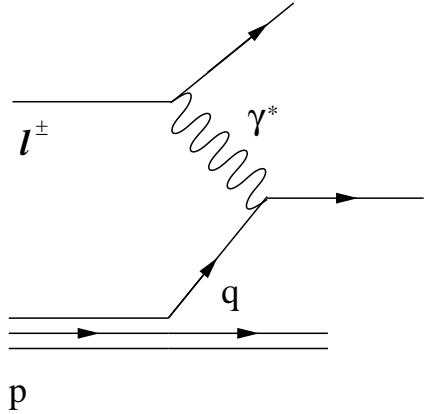}
\caption{Deep Inelastic Scattering in the Quark-Parton Model}
\label{dis}
\end{figure}

The access to nucleon structure through lepton scattering can 
best be seen within the Quark-Parton Model (QPM). 
An example of a deep-inelastic
scattering process is shown in Fig.~\ref{dis}. In this picture a virtual 
photon of four-momentum $q^\mu$(with energy $\nu$ and 
four-momentum transfer $Q^2\equiv-q^2$) 
strikes an asymptotically free quark in the nucleon.
We are interested in the deep-inelastic (Bjorken) limit in 
which $Q^2$ and $\nu$ are large, while the Bjorken scaling
variable $x_B=Q^2/2M\nu$ is
kept fixed ($M$ is the nucleon mass). For unpolarized scattering the 
quark ``momentum'' distributions---$ q_i(x) = u(x), 
d(x), s(x),$ ... --- are probed in this reaction, where  
$x=x_B$ is the quark's momentum fraction.
From the cross section for this process, the structure function $F_1(x)$ can
be extracted. In the quark-parton model this structure function is
related to the unpolarized quark distributions via
\begin{equation}
F_1(x) = {1 \over 2} \sum_i e_i^2 q_i(x) \ , 
\label{eq:f1}
\end{equation}
where the sum is over both quark and anti-quark flavors.
With polarized beams and targets the quark spin distributions can be 
probed. This sensitivity results from the requirement that the quark's 
spin be anti-parallel to the virtual photon's spin in order for the
quark to absorb the virtual photon. With the assumption of nearly 
massless and collinear quarks, angular momentum would not be conserved 
if the quark absorbs a photon when its spin is parallel to the 
photon's spin. Thus measurements of the spin-dependent 
cross section allow the extraction
of the spin-dependent structure function $g_1(x)$. 
Again in the quark-parton model this structure function is related 
to the quark {\it
spin} distributions via 
\begin{equation}
g_1(x) = {1 \over 2} \sum_i e_i^2 \Delta q_i(x) \ .
\label{eq:g1}
\end{equation}
The structure 
function $g_1$ is extracted from the measured asymmetries 
of the scattering cross section as the beam or target spin is reversed. 
These asymmetries are measured with longitudinally polarized beams and
longitudinally ($A_{||})$ and transversely ($A_\perp$) polarized targets
(see Sec.~\ref{sec:total}).

Beyond the QPM, QCD introduces a momentum scale
($Q^2$) dependence into the structure functions (eg. $F_1(x,Q^2)$ and 
$g_1(x,Q^2)$). The calculation of this $Q^2$ dependence is based
on the Operator Product Expansion (OPE) and the renormalization
group equations (see eg. Ref.~\cite{sv,jaffeji91,ael95}). We
will not discuss this in detail, but we will use some elements
of the expansion. In particular, the expansion can be written in
terms of ``twist'' which is the difference between the dimension
and the spin of the operators that form the basis for the expansion.
The matrix elements of these operators cannot be calculated in
perturbative QCD, but the corresponding $Q^2$-dependent coefficients are
calculable. The lowest order coefficients (twist-two) remain
finite as $Q^2 \rightarrow\infty$ while the higher-twist coefficients
vanish as $Q^2 \rightarrow\infty$ (due to their $1/Q^2$ dependence).
Therefore, the full $Q^2$ dependence includes both QCD radiative corrections
(calculated to next-to-leading-order (NLO) at present) and higher-twist 
corrections. The NLO corrections will be discussed in 
Sec.~\ref{subsec:nlo}.

\subsection{Theoretical Introduction \label{subsec:theory}}

To go beyond the simple picture of nucleon spin structure 
discussed above we must address the spin structure within the context
of QCD. We discuss several of these issues in the following sections.

\subsubsection{Quark Helicity Distributions and $g_1(x, Q^2)$ 
\label{subsubsec:quark} }

In polarized DIS, the antisymmetric part of the nucleon tensor
is measured, 
\begin{equation}
	W_{\mu\nu} = {1\over 4\pi}
\int e^{i\xi \cdot q} d^4\xi \langle PS|[J_\mu(\xi), J_\nu(0)|PS\rangle \ , 
\end{equation}
where $|PS\rangle$ is the ground state of the nucleon
with four-momentum $P^\mu$ and polarization $S^\mu$ ($P\cdot S=0$), and $J^\mu$
is the electromagnetic current. The antisymmetric part
can be expressed in terms of two invariant structure functions,
\be
      W^{[\mu\nu]} = -i\epsilon^{\mu\nu\alpha\beta}
  q_\alpha \left[G_1(\nu, Q^2)S_\beta/M^2 
  + G_2(\nu, Q^2)(S_\beta \nu M - P_\beta S\cdot q)/M^4\right] \ .
\ee
In the Bjorken limit, we obtain two scaling functions,
\begin{eqnarray}
   g_1(x,Q^2) &=& \left({\nu \over M}\right)G_1(\nu, Q^2) \ \rightarrow \  g_1(x)\ , \nonumber\\
   g_2(x,Q^2) &=& \left({\nu \over M}\right)^2G_2(\nu, Q^2) \ \rightarrow \  
   g_2(x) \ ,  
\end{eqnarray}
which are non-vanishing.

If the QCD radiative corrections are neglected, $g_1(x, Q^2)$
is related to the polarized quark distributions $\Delta q(x)$ as shown in 
Eq. (\ref{eq:g1}). In QCD, the distribution can be expressed as the Fourier
transform of a quark light-cone correlation, 
\begin{equation}
      \Delta q(x, \mu^2) = {1\over 2} \int  {d\lambda\over 2\pi} 
          e^{i\lambda x} \langle PS|\overline{\psi}(0)
   U(0,\lambda n)\not\! n \gamma_5 \psi(\lambda n) |PS\rangle
\end{equation}
where $n$ is a light-cone vector (eg. $n = (1,0,0,-1)$) and $\mu^2$ is 
a renormalization scale. $U(0, \lambda n) = 
\exp(-ig\int^0_\lambda n\cdot A(\mu n) d\mu)$ is a path-ordered 
gauge link making the operator gauge invariant. 
When QCD radiative corrections are taken into 
account, the relation between $g_1(x, Q^2)$ and $\Delta q(x, \mu^2)$
is more complicated (see Sec.~\ref{subsec:nlo}). When $Q^2$ is not too
large ($<5$ GeV$^2$), one must take into account the higher-twist
contributions to $g_1(x, Q^2)$, which appear as $1/Q^2$ power 
corrections. Some initial theoretical estimates of these power corrections 
have been performed~\cite{bal90,jiunrau}. 

Integrating the polarized quark distributions over $x$ yields
the fraction of the nucleon spin carried by quarks, 
\begin{equation}
    \Delta \Sigma = \int^1_0 dx \sum_ i \left( \Delta q_i(x)
           +\Delta \bar q_i(x)\right) \ .
\end{equation}
The individual quark contribution $\Delta q$ is also called the 
axial charge because it is related to the matrix element of the
axial current $\bar\psi\gamma_\mu\gamma_5\psi$ in the nucleon state.
$\Delta \Sigma$ is the singlet axial charge. Because of the axial anomaly,
it is a scale-dependent quantity. 

\subsubsection{The Nucleon Spin Sum Rule \label{subsubsec:nucleon} }

To understand the spin structure of the nucleon 
in the framework QCD, 
we can write
the QCD angular momentum operator 
in a gauge-invariant form~\cite{jiprl97}
\begin{equation}
    \vec{J}_{\rm QCD} = \vec{J}_{q} + \vec{J}_g \, ,
\end{equation}
where
\begin{eqnarray}
     \vec{J}_q &=& \int d^3x ~\vec{x} \times \vec{T}_q \nonumber \\
                 &=& \int d^3x ~\left[ \psi^\dagger
     {\vec{\Sigma}\over 2}\psi + \psi^\dagger \vec{x}\times
          (-i\vec{D})\psi\right]
     \, ,  \nonumber \\
     \vec{J}_g &=& \int d^3x ~\vec{x} \times (\vec{E} \times \vec{B}) \, .   
\label{ang}
\end{eqnarray}
(The angular momentum operator in a gauge-variant form
has also motivated a lot of theoretical work, but is
unattractive both theoretically and experimentally
\cite{gdependent}.)
The quark and gluon components of the angular momentum
are generated from the quark and gluon
momentum densities $\vec {T}_q$ and $\vec{E}\times \vec{B}$,
respectively. $\vec{\Sigma}$ is
the Dirac spin matrix and the corresponding term
is the quark spin contribution. $\vec{D}= \vec{\nabla}-
ig\vec{A}$ is the covariant derivative and the
associated term is the gauge-invariant quark
orbital angular momentum contribution.

Using the above expression, one can easily construct
a sum rule for the spin of the nucleon. Consider
a nucleon moving in the $z$ direction, and polarized
in the helicity eigenstate $\lambda = 1/2$. The total
helicity can be evaluated as an expectation value of
$J_z$ in the nucleon state,
\begin{equation}
        {1\over 2} = {1\over 2}\Delta \Sigma (\mu)
    + L_q(\mu) + J_q(\mu) \, ,     
\end{equation}
where the three terms denote the matrix elements
of three parts of the angular momentum operator
in Eq.~\ref{ang}. The physical significance
of each term is obvious, modulo the momentum transfer scale
$Q^2$ and scheme dependence (see Sec.~\ref{subsec:nlo}) indicated by $\mu$.
There have been attempts
to remove the scale dependence in $\Delta \Sigma$ 
by subtracting
a gluon contribution~\cite{altarelli}. Unfortunately, such
a subtraction is by no means unique. Here we adopt
the standard definition of $\Delta \Sigma (\mu)$
as the matrix element of the multiplicatively
renormalized quark spin operator.  
As has been discussed above, $\Delta \Sigma(\mu)$ can be 
measured from polarized deep-inelastic scattering and the measurement
of the other terms will be discussed in later sections. 
Note that
the individual terms in the above equation are
independent of the nucleon velocity~\cite{jiinvariant}.
In particular, the equation applies when the nucleon
is traveling with the speed of light (the infinite
momentum frame).         

The scale dependence of the quark and gluon 
contributions can be calculated in perturbative QCD. 
By studying renormalization of the nonlocal
operators, one can show~\cite{jiprl97,gdependent}
\begin{equation}
{\partial \over \partial \ln \mu^2}
  \left(\begin{array}{c}
         J_q(\mu) \\
          J_g(\mu)
    \end{array} \right)
   = {\alpha_s (\mu)\over 2\pi}
    {1\over 9}\left( \begin{array}{rr}
        -16 & 3n_F  \\
        16 & -3n_F  \\
      \end{array} \right)
        \left( \begin{array}{c}  
           J_q(\mu) \\
            J_g(\mu)
        \end{array} \right) \ .
\end{equation}
As $\mu\rightarrow \infty$, there exists a fixed-point
solution
\begin{eqnarray}
       J_q(\infty) &=& {1\over 2} {3n_f\over 16 + 3n_f} \ ,  \nonumber
\\
       J_g(\infty) &=& {1\over 2} {16\over 16 + 3n_f} \ .
\end{eqnarray}
Thus as the nucleon is probed at an infinitely small distance
scale, approximately one-half of the spin is carried by   
gluons. A similar result has been obtained by Gross and
Wilczek in 1974 for the quark and gluon contributions to
the momentum of the nucleon~\cite{gro}. Strictly speaking,
these results reveal little about the nonperturbative
structure of bound states. However, experimentally
it is found that about half of the nucleon momentum is
carried by gluons even at relatively low energy scales (see 
eg.~\cite{cteq}). Thus the gluon degrees of freedom not only
play a key role in perturbative QCD, but also are a major
component of nonperturbative states as expected.
An interesting question is then, how much of the nucleon 
spin is carried by the gluons at low energy scales? 
A solid answer from the fundamental theory is not 
yet available. Balitsky and Ji have
made an estimate using the QCD sum rule approach
\cite{bal1}:
\begin{equation}
    J_g(\mu\sim 1 {\rm GeV}) \simeq {8\over 9} {e<\bar u\sigma Gu>
     <\bar uu> \over M_{1^{-+}}^2\lambda_N^2}
\end{equation}
which yields approximately
0.25. Based on this calculation, the spin structure of the nucleon
would look approximately like 
\begin{equation}
       {1\over 2} = 0.10({\rm from~} {1\over2}\Delta \Sigma)
        + 0.15({\rm from~} L_q) + 0.25({\rm from~} J_g) \ .
\end{equation}
Lattice~\cite{liuang} and quark model~\cite{barone}
calculations of $J_q$ have yielded similar results.

While $\Delta \Sigma$ has a simple parton
interpretation, the gauge-invariant orbital angular
momentum clearly does not. Since we are addressing the 
structure of the nucleon, it is not required that a physical
quantity have a {\it simple} parton interpretation. 
The nucleon mass, magnetic moment, and charge radius  
do not have simple parton model
explanations. The quark orbital angular momentum
is related to the transverse momentum of the 
partons in the nucleon. It is well known that transverse
momentum effects are beyond the naive parton
picture. As will be discussed later, however,
the orbital angular momentum does have a more 
subtle parton interpretation (see Sec.~\ref{sec:off}).

In the literature, 
there are suggestions that $\vec{r}\times
(-i\nabla)$ be considered 
the orbital angular momentum~\cite{gdependent}. 
This quantity is clearly not gauge invariant and $-i\nabla$
does not correspond to the velocity in classical mechanics
\cite{feynman}. 
Under scale evolution, this operator mixes with an 
infinite number of other operators in light-cone gauge~\cite{hjw99}. 
More importantly, there is no known way to measure such 
``orbital angular momentum.''

\subsubsection{Gluon Helicity Distribution $\Delta G(x,Q^2)$ 
\label{subsubsec:gluon}}

In a longitudinally-polarized nucleon the polarized 
gluon distribution $\Delta G(x, Q^2)$ 
contributes to spin-dependent scattering processes 
and hence various experimental spin asymmetries. 
In QCD, using the infrared factorization of hard
process, $\Delta G(x,Q^2)$ with $-1<x<1$ can be expressed as 
\begin{equation}
\Delta G(x, \mu^2)
 = {i\over 2} \int {d\lambda\over 2\pi}e^{i\lambda x}
    \langle PS|F^{+\alpha}(0) U(0,\lambda n)
  \tilde F^+_{~~\alpha}(\lambda n) |PS\rangle \ , 
\label{gdis}
\end{equation}
where $\tilde F_{\alpha\beta} 
= (1/2) \epsilon_{\alpha\beta\mu\nu} F^{\mu\nu}$. 
Because of the
charge conjugation property of the operator, 
the gluon distribution is symmetric in $x$: $\Delta G(x,Q^2)
= \Delta G(-x,Q^2)$. 

The even moments of $\Delta G(x, Q^2)$ are directly 
related to matrix elements of charge conjugation even local operators. 
Defining
\begin{equation} 
   \int^1_{-1}dx x^{n-1} \Delta G(x, Q^2)
  = a_n(\mu^2) \ , ~~~(n=3, 5 \cdots) 
\end{equation}
we have 
\begin{eqnarray}
   \langle PS|F^{\mu_1\alpha}iD^{\mu_2}
   \cdots iD^{\mu_{n-1}} i\tilde F^{\mu_n}_{~~\alpha}
  |PS\rangle &= 2 a_n(\mu^2) 
   S^{\mu_1}P^{\mu_2}\cdots P^{\mu_n}\\
 &(n=3, 5 \cdots) \nonumber\ . 
\end{eqnarray}
The renormalization scale dependence is directly
connected to renormalization of the local operators.
Because Eq. (\ref{gdis}) involves directly the
time variable, it is difficult to evaluate the distribution
on a lattice. However, the matrix elements of local operators
are routinely calculated in lattice QCD, hence the moments of
$\Delta G(x,Q^2)$ are, in principle, calculable.

From the above equations, it is clear that the first-moment 
($n=1$) of $\Delta G(x)$ does not correspond to a gauge-invariant 
{\it local} operator. In the axial gauge $n\cdot A=0$, 
the first moment
of the nonlocal operator can be reduced to 
a local one, $\vec{E}\times \vec{A}$, which can be 
interpreted as the gluon spin density operator. 
As a result, the first
moment of $\Delta G(x, \mu^2)$ represents the gluon
spin contribution to the nucleon spin in the
axial gauge. In any other gauge, however, it cannot 
be interpreted as such. Thus
one can formally write $J_g = \Delta G + L_g$ in the axial gauge, where
$L_g$ is then the gluon orbital contribution the nucleon spin. 
There is no known way to measure $L_g$ directly from experiment
other than defining it as the difference between $J_g$ and $\Delta G$. 

\section{Experimental Overview \label{sec:exp} }

A wide variety of experimental approaches have been applied to the 
investigation of the nucleon's spin structure. The experiments 
complement each other in their kinematic coverage and in their 
sensitivity to possible systematic errors associated with the 
measured quantities. A summary of the spin structure measurements
is shown in 
Table~\ref{tab:exps}, where the beams, targets, and typical
energies are listed for 
each experiment. The kinematic coverage of each experiment
is indicated in the table by its average four-momentum transfer 
($Q^2$) and Bjorken
$x$ range (for $Q^2 > 1$ GeV$^2$). 
Also given are the average or typical beam and
target polarizations as quoted by each experimental group in their
respective publications (or in their proposals for the experiments 
that are underway). The column labeled $f$ lists the dilution
factor, which is the fraction of scattered events 
that result from the polarized atoms of interest, and the column 
labeled $\mathcal{L}$ is an estimate of the total nucleon luminosity
(\# of nucleons/cm$^2 \times$ \# of beam particles/s) in units of
$10^{32}$ nucleons/cm$^2$/s for each experiment. 

In an effort to eliminate possible sources of unknown systematic error 
in the measurements the experiments have been performed with
significantly different experimental techniques. 
Examples of the large range of experimental parameters
for the measurements include variations in the beam polarization of 
$40 - 80\%$, in the target polarization of $30 - 90\%$ and in the 
correction 
for dilution of the experimental asymmetry due to unpolarized material 
of $0.1 - 1$. 

We now present an overview of the individual experimental techniques with
an emphasis on the different approaches taken by the various experiments.

\begin{landscape}
\begin{table}[ht]
\caption{Summary of High Energy Spin Structure Function Measurements.
}
\vspace{0.01cm}
\begin{center}
\begin{tabular}{|c|c|c|c|c|c|c|c|c|c|c|}
\hline
 Lab & Exp. & Year & Beam & $\langle Q^2\rangle$   & $x$ & $P_B$ & Target &$P_T$ & $f$ &$\mathcal{L}\times 10^{-32}$\\ 
     &      &      &      & GeV$^2$ &     &       &        &      &  &cm$^{-2}$-s \\
\hline
SLAC & E80   &   75 & 10-16 GeV $e^-$    & 2   &  0.1 -- 0.5  & 85\% & H-butanol & 50\% & 0.13 & 400 \\
     & E130  &   80 & 16-23 GeV $e^-$    & 5   &  0.1 -- 0.6  & 81\% & H-butanol & 58\% & 0.15 & 400 \\
     & E142  &   92 & 19-26 GeV $e^-$    & 2   & 0.03 -- 0.6  & 39\% & $^3$He    & 35\% & 0.35 & 2000 \\
     & E143  &   93 & 10-29 GeV $e^-$    & 3   & 0.03 -- 0.8  & 85\% & NH$_3$    & 70\% & 0.15 & 1000 \\ 
     &       &      &                    &     &              &      & ND$_3$    & 25\% & 0.24 & 1000 \\
     & E154  &   95 & 48 GeV $e^-$       & 5   & 0.01 -- 0.7  & 82\% & $^3$He    & 38\% & 0.55 & 3000 \\
     & E155  &   97 & 48 GeV $e^-$       & 5   & 0.01 -- 0.9  & 81\% & NH$_3$    & 90\% & 0.15 & 1000 \\
     &       &      &                    &     &              &      & LiD       & 22\% & 0.36 & 1000 \\
     & E155' &    99& 30 GeV $e^-$       & 3   & 0.02 -- 0.9  & 83\% & NH$_3$    & 75\% & 0.16 & 1000 \\
     &       &      &                    &     &              &      & LiD       & 22\% & 0.36 & 1000 \\
\hline
CERN & EMC  &    85 &100-200 GeV $\mu^+$ & 11  & 0.01 -- 0.7  & 79\% & NH$_3$    & 78\% & 0.16 & 0.3 \\
     & SMC  &    92 &100 GeV $\mu^+$     & 4.6 & 0.006 -- 0.6 & 82\% & D-butanol & 35\% & 0.19 & 0.3 \\
     &      &    93 &190 GeV $\mu^+$     & 10  & 0.003 -- 0.7 & 80\% & H-butanol & 86\% & 0.12 & 0.6 \\
     &      & 94-95 &                    &     &              & 81\% & D-butanol & 50\% & 0.20 & 0.6 \\
     &      &    96 &                    &     &              & 77\% & NH$_3$    & 89\% & 0.16 & 0.6 \\
\hline
DESY &HERMES&    95 & 28 GeV $e^+$       & 2.5 & 0.02 -- 0.6  & 55\% & $^3$He    & 46\% & 1.0 & 1 \\
     &      & 96-97 &                    &     &              & 55\% & H         & 88\% & 1.0 & 0.1 \\
     &      & 98    & 28 GeV $e^-$       &     &              & 55\% & D         & 85\% & 1.0 & 0.2 \\
     &      & 99-00 & 28 GeV $e^+$       &     &              & 55\% & D         & 85\% & 1.0 & 0.2 \\
\hline
CERN &COMPASS&   01 &190 GeV $\mu^+$     & 10  & 0.005 -- 0.6 & 80\% & NH$_3$    & 90\% & 0.16 & 3 \\
     &       &      &                    &     &              &      &  LiD      & 40\% & 0.50 & 3 \\
\hline
BNL  & RHIC &    02 & 200 GeV p - p      & $\sim 100$  & 0.05 -- 0.6 & 70\% & Collider  & 70\% & 1.0 & 2 \\
\hline
DESY & ZEUS/H1 & ?? & $28\times 800$ GeV e - p   & 22  &0.00006 -- 0.6& 70\% & Collider  & 70\% & 1.0 & 0.2 \\
\hline
\end{tabular}
\end{center}
\label{tab:exps}
\end{table}
\end{landscape}

\subsection{SLAC Experiments \label{subsec:slac} }

The SLAC program has focused on high statistics measurements of the
inclusive asymmetry. 
The first pioneering experiments on the proton spin structure were
performed at SLAC in experiments E80~\cite{e80} and E130~\cite{e130}.
These experiments are typical of the experimental approach of the 
SLAC spin program. Polarized electrons are injected into the SLAC
linac, accelerated to the full beam energy and impinge on fixed
targets in End Station A. The polarization of the electrons is 
measured at low energies at the injector using Mott scattering
and at high energies in the End Station using Moller scattering.
Target polarization is typically measured using NMR techniques.
The scattered electrons are detected with magnetic spectrometers where 
electron identification is usually done with Cerenkov detectors and 
Pb-Glass calorimeters. 

For E80 and E130, electrons were produced by photoionization 
of $^6$Li produced in an atomic beam source. Electron polarization is
produced by Stern-Gerlach separation in an inhomogeneous magnetic
field. Polarized protons were produced by dynamic polarization of
solid-state butanol doped with a paramagnetic substance. 
Depolarization effects in the target limited the average beam currents 
to $\sim$ 10 nA. In these experiments a considerable amount of 
unpolarized material is present in the target resulting in a dilution
of the physics asymmetry. For E80 and E130 this dilution reduced the
asymmetry by a factor of $\sim$ 0.15.

\begin{figure}
\includegraphics*[angle=-90,width=4.5in]{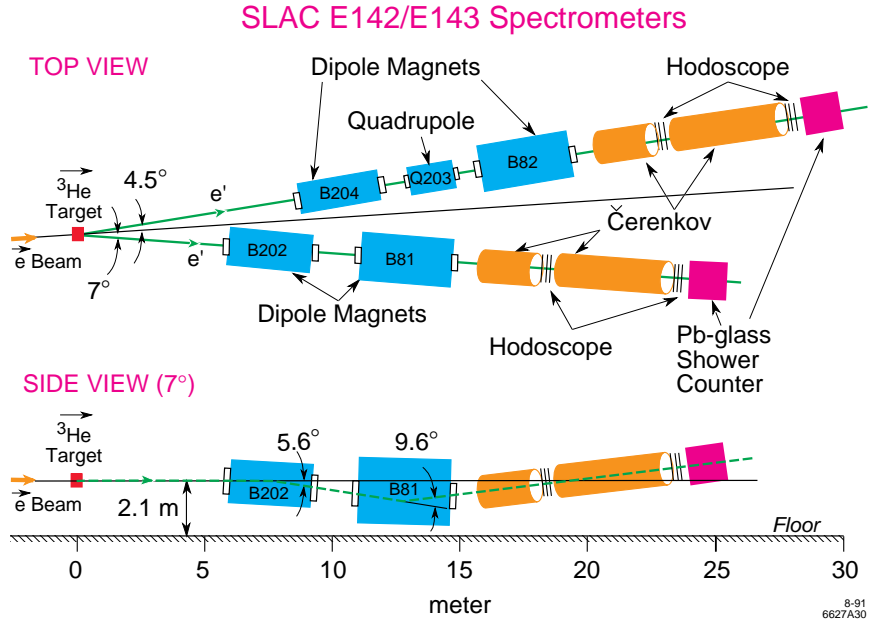}
\caption{Schematic diagram of SLAC E142/E143 spectrometers}
\label{slacspec}
\end{figure}

Over the last ten years a second generation of high precision measurements
have been performed at SLAC. Information on the neutron spin structure
has been obtained using polarized $^3$He in experiments 
E142~\cite{e142n1,e142fin}
and E154~\cite{e154n1}. Here the polarized $^3$He behaves approximately as
a polarized neutron due to the almost complete pairing off of the 
proton spins. The nuclear correction to the neutron asymmetry is 
estimated to be $\sim$ 5 - 10 \%. Beam currents were typically .5 - 2
$\mu$A and the polarization was significantly improved for the E154
experiment using new developments in strained gallium-arsenide 
photocathodes~\cite{strain}. A schematic diagram of the spectrometers
used for E142 is shown in Fig.~\ref{slacspec}.

Additional data on the neutron and more precise data on the proton
has come from E143~\cite{e143p1,e143d,e143fin} 
and E155~\cite{e155d1,e155pd} where both $^2$H 
and H polarized targets using polarized ammonia (NH$_3$ and ND$_3$)
and $^6$LiD were employed.
The main difference between these two experiments was again an 
increase in beam energy from 26 - 48 GeV and an increase in 
polarization from 40 \% to 80 \%. 

\subsection{CERN Experiments \label{subsec:cern} }

Following the early measurements at SLAC, the EMC 
(European Muon Collaboration) experiment~\cite{emc1,emc2} performed
the first measurements at $x < 0.1$. Polarized muon beams were 
produced by pion decay yielding beam intensities of $10^7 \mu$/s.
The small energy loss rate of the muons allowed the use of 
very thick targets ($\sim$ 1 m) of butanol and methane. The
spin structure measurements
by EMC came at the end of a series of measurements of unpolarized
nucleon and nuclear structure functions, but the impact of the EMC
spin measurements was significant. Their low $x$ measurements, accessible
due to the high energy of the muons, suggested the breakdown of 
the naive parton picture that quarks provide essentially all of the 
spin of the nucleon.

\begin{figure}
\includegraphics*[angle=0,width=4.5in]{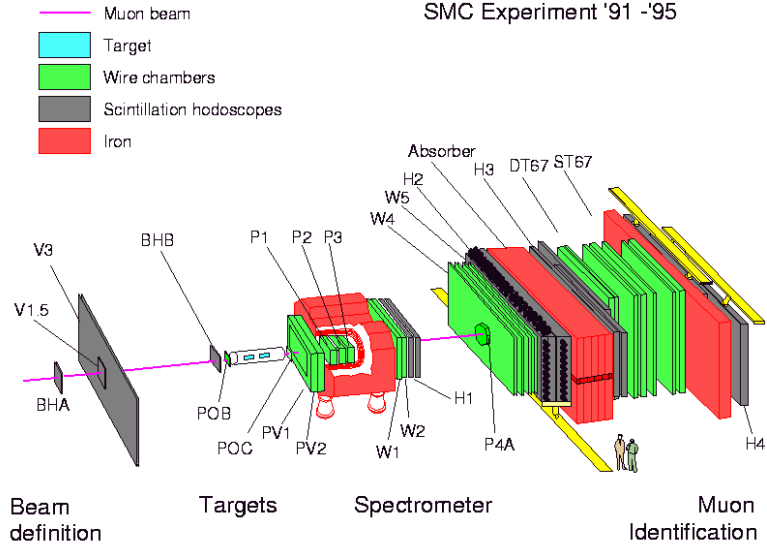}
\caption{Schematic diagram of SMC spectrometer}
\label{smcspec}
\end{figure}

The SMC (Spin Muon Collaboration)
experiment~\cite{smcd1,smcd2,smcp1,smcp2,smcp2big} 
began as a dedicated follow-on experiment to 
the EMC spin measurements using an upgraded apparatus. An extensive program
of measurements with polarized $^1$H and $^2$H targets was undertaken 
over a period of ten years. Improvements in target and beam performance
provided high precision data on inclusive spin-dependent structure 
functions. The large acceptance of the SMC spectrometer in the forward
direction (see Fig.~\ref{smcspec}) 
allowed them to present the first measurements of spin structure using
semi-inclusive hadron production. As with EMC, the high energy of the
muon beam provided access to the low $x$ regime ($x<0.01$).

A new experiment is underway at CERN whose goal is to provide direct 
information on the gluon polarization. The COMPASS~\cite{compass} 
(COmmon Muon Proton Apparatus for Structure and Spectroscopy)
experiment will use a large acceptance spectrometer with full particle
identification to generate a high statistics sample of 
charmed particles. Using targets similar to those used in SMC and an 
intense muon beam ($\sim 10^8 \mu$/s) improved measurements of
other semi-inclusive asymmetries will also be possible.

\subsection{DESY Experiments \label{subsec:desy} }

Using very thin gaseous targets of pure atoms ($^1$H, $^2$H, $^3$He) and 
very high currents ($\sim$ 40 mA) of stored, circulating positrons or 
electrons HERMES (HERa MEasurement of Spin) 
has been taking data at DESY since 1995. HERMES is a fixed 
target experiment that uses the stored $e^\pm$ beam of the HERA collider.
The polarization of the beam is achieved through the Sokolov-Ternov 
effect~\cite{sok64}, whereby the
beam becomes transversely polarized due to a small-spin dependence in
the synchrotron radiation emission. The transverse polarization is
rotated to the longitudinal direction by a spin rotator - a sequence
of horizontal and vertical bending magnets that takes advantage of the
$g-2$ precession of the $e^\pm$. The beam polarization is measured with 
Compton polarimeters~\cite{hermtp}. 

\begin{figure}
\includegraphics*[angle=0,width=4.5in]{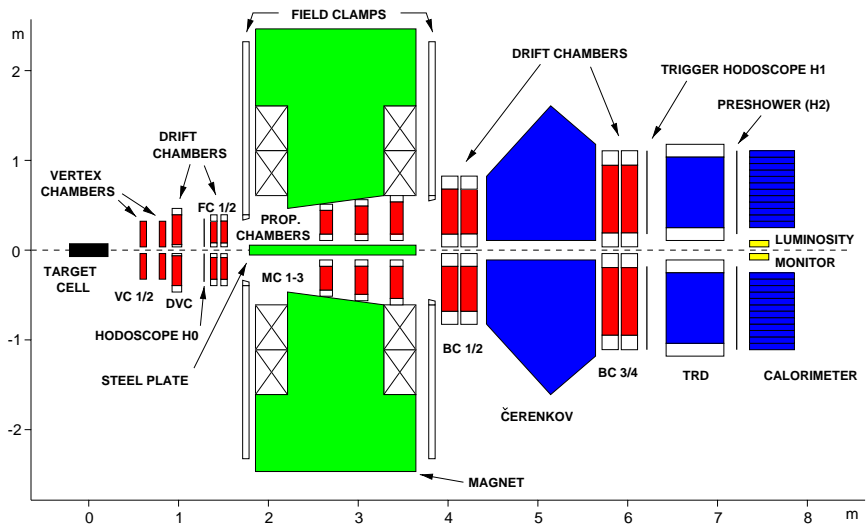}
\caption{Schematic diagram of HERMES spectrometer}
\label{hermspec}
\end{figure}

HERMES has focused its efforts on measurements of semi-inclusive 
asymmetries, where the scattered $e^\pm$ is detected in coincidence with
a forward hadron. This was achieved with a large acceptance magnetic
spectrometer~\cite{hermspec} as shown in Fig.~\ref{hermspec}. Initial 
measurements allowed some limited pion identification with a gas threshold
Cerenkov detector and a Pb-glass calorimeter. Since 1998 a Ring Imaging
Cerenkov (RICH) detector has been in operation allowing full 
hadron identification over most of the momentum acceptance of the 
spectrometer. 

Up to the present, HERMES has taken data only with a longitudinally
polarized target. Future runs will focus on high statistics measurements
with a transversely polarized target to access eg. transversity (see Sect. 6.1)
and $g_2$ (see Sect. 6.2). 

Promising future spin physics options also exist at DESY if polarized
protons can be injected and accelerated in the HERA ring. The 
HERA-$\vec{\mbox{N}}$~\cite{kn97}
program would use the stored 820 GeV proton beam 
and a fixed target of gaseous polarized nucleons. This would allow 
measurements of quark and gluon polarizations at
$\sqrt{s}\sim 50$ GeV, complimenting the higher energy measurements 
possible in the RHIC spin program. 

A stored polarized proton beam in HERA would also allow $\vec{e} - \vec{p}$
collider measurements~\cite{phera}
with the existing H1 and ZEUS detectors. Inclusive
polarized DIS could be measured to much higher $Q^2$ and lower $x$ than
existing measurements. This would allow improved extraction of the 
gluon polarization via the scaling violations of the spin-dependent
cross section. Heavy quark and jet production as
well as charged-current vs. neutral-current 
scattering would also allow improved measurements 
of both quark and gluon polarizations. 

\subsection{RHIC Spin Program \label{subsec:rhic} }

The Relativistic Heavy-Ion Collider (RHIC)~\cite{rhic}
at the Brookhaven National
Laboratory recently began operations. This collider was designed to 
produce high luminosity collisions of high-energy heavy ions as
a means to search for a new state of matter known at the quark-gluon
plasma. The design of the accelerator also allows the acceleration
and collision of high energy beams of polarized protons and a fraction
of accelerator operations will be devoted to spin physics with
colliding $\vec p - \vec p$. Beam polarizations of 70\% and 
center-of-mass energies of $\sqrt{s} = 50 - 500$ are expected. 

\begin{figure}[ht]
\includegraphics*[angle=0,width=4.5in]{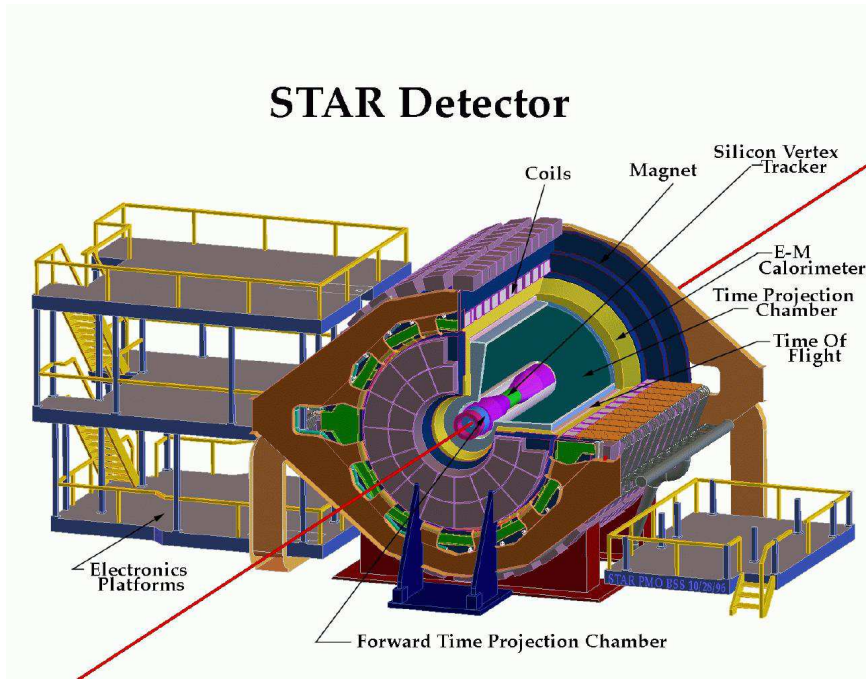}
\caption{Schematic diagram of the STAR detector}
\label{star}
\end{figure}

Two large collider detectors, PHENIX~\cite{phenix} and STAR~\cite{star}, 
along with several smaller experiments, BRAHMS~\cite{brahms}, 
PHOBOS~\cite{phobos} and PP2PP, will participate in
the RHIC spin program. As an example a schematic diagram of the STAR 
detector is shown in Fig.~\ref{star}. Longitudinal beam polarization
will be available for the PHENIX and STAR detectors enabling measurements
of quark and gluon spin distributions (see Sects. 4.2 and 5.74). 

\section{Total Quark Helicity Distribution \label{sec:total} }

A large body of data has been accumulated over the past ten years 
on inclusive polarized lepton scattering from polarized targets. These
data allow the extraction of the spin structure functions 
$g_1^{p,n}(x,Q^2)$
and the nearly model-independent determination of the total quark 
contribution to the nucleon spin $\Delta \Sigma = 
(\Delta u + \Delta\bar u) + (\Delta d + \Delta\bar d) + 
(\Delta s + \Delta\bar s)$. Inclusive data combined with assumptions
about flavor symmetry, $SU(3)_f$, and results from beta decay 
provide some model-dependent information on the individual flavor 
contributions to the nucleon spin. Studies of the $Q^2$ dependence of
$g_1(x,Q^2)$ allow a first estimate of the gluon spin contribution 
albeit with fairly large uncertainties. These results are discussed in 
the following sections. 

\subsection{Virtual Photon Asymmetries \label{subsec:virtual} }

Virtual photon asymmetries can be defined in terms of a helicity 
decomposition of the virtual photon-nucleon scattering cross sections.
For a transversely polarized virtual photon (eg. with helicity $\pm 1$)
incident on a longitudinally polarized nucleon there are 
two helicity cross sections $\sigma_{1\over 2}$ and $\sigma_{3\over 2}$
and the longitudinal asymmetry is given by 
\be
A_1 = {\sigma_{1\over 2} - \sigma_{3\over 2} \over \sigma_{1\over2} + \sigma_{3\over 2}} \ .
\label{eq:a1}
\ee
$A_2$ is a virtual photon asymmetry that results from an interference
between transverse and longitudinal virtual photon-nucleon 
amplitudes:
\be
A_2 = {2 \sigma_{LT}\over \sigma_{1\over2} + \sigma_{3\over 2}} \ .
\label{eq:a2}
\ee
These virtual photon asymmetries, in general a function of $x$ and $Q^2$,
are related to the nucleon spin structure functions $g_1(x,Q^2)$ and 
$g_2(x,Q^2)$ via
\bea
A_1(x,Q^2) &=& {g_1(x,Q^2) - \gamma^2 g_2(x,Q^2)\over F_1(x,Q^2)} \ , \nonumber\\
A_2(x,Q^2) &=& {\gamma\left[ g_1(x,Q^2) + g_2(x,Q^2)\right]\over F_1(x,Q^2)} \ ,
\label{eq:a1-g1}
\eea
where $\gamma=2Mx/\sqrt{Q^2}$.

These virtual photon asymmetries can be related to measured lepton asymmetries
through polarization and kinematic factors.
The experimental 
longitudinal and transverse lepton asymmetries are defined as
\bea
A_{||} &=& {\sigma^{\uparrow\downarrow} - \sigma^{\uparrow\uparrow} \over \sigma^{\uparrow\downarrow}  + \sigma^{\uparrow\uparrow}} \nonumber \\
A_\perp &=& {\sigma_{\downarrow\rightarrow} - \sigma_{\uparrow\rightarrow}\over
\sigma_{\downarrow\rightarrow} + \sigma_{\uparrow\rightarrow}} \ ,
\eea
where $\sigma^{\uparrow\uparrow}$ ($\sigma^{\uparrow\downarrow}$) 
is the cross section for the lepton and
nucleon spins aligned (anti-aligned) longitudinally, 
while $\sigma_{\downarrow\rightarrow}$
($\sigma_{\uparrow\rightarrow}$) is the cross section for longitudinally
polarized lepton and transversely polarized nucleon.
The lepton asymmetries are then given in terms of 
the virtual photon asymmetries
through
\bea
A_{||} &=& D\left( A_1 + \eta A_2\right) \nonumber \\
A_\perp &=& d\left(A_2 - \zeta A_1\right) \ .
\eea
The virtual photon (de)polarization factor $D$ is approximately equal to 
$y\equiv\nu/E$ (where $\nu$ is the energy of the virtual photon and
$E$ is the lepton energy), but is given explicitly as 
\begin{equation}
D = [1-(1-y)\epsilon]/(1+\epsilon R) \ ,
\end{equation}
where $\epsilon$ is the magnitude of the virtual photon's transverse
polarization 
\begin{equation}
\epsilon = [4(1-y) - \gamma^2 y^2]/[2y^2+4(1-y)+\gamma^2 y^2] \ .
\end{equation}
and 
\begin{equation}
R = \sigma_L/\sigma_T 
\end{equation}
is the ratio of longitudinal to transverse virtual photon cross sections.

The other factors are given by 
\be
\eta = \epsilon \gamma y/[1-\epsilon(1-y)]
\ee
\be
d = D \sqrt{{2\epsilon\over 1+\epsilon}}
\ee
\be
\zeta = \eta\left({1+\epsilon\over 2\epsilon}\right) \ .
\ee

\subsection{Extraction of $g_1(x,Q^2)$ \label{subsec:extract} }

The nucleon structure function is extracted from measurements of the
lepton-nucleon longitudinal asymmetry (with longitudinally polarized
beam and target)
\be
A_{||} = {\sigma^{\uparrow\downarrow} - \sigma^{\uparrow\uparrow} \over \sigma^{\uparrow\downarrow}  + \sigma^{\uparrow\uparrow}} \ ,
\ee
where $\sigma^{\uparrow\uparrow}$ ($\sigma^{\uparrow\downarrow}$) 
represents the cross section when the electron and
nucleon spins are aligned (anti-aligned). These cross sections can also
be expressed in terms of spin-independent $\sigma_U$ 
and spin-dependent $\sigma_P$ cross sections
\bea
\sigma^{\uparrow\uparrow}=&\sigma_U + \sigma_P \\
\sigma^{\uparrow\downarrow}=&\sigma_U - \sigma_P \ .
\eea

In the limit of stable beam currents, target densities and polarizations,
the experimentally measured asymmetry $A_{\rm exp}$ 
is usually expressed in terms of the measured count rates $N$ and
the number of incident electrons $N_B$
\be
A_{\rm exp} = {N^{\uparrow\downarrow}/N_B^{\uparrow\downarrow} - N^{\uparrow\uparrow}/N_B^{\uparrow\uparrow} \over N^{\uparrow\downarrow}/N_B^{\uparrow\downarrow}  + N^{\uparrow\uparrow}/N_B^{\uparrow\uparrow}} \ .
\ee
$A_{||}$ is then determined via
\be
A_{||} = {A_{\rm exp}\over P_B P_T f} + \Delta_{\rm RC} \ ,
\ee
where $P_B$ and $P_T$ are the beam and target polarizations respectively,
$f$ is a dilution factor due to scattering from unpolarized material and
$\Delta_{RC}$ accounts for QED radiative effects~\cite{radcor}. 

If however there is a time variation of the beam or target polarization
or luminosity, the asymmetry
should be determined using
\be
A_{||} = {N^{\uparrow\downarrow}L^{\uparrow\uparrow} - N^{\uparrow\uparrow}L^{\uparrow\downarrow} \over f(N^{\uparrow\downarrow}L^{\uparrow\uparrow}_P + N^{\uparrow\uparrow}L^{\uparrow\downarrow}_P)} + \Delta_{\rm RC} \ ,
\ee
since in this case the measured count rates can be written in terms
of $\sigma_U$ and $\sigma_P$
\be
\begin{array}{rcl}
N^{\uparrow\uparrow} =& \sigma_U\int n_B^{\uparrow\uparrow}(t)dt + \sigma_P\int n_B^{\uparrow\uparrow}(t)P_B(t)P_T(t)dt\equiv \sigma_U L^{\uparrow\uparrow}+\sigma_P  L^{\uparrow\uparrow}_P\\
N^{\uparrow\downarrow} =& \sigma_U\int n_B^{\uparrow\downarrow}(t)dt - \sigma_P\int n_B^{\downarrow\uparrow}(t)P_B(t)P_T(t)dt\equiv \sigma_U L^{\uparrow\downarrow}+\sigma_P  L^{\uparrow\downarrow}_P \ ,
\label{asymform}
\end{array}
\ee
where now $n_B$ represents the product of beam current and target areal 
density - the luminosity.
In Eq.~\ref{asymform} we have ignored a factor accounting for the acceptance
and solid angle of the apparatus which is assumed to be independent of time.

The spin structure function $g_1(x,Q^2)$ can then be determined from the
longitudinal asymmetry $A_{||}(x,Q^2)$,
\begin{equation}
g_1 = {F_1 \over (1 + \gamma^2)}[ A_{||}/D + (\gamma - \eta) A_2] \ ,
\label{g1eq}
\end{equation}
where $F_1\equiv F_1(x,Q^2)$ is the unpolarized structure function. 
The unpolarized structure function $F_1$ is usually determined from 
measurements of the unpolarized structure function $F_2$ and $R$
using
\begin{equation}
F_1 = F_2 (1+\gamma^2)/(2x(1+R))  \ .
\end{equation}
To use the above equation we need an estimate for $A_2$.
$|A_2|$ is constrained to be less than $\sqrt{R}$~\cite{posit}, but
$A_2$ can also be determined from measurements (see Sec.~\ref{subsec:g2}) 
with a 
longitudinally polarized lepton beam and a transversely
polarized nucleon target (when combined with the longitudinal 
asymmetry). 

As a guide to the relative importance of various kinematic terms in the
above equations we present examples of the magnitude of these
terms in Table~\ref{tab:kin}, typical for the 
SMC and HERMES experiments. 

\begin{table}[!hbp]
\caption{Typical kinematic factors entering into the extraction of $g_1$. 
Examples from the SMC and HERMES experiments are given.}
\vspace{0.01cm}
\begin{center}
\begin{tabular}{|c|c|c|c|c|c|c|}
\hline
SMC \\
\hline
$<x>$ & $<Q^2>$ & $y$ & $\gamma$ & $\epsilon$ & $\gamma-\eta$ & $D$
\\ \hline
0.005&  1.30& 0.729& 0.008& 0.505& 0.005& 0.721\\
 0.008&  2.10& 0.736& 0.010& 0.493& 0.006& 0.745\\
 0.014&  3.60& 0.721& 0.014& 0.517& 0.008& 0.748\\
 0.025&  5.70& 0.639& 0.020& 0.638& 0.009& 0.671\\
 0.035&  7.80& 0.625& 0.024& 0.657& 0.011& 0.666\\
 0.049& 10.40& 0.595& 0.029& 0.695& 0.012& 0.643\\
 0.077& 14.90& 0.543& 0.037& 0.756& 0.014& 0.592\\
 0.122& 21.30& 0.490& 0.050& 0.809& 0.016& 0.545\\
 0.173& 27.80& 0.451& 0.062& 0.843& 0.018& 0.508\\
 0.242& 35.60& 0.413& 0.076& 0.873& 0.020& 0.468\\
 0.342& 45.90& 0.376& 0.095& 0.897& 0.022& 0.428\\
 0.480& 58.00& 0.339& 0.118& 0.919& 0.024& 0.384\\
\hline
HERMES \\
\hline
$<x>$ & $<Q^2>$ & $y$ & $\gamma$ & $\epsilon$ & $\gamma-\eta$ & $D$
\\ \hline
 0.023&  0.92& 0.775& 0.045& 0.427& 0.029& 0.778\\
 0.033&  1.11& 0.652& 0.059& 0.620& 0.028& 0.635\\
 0.047&  1.39& 0.573& 0.075& 0.721& 0.030& 0.547\\
 0.067&  1.73& 0.500& 0.096& 0.798& 0.032& 0.476\\
 0.095&  2.09& 0.426& 0.123& 0.861& 0.034& 0.405\\
 0.136&  2.44& 0.348& 0.163& 0.913& 0.035& 0.329\\
 0.193&  2.81& 0.282& 0.216& 0.945& 0.037& 0.268\\
 0.274&  3.35& 0.237& 0.281& 0.962& 0.040& 0.227\\
 0.389&  4.25& 0.212& 0.354& 0.969& 0.047& 0.208\\
 0.464&  4.80& 0.200& 0.397& 0.972& 0.050& 0.198\\
 0.550&  5.51& 0.194& 0.440& 0.973& 0.055& 0.195\\
 0.660&  7.36& 0.216& 0.457& 0.965& 0.065& 0.224\\
\hline
\end{tabular}
\end{center}
\label{tab:kin}
\end{table}

For extraction of the neutron structure function $g_1^n$ from nuclear targets,
eg. $^2$H and $^3$He, additional corrections must be applied. For the 
deuteron, the largest contribution is due to the polarized proton in
the polarized deuteron which must be subtracted. In addition 
a D-state admixture into the $p-n$ wave function will reduce
the deuteron spin structure function due to the opposite alignment of the
$p-n$ spin system in this orbital state; thus
\be
g_1^n(x,Q^2) = {2 g_1^d(x,Q^2)\over (1 - 1.5 \omega_D)} - g_1^p(x,Q^2)
\ee
where $\omega_D$ is the D-state probability of the deuteron. Typically
a value of $\omega_D = 0.05\pm 0.01$~\cite{lac81} is used for this correction.

For polarized $^3$He, a wavefunction correction for the neutron and proton
polarizations is applied using
\be
g_1^n(x,Q^2) = {1\over\rho_n}(g_1^{^3He} - 2 \rho_p g_1^p) \ ,
\ee 
where $\rho_n = (86\pm 2)\%$ and $\rho_p=(-2.8\pm 0.4)\%$ as taken 
from a number of calculations~\cite{friar,nuclcorr}. Additional
corrections due to the neutron binding energy and Fermi motion
have also been investigated~\cite{blank,nuclcorr,shulz} and shown 
to be relatively small.

\subsection{Recent Results for $g_1(x,Q^2)$ \label{subsec:recent} }

\begin{figure}[ht]
\includegraphics*[angle=0,width=4.in]{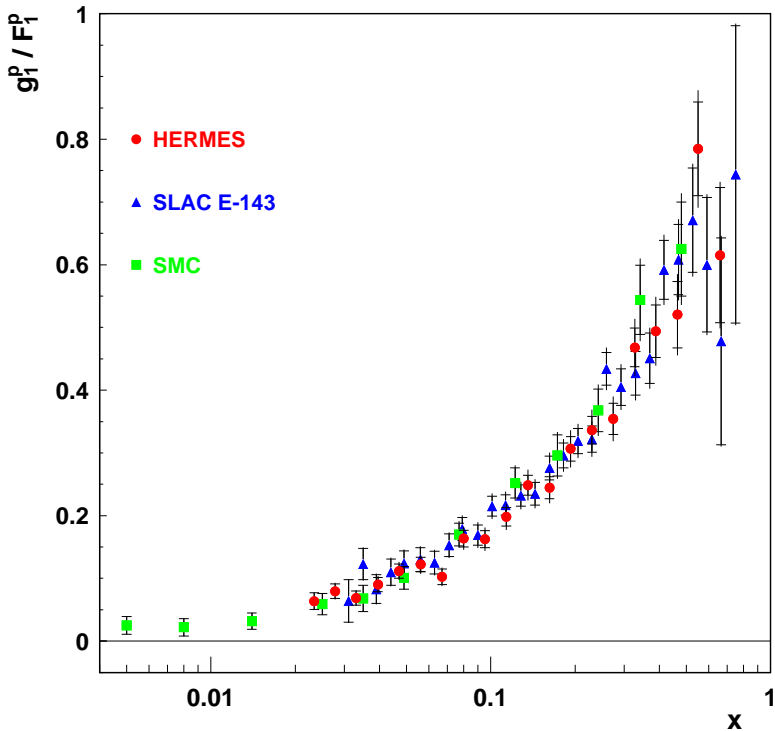}
\caption{Ratio of polarized to unpolarized proton structure function
from the SMC, E143 and HERMES experiments.}
\label{g1_f1}
\end{figure}

Most of the experiments listed in Table~\ref{tab:exps} have contributed
high precision data on the spin structure function $g_1(x,Q^2)$. Where
there is overlap (in $x$ and $Q^2$), the agreement between the 
experiments is extremely good. This can be seen in Fig.~\ref{g1_f1},
where the ratio of the polarized to unpolarized proton structure function
$g_1^p/F_1^p$ is shown. Analysis of the $Q^2$ dependence of this
ratio~\cite{e143q2} has shown that it is consistent experimentally 
with being independent of 
$Q^2$ within the range of existing experiments, although this behavior
is not expected to persist for all $Q^2$. 

\begin{figure}[!hbp]
\includegraphics*[angle=0,width=4.5in]{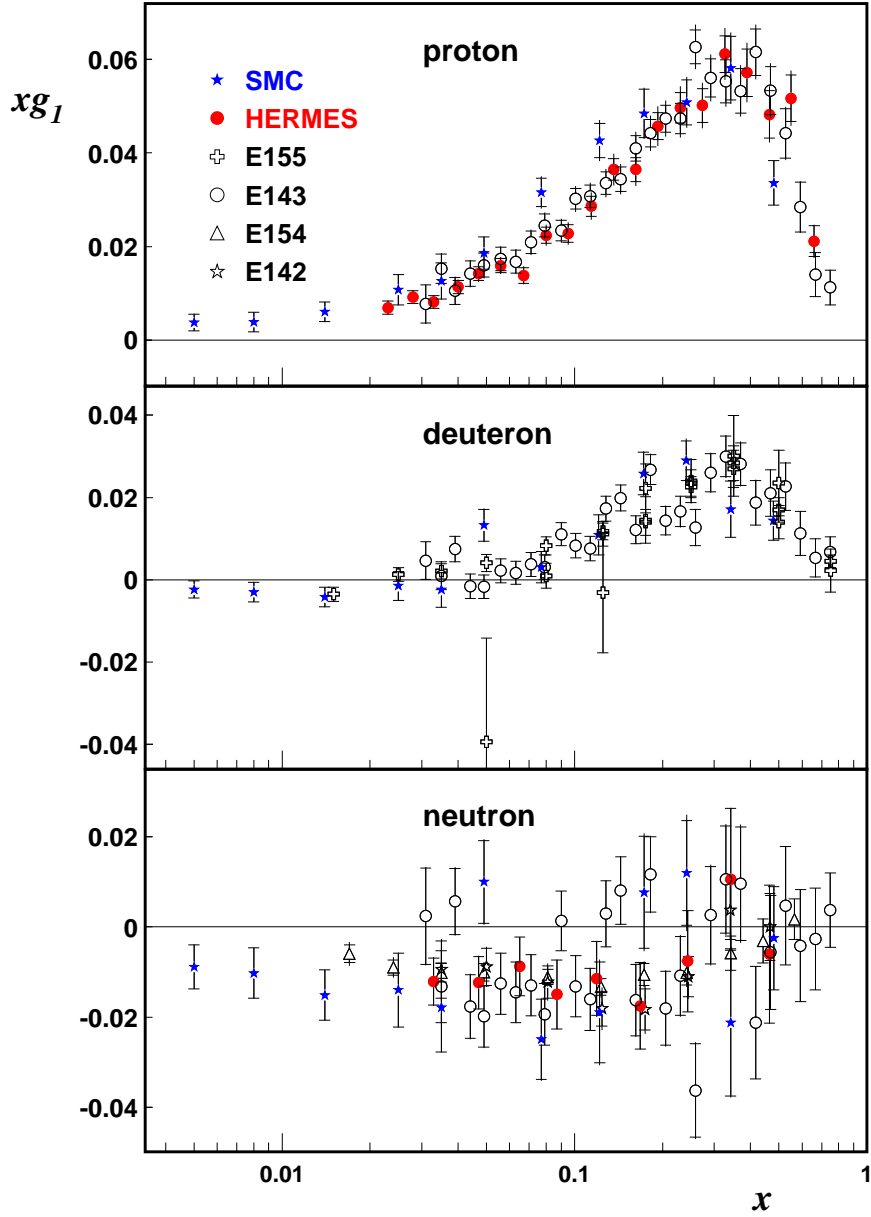}
\caption{Compilation of the world's data on $g_1^p, g_1^d, g_1^n$}
\label{g1world}
\end{figure}

A comparison of the spin structure functions $g_1^{p,d,n}$ are shown 
in Fig.~\ref{g1world}. Some residual $Q^2$ dependence is visible in 
the comparison of the SMC data with the other experiments. The
general $Q^2$ dependence of $g_1$ will be discussed in Sect. 3.5.

While the results shown in Figs.~\ref{g1_f1},\ref{g1world} correspond to 
$Q^2 > 1$ GeV$^2$, data also exists at lower $Q^2$, because of the 
large kinematic acceptance in many of the experiments. Much of this
data~\cite{e143q2,smcloq,hermloq},
when expressed as $g_1/F_1$ appears to be largely independent
of $Q^2$. 

\subsection{First Moments of $g_1(x,Q^2)$ \label{subsec:first} }

The initial interest in measurements of $g_1(x,Q^2)$ was in
comparing the measurements to several predicted sum rules, specifically
the Ellis-Jaffe and Bjorken sum rules. These sum rules relate 
integrals over the measured structure functions to measurements of 
neutron and hyperon beta-decay. 

The Ellis-Jaffe sum rule~\cite{ejsum} starts with the leading-order QPM result
for the integral of $g_1(x)$:
\be
\int_0^1 g_1(x) dx = {1\over 2}\int_0^1 \sum_i e^2_i \Delta q_i(x) dx \ ,
\label{g1lo} 
\ee
where the sum is over $u,d,s,\bar u, \bar d, \bar s$ for three active
quark flavors and the $Q^2$ dependence has been suppressed as it is 
absent in the simple QPM. Introducing the $SU(3)_f$ nucleon axial charges:
\bea 
a_0 = &(\Delta u + \Delta\bar u) + (\Delta d + \Delta\bar d) + 
(\Delta s + \Delta\bar s)\\
a_3 = &(\Delta u + \Delta\bar u) - (\Delta d\ + \Delta\bar d)\\
a_8 = &(\Delta u + \Delta\bar u) + (\Delta d  + \Delta\bar d) - 
2(\Delta s + \Delta\bar s) \ ,
\eea
where $\Delta q_i = \int_0^1 \Delta q_i (x) dx$, the Ellis-Jaffe sum 
rule then assumes that the strange quark and sea polarizations are
zero ($\Delta s = \Delta\bar q_i = 0$). 
Then for the proton and neutron integrals the Ellis-Jaffe sum
rule gives:
\bea
\Gamma_1^p \equiv \int_0^1 g_1^p(x) = &{3\over 36}a_3 + {1\over 36}a_8 + {4\over 36}a_0=0.186\pm0.004\nonumber \\ 
\Gamma_1^n \equiv \int_0^1 g_1^p(x) = &-{3\over 36}a_3 + {1\over 36}a_8 + {4\over 36}a_0=-0.025\pm0.004 \ .
\label{eq:ejsum}
\eea
To evaluate the integrals it is assumed that $a_0 = a_8$ which is true
if $\Delta s = 0$. Then $a_3$ is determined from 
the ratio of axial-vector to vector coupling 
constants in neutron decay $a_3 = -g_A/g_V = 1.2670 \pm 0.0035$~\cite{pdg00}.
A value for $a_8$ can be estimated with the additional assumption of SU(3) 
flavor symmetry which allows one to express $g_A/g_V$ for 
hyperon beta decays in terms of $a_3$ and $a_8$ (see Table~\ref{betadecay}),
giving $a_8 = 0.58\pm0.03$. Nucleon and hyperon beta decay is sometimes
parameterized in terms of the $F$ and $D$ coefficients. These coefficients
are related to the axial charges $a_3$ and $a_8$ with
\bea
a_3 &=& F + D
\nonumber\\
a_8 &=& 3 F - D \ . 
\eea

\begin{table}[ht]
\caption{Relation of the neutron and hyperon beta decays to the nucleon's
axial charges (as defined in the text) assuming $SU(3)_f$ symmetry.}
\vspace{0.01cm}
\begin{center}
\begin{tabular}{|c|c|c|}
\hline
Decay & $g_A/g_V$ in terms of Axial charges & Experimental~\cite{pdg00}
\\ \hline
$n\rightarrow p e_- \bar \nu_e$ & $-a_3$ &$ -1.2670\pm 0.0035$\\
$\Lambda\rightarrow p e^- \bar \nu_e$& $-{1\over 2}a_3 - {1\over 6}a_8$& $-0.718\pm0.015$\\
$\Sigma^- \rightarrow n e^- \bar \nu_e$& ${1\over 2}a_3 - {1\over 2}a_8$& $0.340\pm0.017$\\
$\Xi^-\rightarrow \Lambda e^- \bar \nu_e$& $- {1\over 3}a_8$& $-0.25\pm0.05$\\

\hline
\end{tabular}
\end{center}
\label{betadecay}
\end{table}

The assumptions implicit in the Ellis-Jaffe sum rule, eg. 
$\Delta s = \Delta\bar q_i = 0$ and $SU(3)_f$ symmetry, 
may be significantly violated. On the
contrary, the Bjorken sum rule~\cite{bjsum}
\be
\int_0^1 \left[ g_1^p(x) - g_1^n(x)\right] dx = {1\over 6}a_3 = 0.211\pm 0.001
\label{eq:bjsum}
\ee
requires only current algebra and isospin symmetry
(eg. $\Delta u^p = \Delta d^n$) in its derivation. Note that both
the Ellis-Jaffe and Bjorken sum rules must be 
corrected for QCD radiative corrections. For example, these corrections
have been evaluated up to order $\alpha_s^3$
~\cite{larver} and amount to $\sim 10\%$ correction for the
Ellis-Jaffe sum rule and $\sim 15\%$ correction for the Bjorken sum
rule at $Q^2 = 5$ GeV$^2$. 

Comparison of these predictions with experiment requires forming the
integrals of the measurements of $g_1(x,Q^2)$ over the full
$x$ range from $0\rightarrow 1$
at a fixed $Q^2$. Thus extrapolations are necessary in order to include 
regions of unmeasured $x$, both at high and low $x$. For the large $x$ 
region this is straightforward: since $g_1(x)$ is proportional to a
difference of quark
distributions it must approach zero as $x \rightarrow 1$ as this is
the observed behavior of the unpolarized distributions. However the
low $x$ region is problematic, as there is no clear dependence expected. 
In the first analyses simple extrapolations based on Regge 
parameterizations~\cite{regge,ek88} were used. Thus $g_1(x)$ was
assumed to be nearly constant for $x\rightarrow 0$. Later,
Next-to-Leading-Order (NLO) QCD calculations~\cite{ek95}
(see Sec.~\ref{subsec:nlo}) 
suggested that these parameterizations likely underestimated
the low $x$ contributions. The NLO calculations cannot predict the
actual $x$ dependence of the structure function, but can only take a 
given $x$ dependence and predict its dependence on $Q^2$. Thus by 
using the Regge parameterizations for low $Q^2\ltorder 1$, they can
give the low $x$ behavior at the $Q^2$ of the experiments, eg. 
$Q^2 \sim 3 - 10$ GeV$^2$. 

Evaluating the experimental integrals at a fixed $Q^2$ requires 
an extrapolation of the measured structure function. 
In general, for each experiment, 
the experimental acceptance imposes a correlation between $x$ and
$Q^2$ preventing a single experiment from measuring the
full range in $x$ at a constant value of $Q^2$. 
Thus the data must be QCD-evolved to a fixed value
of $Q^2$. This has often been done by exploiting the observed 
$Q^2$ independence of $g_1(x,Q^2)/F_1(x,Q^2)$ (see Fig.~\ref{g1_f1}). 
In this case most
of the $Q^2$ dependence of $g_1(x,Q^2)$ results from the
$Q^2$ dependence of the unpolarized structure function $F_1(x,Q^2)$
which is well measured in other experiments. Alternatively, NLO
QCD fits (as described in the next section) can be used to 
evolve the data sets to a common $Q^2$.

The E155 collaboration has recently reported~\cite{e155pd} 
a global analysis of 
spin structure function integrals. They have evolved the world data
set on $g_1^p(x,Q^2)$ and $g_1^n(x,Q^2)$ to
$Q^2 = 5$ GeV$^2$ and have extrapolated to low and to high $x$ using a NLO
fit to the data. Their results are compared in Table~\ref{sums}
with the predictions
for the Ellis-Jaffe and Bjorken sum rules 
(Eqs.~\ref{eq:ejsum},\ref{eq:bjsum})
including QCD radiative
corrections for $Q^2 = 5$ GeV$^2$ up to order $\alpha_s^3$ using the
calculations of Ref.~\cite{larver} and world-average for
$\alpha_s$~\cite{pdg00}. 
\begin{table}[ht]
\caption{Comparison of Sum Rule predictions including corrections up to
order $\alpha_s^3$ with a global analysis of the experiments.
}
\vspace{0.01cm}
\begin{center}
\begin{tabular}{|c|c|c|}
\hline
Sum Rule & Calculation& Experiment~\cite{e155pd} 
\\ \hline
EJ Sum $\Gamma_1^p(Q^2=5$ GeV$^2$) & $0.163\pm 0.004$  & $0.118\pm 0.004\pm 0.007$\\
EJ Sum $\Gamma_1^n(Q^2=5$ GeV$^2$) & $-0.019\pm 0.004$ & $-0.058\pm 0.005\pm 0.008$\\
Bj  $\Gamma_1^p-\Gamma_1^n(Q^2=5$ GeV$^2$) & $0.181\pm 0.005$ & $0.176\pm 0.003\pm 0.007$\\ 

\hline
\end{tabular}
\end{center}
\label{sums}
\end{table}

As seen in Table~\ref{sums} the Bjorken sum rule is well verified. In fact
some analyses~\cite{ek95} have assumed the validity of the Bjorken sum
rule and used the $Q^2$ dependence of $\Gamma_1^p - \Gamma_1^n$
to extract a useful value for $\alpha_s$. In contrast there is a strong 
violation of the Ellis-Jaffe sum rules. Many early analyses of these
results interpreted the violation in terms of a non-zero
value for $\Delta s$ (in which case $a_0 \ne a_8$), using only the
leading order QPM. However modern analyses have demonstrated that
a full NLO analysis is necessary in order to interpret the results.
This analysis will be described in the next section. Here, for 
completeness, we give the leading order QPM result. 

Within the leading order QPM,
$\Delta u+\Delta\bar u, \Delta d+\Delta\bar d$ and $\Delta s+\Delta\bar s$ 
can be determined by 
using Eqs.~\ref{eq:ejsum} with the experimental values from
Table~\ref{sums}. Dropping the assumption of
$a_0 = a_8$, but retaining the $SU(3)_f$ assumption to determine
$a_8$, one finds:
\be
\Delta u+\Delta\bar u = 0.78\pm0.03, \,\,\, \Delta d+\Delta\bar d = -0.48\pm0.03, \,\,\, \Delta s+\Delta\bar s = -0.14\pm 0.03
\ee
after applying the relevant QCD radiative corrections to the
terms in Eq.~\ref{eq:ejsum} (corresponding to a factor of 0.859 
multiplying the triplet $a_3$ and octet $a_8$ charges and a factor 
of 0.878 multiplying the singlet $a_0$ charge for $Q^2 = 5$ GeV$^2$). 
This then gives a very small value for the total quark contribution
to the nucleon's spin, $\Delta\Sigma = 0.16\pm 0.08$. 
Note that the quoted uncertainties reflect
only the uncertainty in the measured value of $\Gamma_1$ and not
possible systematic effects due to the assumption of $SU(3)_f$ 
symmetry and NLO effects. Studies of the effect of $SU(3)_f$
symmetry violations have been estimated~\cite{e143fin} to have little
effect on the uncertainty in $\Delta u$ and $\Delta d$, but can
increase the uncertainty on $\Delta s$ by a factor of two to three. 
NLO effects are the subject of the next section.

\subsection{Next-to-Leading Order Evolution of $g_1(x,Q^2)$ 
\label{subsec:nlo} }

As discussed above the spin structure functions possess a significant
$Q^2$ dependence due to QCD radiative effects. It is important to 
understand these effects for a number of reasons, including
comparison of different experiments, forming structure function 
integrals, parameterizing the data and obtaining sensitivity to
the gluon spin distribution. As the experiments are taken at
different accelerator facilities with differing beam energies the data 
span a range of $Q^2$. In addition, 
because of the extensive data set that has been accumulated and
the recently computed higher-order QCD corrections, it
is possible to produce parameterizations of the data based on
Next-to-Leading-Order (NLO) QCD fits to the data. This provides
important input to future experiments utilizing polarized beams
(eg. the RHIC spin program). 
These fits have also yielded some initial information
on the gluon spin distribution, because of the radiative effects that
couple the quark and gluon spin distributions at NLO. 

At NLO the QPM expression for the spin structure function becomes
\be
g_1(x,Q^2)={1\over 2}\sum_i e_i^2 C_q(x,\alpha_s)\otimes \Delta q_i(x,Q^2)
+{1\over N_f}C_g(x,\alpha_s)\otimes \Delta G(x,Q^2) \ ,
\label{g1nlo} 
\ee
where for three active quark flavors ($N_f = 3$) 
the sum is again over quarks and antiquarks: $u,d,s,\bar u,\bar d,\bar s$.
$C_q(x,\alpha_s)$ and $C_g(x,\alpha_s)$ are Wilson coefficients
and correspond to the polarized photon-quark
and photon-gluon hard scattering cross section respectively. 
The convolution $\otimes$ is defined as
\be
C(x,\alpha_s) \otimes q(x,Q^2) = \int_x^1 {dy\over y} C ({x\over y},\alpha_S) q(x,Q^2) \ . 
\ee
The explicit dependence of the nucleon spin structure function
on the gluon spin distribution is apparent in Eq.~\ref{g1nlo}. 
At Leading Order (LO) 
$C_q^0 = \delta(1-x)$ and $C_g^0 = 0$ and the usual dependence 
(Eq.~\ref{g1lo})
of the
spin structure function on the quark spin distributions emerges. 
At NLO however, the factorization between the 
quark spin distributions and coefficient functions 
shown in Eq.~\ref{g1nlo} cannot
be defined unambiguously. This is known as factorization 
scheme dependence and results from an ambiguity in how the
perturbative physics 
is divided between the definition of the
quark/gluon spin distributions and the coefficient functions. 
There are also ambiguities 
associated with the definition of the $\gamma_5$ matrix in $n$ 
dimensions~\cite{hv72} and in how to include the axial anomaly. 
This has lead to a variety of 
factorization schemes that deal with these ambiguities by different
means. 

We can classify the factorization schemes in terms of their treatment
of the higher order terms in the expansion of the 
coefficient functions. The $Q^2$ dependence of this expansion can
be written as:
\be
C_i(x,\alpha_s) = C_i^0(x) + {\alpha_s(Q^2)\over 2\pi}C_i^{(1)}+\cdots \ .
\label{eq:coeff}
\ee
In the so-called Modified-Minimal-Subtraction ($\overline{\rm MS}$) 
scheme~\cite{mn96,vog96}
the first moment of the NLO correction to $C_g$ vanishes
(i.e. $\int_0^1 C_g^{(1)} (x)dx = 0$), such that $\Delta G$ does not 
contribute to the first moment of $g_1$. In the Adler-Bardeen~\cite{bfr96,qcd5}
scheme (AB) the treatment of the 
axial anomaly causes the first moment of $C_g^{(1)}$ to be non-zero, leading
to a dependence of $\int g_1(x)dx$ on $\int \Delta G(x) dx$. This
then leads to a difference in the singlet quark distribution 
$\Delta \Sigma$ in the two schemes:
\bea
\Delta \Sigma(x,Q^2)_{\rm AB} &=& \Delta\Sigma(x,Q^2)_{\overline{\rm MS}} + N_f{\alpha_s(Q^2)\over 2\pi}\int_x^1 {dy\over y}\Delta G(y,Q^2)\nonumber\\ 
\Delta G(x,Q^2)_{\rm AB} &=& \Delta G(x,Q^2)_{\overline{\rm MS}} \ .
\eea

A third scheme, sometimes called the JET scheme~\cite{ccm88,lss98} or 
chirally invariant (CI) scheme~\cite{che98}, is also used. 
This scheme attempts to include all perturbative anomaly effects
into $C_g$. Of course any physical observables (eg. $g_1(x,Q^2)$) are 
independent of the choice of scheme. There are also straightforward 
transformations~\cite{qcd5,mt97,lss2-98} that relate the schemes and 
their results to one another. 

Once a choice of scheme is made the $Q^2$ dependence of $g_1$ can
be calculated using the Dokahitzer-Gribov-Lipatov-Altarelli-Parisi
(DGLAP)~\cite{dglap} equations. These equations
characterize the evolution of the spin distributions in terms of 
$Q^2$-dependent splitting functions $P_{ij}(x,\alpha_s)$:
\bea
{d\over d \ln Q^2}\Delta q_{NS}(x,Q^2)&=&{\alpha_s(Q^2)\over 2\pi}P_{qq}^{NS}
\otimes\Delta q_{NS}\nonumber \\
{d\over d \ln Q^2}
\left( 
\begin{array}{c}
\Delta \Sigma \\ \Delta G
\end{array}
\right ) &=& {\alpha_s(Q^2)\over 2\pi}\left (
\begin{array}{cc}
P_{qq}&P_{qg}\\ P_{gq} & P_{gg}
\end{array} \right )
\otimes
\left( 
\begin{array}{c}
\Delta \Sigma \\ \Delta G
\end{array}
\right ) \ ,
\eea
where the non-singlet quark distributions $\Delta q_{NS}(x,Q^2)$ for three
quark flavors are defined with
\be
\Delta q_{NS}(x,Q^2) = (\Delta u + \Delta\bar u) - 
{1\over 2}(\Delta d + \Delta\bar d) -
{1\over 2}(\Delta s + \Delta\bar s) \ .
\ee
The splitting functions $P_{ij}$
can be expanded in a form similar to that for
the coefficient functions $C_i(x,\alpha_s)$ in Eq.~\ref{eq:coeff} and
have been recently evaluated~\cite{mn96,vog96} in NLO. 

The remaining ingredients in providing a fit to the data are the choice
of starting momentum scale $Q_0^2$ and the form of the parton distributions
at this $Q_0^2$.
The momentum scale is usually chosen to be $\leq 1$ GeV$^2$ so that the
quark spin distributions are dominated by the valence quarks and
the gluon spin distribution is likely to be small. Also, as discussed
above, at lower momentum transfer some models for the $x$ dependence
of the distributions (eg. Regge-type models for the low $x$ region) 
are more reliable. The form of the polarized parton distributions at the
starting momentum scale are parameterized by a variety of $x$
dependences with various powers. This parameterization is the source
of some of the largest uncertainties as the $x$ dependence at low 
values of $x\le 0.003$ is largely unconstrained by the measurements. 
As an example, Ref.~\cite{qcd5} assumes for one of its fits that
the polarized parton 
distributions can be parameterized by
\be
\Delta q_i(x,Q^2_0) = A_i x^{\alpha_i}\left( 1 - x\right)^{\beta_i}
\left(1+\gamma_i x^{\delta_i}\right) \ .
\ee
With such a large number of parameters it is usually required to 
place additional constraints on some of the parameters. Often 
$SU(3)_f$ symmetry is used to constrain the parameters, or the
positivity of the distributions ($|\Delta q_i(x)| \le q_i(x)$)
is enforced (note that this positivity is strictly valid only 
when all orders are included; see Ref.~\cite{afr98}). Thus in other fits, 
the polarized distributions are
taken to be proportional to the unpolarized distributions as in
eg. Ref.~\cite{e155pd}:
\be
\Delta q_i(x,Q^2_0) = A_i x^{\alpha_i} q_i(x,Q^2_0) \ .
\ee

A large number of NLO fits have recently been published~\cite{grsv,
gs,bfr96,qcd5,smcnlo,e154nlo,bou98,gor98,lss2-98,lea98,lss99,
got00,e155pd,semifits}. 
These fits
include a wide variety of assumptions for the forms of the
polarized parton distributions, differences in factorization scheme 
and what data sets they 
include in the fit (only the most recent fits~\cite{got00} include 
all the published inclusive data). Some fits~\cite{semifits}
have even performed a NLO
analysis including information from semi-inclusive scattering (see 
Sec.~\ref{subsec:semi}). A comparison of the results from some 
of these recent fits is shown in Table~\ref{tab:NLOfit}. 
\begin{landscape}
\begin{table}[ht]
\caption{Results from NLO fits to data for first moments of quark and gluon
distributions. Missing data refers to data sets 
that are not included in the fits, PPDF refers to the assumptions for
Polarized Parton Distribution Functions and $Q^2_{ev}$ refers to the
evolved $Q^2$ where the first moments are evaluated. 
}
\vspace{0.01cm}
\begin{center}
\begin{tabular}{|c|c|c|c|c|c|c|c|}
\hline 
Reference & Scheme & $Q^2_0$ & Missing Data & PPDF & $Q^2_{ev}$ & $\Delta\Sigma$ & $\Delta G$ \\ 
          &        & GeV$^2$ &              &      &  GeV$^2$   &                &           
\\ \hline
ABFR98~\cite{qcd5} & AB  & 1 & HERMES(p) & $\Delta q_i\not\propto q_i$ & 1 & $0.41\pm 0.03$ & $0.95 \pm 0.18$ \\
(Fit-A)            &     &   & E155(pd)  &                             &   &                &                 \\
                   &    &   & Semi-inc  &                             &   &                &                 \\
\hline
LSS99~\cite{lss99}  & JET             & 1 & Semi-inc  & $\Delta q_i\propto q_i$     & 1 & $0.39\pm 0.04$ & $0.57 \pm 0.14$ \\
       & AB              & 1 &           &                             &   & $0.41\pm 0.04$ & $0.58 \pm 0.04$ \\
       & $\overline{\rm MS}$ & 1 &           &                             &   & $0.28\pm 0.04$ & $0.07 \pm 0.10$ \\
\hline
GOTO00~\cite{got00}   & $\overline{\rm MS}$ & 1 & Semi-inc  & $\Delta q_i\propto q_i$     & 1 & 0.050 & 0.53 \\
   (NLO-1)            &                     & 1 &           &                             & 5 & 0.054 & 0.86 \\
                      &                 & 1 &           &                             &10 & 0.055 & 1.0 \\
\hline
FS00~\cite{semifits}  & $\overline{\rm MS}$ & 0.5 &  --       & $\Delta q_i\propto q_i$     & 10 & 0.050 & 0.53 \\
   (ii)               & &     &           &                             &  &         &      \\
                      &                 &     &           &                             &  &         &      \\
\hline
\end{tabular}
\end{center}
\label{tab:NLOfit}
\end{table}
\end{landscape}
Note that in the JET and AB schemes $\Delta\Sigma$ includes a contribution
from $\Delta G$. Thus the overriding result of these fits is that 
the quark spin distribution $\Delta\Sigma$ is constrained between
$0.05 - 0.30$ but that the gluon distribution and its first moment
are largely unconstrained. The extracted value for 
$\Delta G(Q^2 = 5 \,{\rm GeV}^2)$
is typically positive but the corresponding uncertainty 
is often $50 - 100\%$ of the value. Note that the uncertainties listed in
Table~\ref{tab:NLOfit} are dependent on the assumptions used in the fits.

Estimates of the contribution from higher twist effects~\cite{bal90,jiunrau}
($1/Q^2$ corrections) suggest that the effects are relatively small at
the present experimental $Q^2$. This is further supported by the 
generally good fits that the NLO QCD calculations can achieve without
including possible higher-twist effects.

Lattice QCD calculations of the first moments and second
moments of the polarized spin
distributions are underway~\cite{fuk95,don95,goc96,gus99}. Agreement
with NLO fits to the data is reasonable for the quark contribution, 
although the Lattice calculations are not yet able to calculate the
gluon contribution.

\section{Individual Quark Helicity Distributions \label{sec:indiv} }

As shown in the last section, the inclusive lepton asymmetries 
generally provide spin structure information only for the sum over 
quark flavors. Access to the individual flavor contributions to
the nucleon spin requires assumptions including $SU(3)_f$ 
symmetry in the weak decay of the octet baryons (nucleons and
strange hyperons). 

Potentially more direct information on the individual contributions
of $u, d$, and $s$ quarks as well as the separate contributions of
valence and sea quarks is possible via semi-inclusive scattering. 
Here one or more hadrons in coincidence with the scattered lepton 
are detected. The charge of the hadron and its valence quark
composition provide sensitivity to the flavor of the struck quark
within the Quark-Parton Model (QPM). 

Semi-inclusive asymmetries also allow access to the third 
leading-order quark distribution $\delta q$ called transversity. 
Because of the 
chiral odd structure of this distribution function it is not
measurable in inclusive DIS. Transversity will be discussed in 
Sec.~\ref{subsec:trans}. Additionally, semi-inclusive asymmetries 
can provide a degree of selectivity for different reaction mechanisms 
that are sensitive to the gluon polarization. The sensitivity
of semi-inclusive asymmetries to the gluon polarization will be
discussed in Sect.~\ref{sec:gluon}. The flavor decomposition of the
nucleon spin
using semi-inclusive scattering will be discussed 
in the next two sections. 

\subsection{Semi-Inclusive Polarized Lepton Scattering 
\label{subsec:semi} }

Within the QPM, the cross section for leptoproduction
of a hadron (semi-inclusive scattering) can be expressed as
\be
{d\sigma^h\over dz}=\sigma_{DIS}(x,Q^2)\left[{\sum_i e_i^2 q_i(x,Q^2) D_i^h(z,Q^2)\over \sum_i e_i^2 q_i(x,Q^2)}\right] \ ,
\label{sigsemi}
\ee
where $\sigma_{DIS}$ is the inclusive DIS cross section, the
fragmentation function, $D_i^h(z,Q^2)$, is
the probability that the hadron $h$ originated from the struck quark of 
flavor $i$, $z = E_h/\nu$ is the hadron momentum fraction
and the sums are over quark and antiquark flavors 
$u, d, s, \bar u, \bar d, \bar s$. To maximize the sensitivity to the
struck ``current'' quark, kinematic cuts are imposed on the data
in order to suppress effects from target fragmentation. These cuts
typically correspond to $W^2 > 9 - 10$ GeV$^2$ and $z > 0.2$.

In general the fragmentation functions $D_i^h(z,Q^2)$ depend on both the 
quark flavor and the hadron type. In particular for a given hadron
$D_i^h \ne D_j^h$. This effect can be understood in terms of the
QPM: if the struck quark is a valence quark for a particular hadron,
it is more likely to fragment into that hadron (eg. 
$D_u^{\pi^+} > D_d^{\pi^+}$). A flavor sensitivity is therefore obtained as
is a sensitivity to the antiquarks (eg. $D_{\bar d}^{\pi^+} >
D_{d}^{\pi^+}$). 

Eq.~\ref{sigsemi} displays a factorization of the cross section into separate
$z$ and $x$ dependent terms. This is an assumption of the QPM
and must be experimentally tested. Measurements of unpolarized
hadron leptoproduction~\cite{EMCfrag}
have shown good agreement with the factorization hypothesis. 
Data from $e^+-e^- \rightarrow$ hadrons can also be used
to extract fragmentation functions~\cite{eefrag}. Both the
$Q^2$ and $z$ dependence of the fragmentation functions have been
parameterized within string models of fragmentation~\cite{pythia}
that are in reasonable agreement with the measurements. Recently the
$Q^2$ dependence of the fragmentation functions have been calculated
to NLO~\cite{nlofrag}.

Assuming factorization of the cross section as given in 
Eq.~\ref{sigsemi}, we can write the asymmetry for leptoproduction of
a hadron as 
\be 
A_1^h(x,Q^2) = {\sum_i e_i^2 \Delta q_i(x,Q^2) \int_z D_q^h(z,Q^2) dz \over \sum_i e_i^2 q_i(x,Q^2) \int_z D_i^h(z,Q^2) dz} \ .
\label{a1semi}
\ee

Due to parity conservation the fragmentation functions contain no
spin dependence as long as the final-state polarization of the hadron
is not measured (spin dependent fragmentation can be accessed through
the self-analyzing decay of $\Lambda$ - see Sec.~\ref{subsec:frag}).
By making measurements with $H, D$ and $^3$He targets for different 
final-state hadrons and assuming 
isospin symmetry of the quark distributions and fragmentation 
functions a system of linear equations can be constructed:
\begin{equation}
\left(
\begin{array}{c}
A_{\pi^+}^p \\
A_{\pi^-}^p \\
A_{K^+}^n \\
A_{K^+}^n \\
\vdots 
\end{array} 
\right)
= f[q_i(x),D_i^h]
\left(
\begin{array}{c}
\Delta u \\
\Delta d \\
\Delta s \\
\Delta \bar{u} \\
\vdots
\end{array}
\label{eq:semimat}
\right)
\end{equation}
and solved for the $\Delta q_i$. In these
equations, the unpolarized quark distributions are taken from
a variety of parameterizations (eg. Ref.~\cite{grv,cteq}) and the fragmentation
functions are taken from measurements~\cite{EMCfrag,eefrag} or 
parameterizations~\cite{pythia}.

\begin{figure}
\includegraphics*[angle=0,width=4.in]{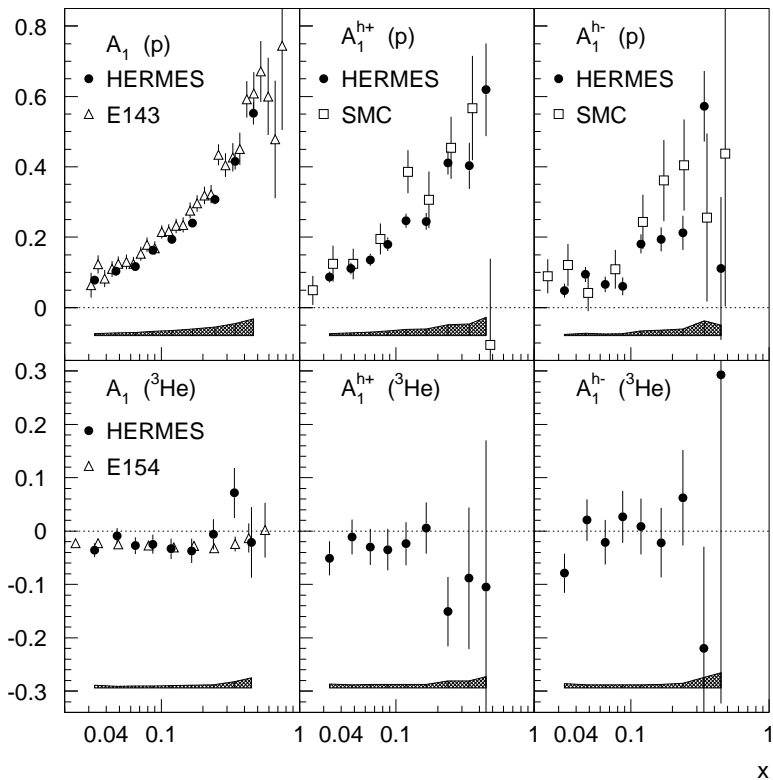}
\caption{Virtual photon asymmetries for semi-inclusive lepton scattering.
Inclusive asymmetries are shown in the leftmost panels for comparison.}
\label{semiasym}
\end{figure}

EMC, SMC and HERMES have made measurements of semi-inclusive asymmetries.
A comparison of the measurements from SMC and HERMES is shown in 
Fig.~\ref{semiasym}.
As the HERMES data are taken at $<Q^2> = 2.5$ GeV$^2$ and the SMC data
at $<Q^2> = 10$ GeV$^2$, these data suggest that the semi-inclusive
asymmetries are also approximately independent of $Q^2$.

It is important to note, especially for the lower $Q^2$ data of HERMES,
that Eq.~\ref{a1semi} must be modified if parameterizations of 
the unpolarized quark distributions are used. In some parameterizations
it is assumed that the unpolarized structure functions are 
related by the Callen-Gross approximation $F_1 = F_2/2x$ rather
than by the complete expression $ F_1 = F_2 (1+\gamma^2)/(2x(1+R))$.
Thus some experimental groups will present Eq.~\ref{a1semi} with
an extra factor of $(1+R)/(1+\gamma^2)$ included. 

Up to now results have only been reported for positively and negatively
charged hadrons (summing over $\pi$, $p$ and $K$) because of the lack of
sufficient particle identification in the experiments. This reduces
the sensitivity to some quark flavors (eg. strangeness) and requires 
additional assumptions about the flavor dependence of the 
sea quark and anti-quark distributions. Two assumptions have been used
to extract information on the flavor and sea dependence of the 
quark polarizations, namely
\be
{\Delta u_s(x)\over u_s(x)} = {\Delta d_s(x)\over d_s(x)} = {\Delta s(x)\over s(x)} = {\Delta \bar u(x)\over \bar u(x)} = {\Delta \bar d(x)\over \bar d(x)} = {\Delta \bar s(x)\over \bar s(x)} 
\label{asum1}
\ee
or
\be
\Delta u_s(x) = \Delta d_s(x) = \Delta s(x) = \Delta \bar u(x) = \Delta \bar d(x) =\Delta \bar s(x) \ .
\label{asum2}
\ee

\begin{figure}[!hbp]
\includegraphics*[angle=0,width=4.in]{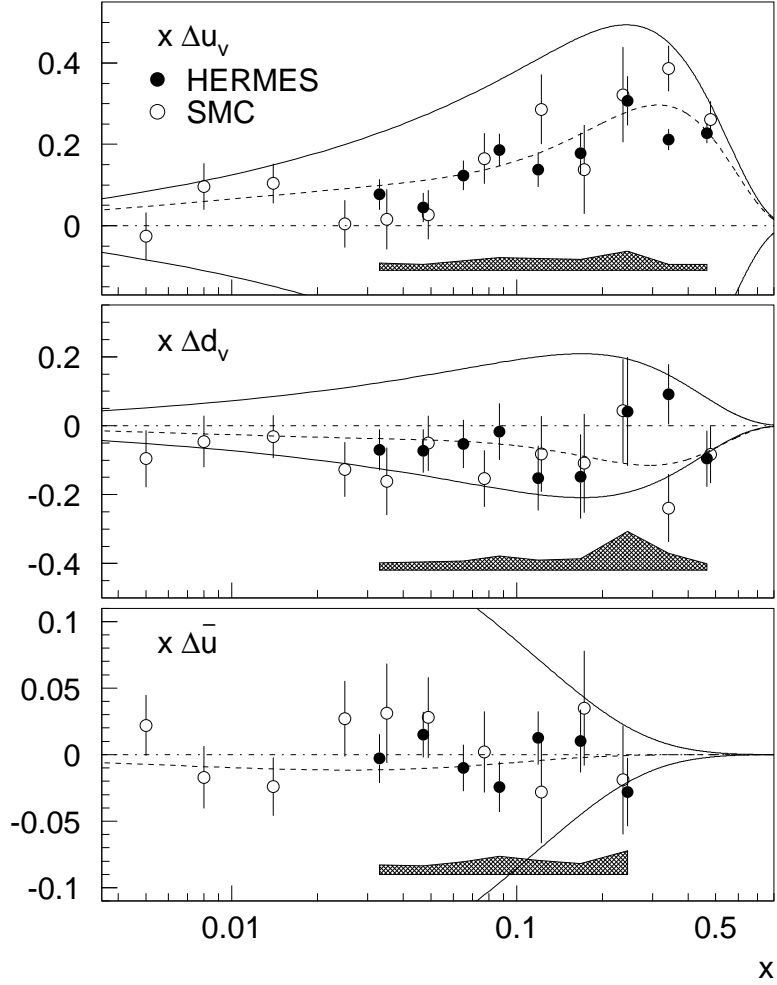}
\caption{Quark flavor spin structure from hadron leptoproduction
from the HERMES~\cite{hermessemi} 
and SMC~\cite{smcsemi1,smcsemi2} experiments. Both data sets
have been evolved to $<Q^2>=2.5$ GeV$^2$. Typical
systematic errors from Ref.~\cite{hermessemi}
are shown in the shaded band. The solid 
lines are the 
positivity limits corresponding to $\Delta q(x) = q(x)$. 
The dashed lines are the parameterization from Gehrmann and Stirling
(Gluon A-LO)~\cite{gs}.}
\label{semidelq}
\end{figure}

Here $\Delta u_s$ and $\Delta d_s$ represent the $u$ and $d$ sea quark 
spin distributions. A comparison of the extracted valence and
sea quark distributions 
from HERMES and SMC is shown in Fig.~\ref{semidelq}. The
valence distributions are defined 
using $\Delta q_v = \Delta q - \Delta \bar q$. 
Typical systematic errors are also shown in Fig.~\ref{semidelq}
and include the difference due to the two assumptions for the sea
distributions given by Eqs.~\ref{asum1}-~\ref{asum2}. The solid lines are 
positivity limits corresponding to $\Delta q(x) = q(x)$. 
The dashed lines are parameterizations from Gehrmann and Stirling
(Gluon A-LO)~\cite{gs}. 

Values for the integrals over the spin distributions from SMC and HERMES
are compared in Table~\ref{tab:deltaq}. The dominant sensitivity to 
$\Delta \bar u$ within the quark sea is due to the factor of two larger 
charge compared to $\bar d$ and $\bar s$. 

\begin{table}[ht]
\caption{Comparison of the first moment of separated quark spin
distributions as determined from semi-inclusive DIS lepton scattering.
Both statistical and systematic uncertainties are given.
}
\vspace{0.01cm}
\begin{center}
\begin{tabular}{|c|c|c|}
\hline
$\int_0^1 \Delta q_i(x)dx$ &  SMC results     & HERMES results\\
                           & $Q^2=10$ GeV$^2$ & $Q^2=2.5$ GeV$^2$
\\ \hline
$\Delta u_v$    & $ ~~~0.77 \pm 0.10 \pm 0.08$ & $ ~~~0.57 \pm 0.05 \pm 0.08$\\
$\Delta d_v$    & $  -0.52 \pm 0.14 \pm 0.09$ & $  -0.22 \pm 0.11 \pm 0.13$\\
$\Delta \bar u$ & $ ~~~0.01 \pm 0.04 \pm 0.03$ & $  -0.01 \pm 0.02 \pm 0.03$\\ 
\hline
\end{tabular}
\end{center}
\label{tab:deltaq}
\end{table}

While the experimental results presented in Table~\ref{tab:deltaq} 
have been extracted through a 
Leading-Order QCD analysis, NLO analyses are possible~\cite{seminlo}
and several such analyses have recently been published~\cite{semifits}.

Future measurements from HERMES and COMPASS will include full particle 
identification providing greater sensitivity to the flavor
separation of the quark spin distributions. In particular, due to
the presence of strange quarks in the $K$ valence quark distribution, 
$K$ identification is expected to give significant sensitity to
$\Delta s(x)$. 

\subsection{High Energy $\vec p - p$ Collisions 
\label{subsec:high} }

The production of weak $W^\pm$ bosons in high energy $\vec p - p$
collisions at RHIC provides unique sensitivity to the quark and
antiquark spin distributions. The maximal parity violation in the
interaction and the dependence of the production on the weak
charge of the quarks can be used in principle to select specific 
flavor and charge for the quarks. Thus the single spin longitudinal
asymmetry for $W^+$ production ($\vec p p \rightarrow W^+ X$) 
can be written~\cite{bs95}

\begin{figure}[!hbp]
\hspace{1in}\includegraphics*[angle=-90,width=3.in]{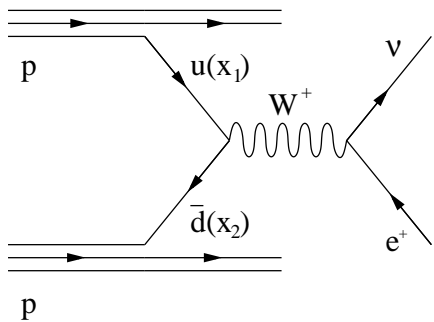}
\caption{W boson production in $\vec p p$ collisions.
}
\label{wprod}
\end{figure}
\begin{figure}[!hbp]
\includegraphics*[angle=0,width=3.in]{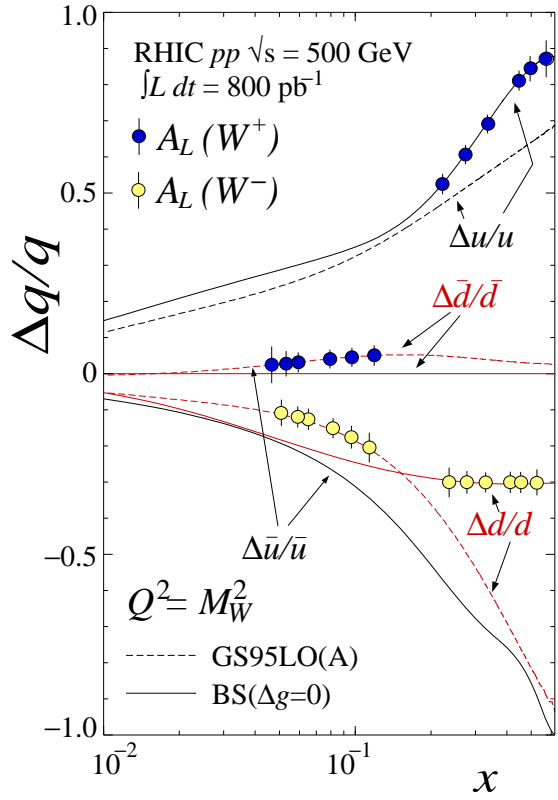}
\caption{Predicted sensitivity of the PHENIX detector at RHIC for 
measurement of the quark flavor spin contributions.}
\label{rhicdelq}
\end{figure}


\be
A_L = {\Delta u(x_1)\bar d(x_2) - \Delta \bar d(x_1)u(x_2)\over 
u(x_1)\bar d(x_2) + \bar d(x_1)u(x_2)} \ ,
\ee
where $x_1$ and $x_2$ refer to the $x$ value of the quark and
antiquark participating in the interaction (see for example 
Fig.~\ref{wprod}). Making the replacement $u\leftrightarrow d$ gives the
asymmetry for $W^-$ production. In the experiments the $W^\pm$
are detected through their decay to a 
charged lepton ($\mu^\pm$ in PHENIX and
$e^\pm$ in STAR) and the $x_1, x_2$ values are determined from
the angles and energies of those detected leptons. Thus
for $W^+$ production with $x_1\gg x_2$ the valence quarks are
selected for $x_1$ and $A_L(W^+)\sim \Delta u(x_1)/u(x_1)$, while for 
$x_1\ll x_2$ valence quarks are selected for $x_2$ and 
$A_L(W^+)\sim \Delta \bar d(x_1)/\bar d(x_1)$. Detection of $W^-$
then gives $\Delta \bar u/\bar u$ and $\Delta d/d$. An example
of the expected sensitivity of the PHENIX experiment after about
four years of data taking is shown in Fig.~\ref{rhicdelq}.

\section{Gluon Helicity Distribution \label{sec:gluon} }

As remarked in the Introduction, the gluon contribution to the
spin of the nucleon can be separated into spin and orbital parts.
As with its unpolarized counterpart, the polarized
gluon distribution is difficult to access 
experimentally. There exists no
theoretically clean and, at the same time,
experimentally straightforward hard scattering process 
to directly measure the distribution. 
In the last decade, many interesting 
ideas have been proposed and some have led to 
useful initial results from the present generation of
experiments; others will be tested soon at 
various facilities around the world. 

In the following subsections, we discuss a few representative
hard-scattering processes in which the gluon spin distribution
can be measured. 

\begin{figure}[t]
\begin{center}
\includegraphics*[angle=0,height=4cm]{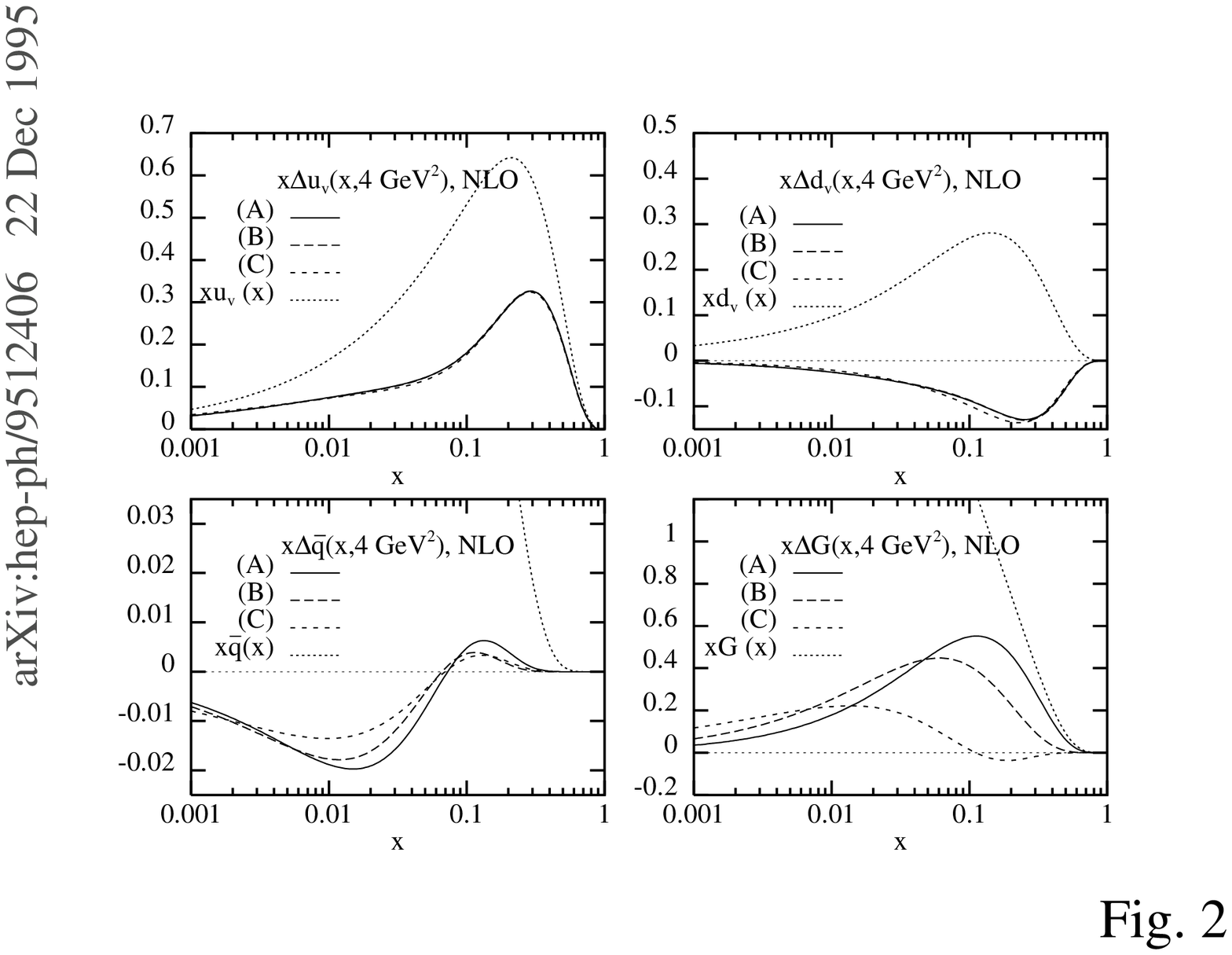}
\end{center}
\caption{Typical gluon helicity distributions~\cite{gs} obtained from 
fits to available polarized DIS data.}
\label{dgdis}
\end{figure}  

\subsection{$\Delta G(x)$ from QCD Scale Evolution \label{subsec:qcd} }

As discussed in Section 3.5, the polarized gluon 
distribution enters in the factorization formula 
for spin-dependent inclusive deep-inelastic scattering. 
Since the $g_1(x, Q^2)$ structure function involves both 
the singlet quark and gluon distributions as shown in Eq.~\ref{g1nlo}, 
only the $Q^2$ dependence of the data can be exploited
to separate them. The $Q^2$ dependence 
results from two different sources: the running coupling
$\alpha_s(Q^2)$ in the coefficient functions and 
the scale evolution of the parton distributions. As the
gluon contribution has its own characteristic $Q^2$ behavior, it
can be isolated in principle from data taken over
a wide range of $Q^2$.

Because the currently available experimental data have rather limited 
$Q^2$ coverage, there presently is a large uncertainty in extracting
the polarized gluon distribution. As described in 
Sec.~\ref{subsec:nlo}, a number of NLO fits to the world data
have been performed to extract the polarized parton
densities. While the results for the polarized quark densities are 
relatively stable, the extracted polarized gluon distribution
depends strongly on the assumptions made about the $x$-dependence
of the initial parameterization. Different fits produce results
at a fixed $x$ differing by an
order of magnitude and even the sign is not well
constrained.

Several sets of polarized gluon distributions
have been used widely in the literature 
for the purpose of estimating outcomes for future
experiments. An example from Ref.~\cite{gs} of 
the range of possible distributions is shown in Fig.~\ref{dgdis}. 
Of course the actual gluon distribution could be very
different from any of these.

\subsection{$\Delta G(x)$ from Di-jet Production
in $e - p$ Scattering \label{subsec:dijet}} 

\begin{figure}[t]
\begin{center}
\includegraphics*[height=4cm,angle=0]{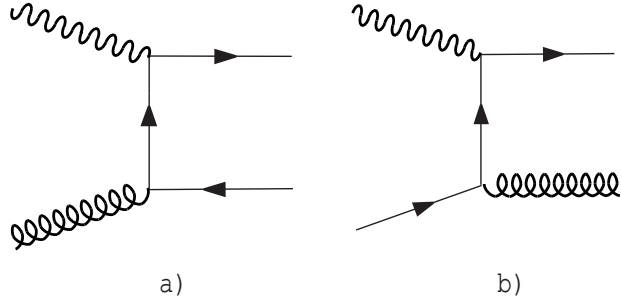}
\caption{Leading-order Feynman diagrams for di-jet
production in DIS: (a) Photon-Gluon Fusion, (b) Photon-Quark 
Compton scattering.}
\label{djfey}
\end{center}
\end{figure}

In lepton-nucleon deep-inelastic scattering, 
the virtual photon can produce two jets with large
transverse momenta from the nucleon target. 
To leading-order in $\alpha_s$, the underlying hard scattering
subprocesses are Photon-Gluon Fusion (PGF) and 
QCD Compton Scattering (QCDC) as shown in Fig.~\ref{djfey}.
If the initial photon has momentum $q$ and the parton from the 
nucleon (with momentum $P$) has momentum $xP$, the invariant mass
of the di-jet is $\hat s=(q+xP)^2$, the  $x$
at which the parton densities are probed is
\begin{equation}
   x_P = x_B\left( 1+ \hat s\over Q^2\right) \ . 
\end{equation}
where $x_B$ is the Bjorken $x$ variable.
Therefore the di-jet invariant mass fixes the parton
momentum fraction. Depending on the relative sizes 
of $\hat s$ and $Q^2$, $x_P$ can be an order of magnitude larger 
than $x_B$. 

If the contribution from the quark initiated subprocess
is small or the quark distribution is known, the two-jet 
production is a useful process to measure the gluon 
distribution. The di-jet invariant
mass provides direct control over the fraction of
the nucleon momentum carried by the gluon ($x_G = x_P$). 
Indeed, di-jet data from HERA have been used by the H1 and ZEUS
collaborations to extract the unpolarized 
gluon distribution~\cite{h195,zeus96}. With a polarized
beam and target, the process is ideal for probing 
the polarized gluon distribution.

The unpolarized di-jet cross section for photon-nucleon
collisions can be written as~\cite{deroeck96}
\begin{equation}
   \sigma_{\rm di-jet}(q,xP) = 
   \sigma_{\rm di-jet}^{\rm PGF}
  + \sigma_{\rm di-jet}^{\rm QCDC} 
  = AG(x) + B q(x) \ , 
\end{equation}
where $G(x)$ and $q(x)$ are the gluon and quark 
densities, respectively, and $A$ and $B$ are the hard scattering
cross sections calculable in perturbative QCD (pQCD).
Similarly, the polarized cross section can be written as
\begin{equation}
  \Delta\sigma_{\rm di-jet}(q,xP) = 
  \sigma^{++}_{\rm di-jet} - \sigma^{+-}_{\rm dijet}
  = a\Delta G(x) + b \Delta q(x) \ , 
\end{equation}
where the first and second $\pm$
refer to the helicities of the photon
and nucleon, respectively. The double spin asymmetry
for di-jet production is then
\begin{equation}
   A_{\rm di-jet} = 
   {\Delta \sigma_{\rm di-jet}
  \over 2\sigma_{\rm di-jet}}
   = {a\over A}{\Delta G(x)\over G(x)}{\sigma_{\rm
di-jet}^{\rm PGF}\over 2\sigma_{\rm di-jet}}
  + {b\over B}{\Delta q(x)\over q(x)} {1\over 2}\left(
  1- {\sigma_{\rm di-jet}^{\rm PGF}\over \sigma_{\rm di-jet}}
  \right) \ . 
\end{equation} 
The experimental asymmetry $A_{\rm exp}$ in DIS
is related to the photon asymmetry by
\begin{equation}
   A_{\rm exp} = P_e P_N D A_1^{\rm di-jet} \ , 
\end{equation}
where $P_e$ and $P_N$ are the electron and nucleon 
polarizations, respectively, and $D$ is the depolarization factor
of the photon. 

At low $x$, the gluon density dominates over the
quark density, and thus the photon-gluon fusion subprocess
dominates. There we simply have
\begin{equation}
A_1^{\rm di-jet} = {a\over A}{\Delta G(x)\over 2G(x)} \ , 
\end{equation}
which provides a direct
measurement of the gluon polarization.
Because of the
helicity selection rule, the photon and gluon must have
opposite helicities to produce a quark and antiquark pair
and hence $a/A=-1$. Therefore, if $\Delta G(x)$
is positive, the spin asymmetry must be negative. 
Leading-order calculations 
in Refs.~\cite{deroeck96,feltesse96,radel97,deroeck99}
show that the asymmetry is large and is strongly 
sensitivitive to the gluon polarization. 

At NLO, the one-loop corrections for the
PGF and QCDC subprocesses must be taken into account. 
In addition, three-jet events with two
of the jets too close to be resolved must be
treated as two-jet production. The sum of 
the virtual ($2\rightarrow 2$ processes with one loop)
and real ($2\rightarrow 3$ leading-order processes) 
corrections are independent of 
the infrared divergence. However, the two-jet
cross section now depends on the scheme in which the jets 
are defined. NLO calculations
carried out in Refs.~\cite{mirkes96,mirkes97,radel99},
show that the strong sensitivity of the cross section to 
the polarized gluon distribution survives. 
In terms of the spin asymmetry, the NLO effects do not 
significantly change the result.

Since the invariant mass of the di-jet is itself a large
mass scale, two-jet production can also be used to 
measure $\Delta G(x)$ even when
the virtuality of the photon is small or zero
(real photon). A great advantage of using nearly-real 
photons is that the cross section is large due to the 
infrared enhancement, and hence the statistics are high.  
An important disadvantage, however, is that there is 
now a contribution from the resolved photons. 
Because the photon is nearly on-shell, it has
a complicated hadronic structure of its own. The
structure can be described by quark and gluon 
distributions which have not yet been well determined
experimentally. Some models of the
spin-dependent parton distributions in the 
photon are discussed in Ref.~\cite{gluck92}. 
Leading-order calculations  
\cite{stratmann97,butterworthxx}
show that there are kinematic regions in which 
the resolved photon contribution is small and 
the experimental di-jet asymmetry can be used favorably 
to constrain the polarized gluon distribution.

\subsection{$\Delta G(x)$ from Large-$p_T$ 
Hadron Production in $e-p$ Scattering \label{subsec:hipt} }

For $e-p$ scattering at moderate center-of-mass energies,
such as in fixed target experiments, jets
are hard to identify because of their large angular
spread and the low hadron multiplicity. However one still
expects that the leading hadrons in the final
state reflect to a certain degree the original parton 
directions and flavors (discounting of course the 
transverse momentum, of order $\Lambda_{\rm QCD}$, from the 
parton intrinsic motion in hadrons and from their fragmentation).
If so, one can try to use the leading high-$p_T$ hadrons  
to tag the partons produced in the hard subprocesses considered 
in the previous subsection. 

Bravar et al.~\cite{bravar98} have proposed to use high-$p_T$ 
hadrons to gain access to $\Delta G(x)$.
To enhance the relative contribution from the
photon-gluon fusion subprocess and hence the
sensitivity of physical observables to the 
gluon distribution they propose a number of selection criteria
for analysis of the data and then test these ``cut'' criteria
in a Monte Carlo simulation of the COMPASS experiment.
These simulations show that these cuts are
effective in selecting the gluon-induced subprocess. 
Moreover, the spin asymmetry for the two-hadron production 
is large (10-20\%) and is strongly
sensitive to the gluon polarization.

Because of the large invariant mass of the hadron pairs, 
the underlying subprocesses can still be described
in perturbative QCD even if the virtuality of the photon is small
or zero~\cite{fontannaz81}. This enhances the data sample but 
introduces additional sub-processes to the high-$p_T$ hadron
production. The contribution from resolved photons, eg. from 
$\gamma\rightarrow \bar q q$ fluctuations, appears not to 
overwhelm the PGF contribution.
Photons can also fluctuate into $\rho$ mesons with
$\rho$-nucleon scattering yielding 
large-$p_T$ hadron pairs. Experimental information on this 
process can be used to subtract its contribution.
After taking into account these contributions, it appears that
the low-virtuality photons can be used as an effective
probe of the gluon distribution to complement the data from 
DIS lepton scattering.

\subsection{$\Delta G(x)$ from Open-charm (Heavy-quark) Production
in $e-p$ Scattering \label{subsec:open} }

\begin{figure}[t]
\begin{center}
\includegraphics*[clip=,height=4cm,angle=0]{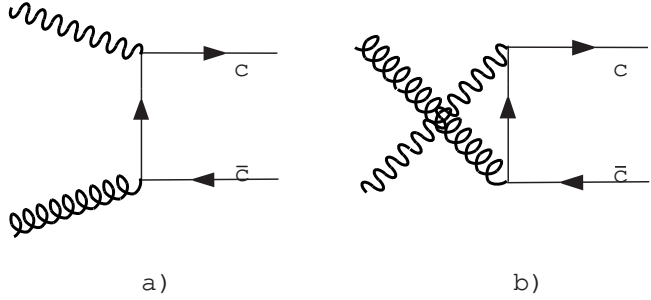}
\end{center}
\caption{Feynman diagrams for charm production via Photon Gluon
Fusion.}
\label{charmfey}
\end{figure}

Heavy quarks can be produced in $e-p$ scattering 
through photon-gluon fusion and can be calculated in 
pQCD (see Fig.~\ref{charmfey}). In the deep-inelastic scattering
region, the charm quark contribution to the $g_1(x, Q^2)$
structure function is known~\cite{gluck91},
\begin{equation}
     g_1^c(x, Q^2)
   = {\alpha_s(\mu^2)\over 9\pi}
  \int^1_{ax} {dy\over y}
  \Delta P\left({x\over y}, Q^2\right)\Delta G(y, \mu^2) \ , 
\end{equation}
where $a=1+4m_c^2/Q^2$, and
\begin{equation}
   \Delta P(x, Q^2)
  = (2x-1)\ln{1+\eta\over 1-\eta} +\eta(3-4x)  \ , 
\end{equation}
with $\eta^2 = 1-4m_c^2x/Q^2(1-x)$. This result assumes that,
because of the large charm quark mass, the direct charm contribution 
(eg. through $\Delta c(x)$) is small and the light-quark fragmentation 
production of charm mesons is suppressed. 
The $x$ dependence of the structure function, 
if measured, can be deconvoluted to give the polarized gluon
distribution. The renormalization scale $\mu$ can be taken 
to be twice the charm quark mass 2$m_c$.  

Following Ref.~\cite{compass}, 
the open charm electro-production cross section is large
when $Q^2$ is small or vanishes and can be written
\begin{equation}
   {d^2\sigma^{\mu N\rightarrow c\bar cX}
  \over dQ^2 d\nu} = \Gamma(E; Q^2, \nu)
   \sigma^{\gamma^*N\rightarrow c\bar c X}(Q^2, \nu) \ . 
\label{pcross}
\end{equation}
where the virtual photon flux is 
\begin{equation}
     \Gamma(E; Q^2, \nu)
   = {\alpha_{\rm em}\over 2\pi}
   { 2(1-y) + y^2 + Q^2/2E^2\over
     Q^2(Q^2+\nu^2)^{1/2}} \ , 
\end{equation}
$E$ and $\nu$ are the lepton and photon energies
and $y=\nu/E$. For a fixed $y$, the flux is inversely
proportional to $Q^2$. The second factor in Eq.~\ref{pcross}
is the photonucleon cross section.

The cross section asymmetry is the simplest 
at the real-photon point $Q^2=0$. The total parton 
cross section for photon-gluon
fusion is 
\begin{equation}
   \sigma(\hat s)= 
     {8\pi \alpha_{\rm em}\alpha_s(\hat s)
   \over 9\hat s}
 \left[-\beta(2-\beta^2) + {1\over 2}(3-\beta^4)
  \ln {1+\beta\over 1-\beta}\right] \ , 
\end{equation}
where $\beta= \sqrt{1-4m_c^2/\hat s}$
is the center-of-mass velocity of the charm quark, and
$\hat s = (q+x_G P)^2$ is the invariant mass of the photon-gluon
system. On the other hand, the spin-dependent cross section is
\begin{equation}
   \Delta \sigma
   = 
  {8\pi\alpha_{\rm em}\alpha_s(\hat s)\over 9\hat s}
   \left[3\beta-\ln {1+\beta\over 1-\beta}\right] \ . 
\end{equation}
The photon-nucleon asymmetry for open charm production can be 
obtained by convoluting the above cross sections with the gluon
distribution, giving 
\begin{equation}
     A_{\gamma N}^{c\bar c}(E,y) = 
      { \Delta \sigma^{\gamma N\rightarrow c\bar cX}
   \over \sigma^{\gamma N\rightarrow c\bar cX} } 
  = {\int_{4m_c^2}^{2MEy} d\hat s \Delta \hat \sigma(\hat s)
          \Delta G(x_G, \hat s)
    \over  \int_{4m_c^2}^{2MEy} d\hat s \hat \sigma(\hat s)
             G(x_G, \hat s) }  \ , 
\end{equation}
where $x_G = \hat s/2M_N E y$ is the gluon momentum
fraction. Ignoring the $Q^2$ dependence, 
the $l-P$ spin asymmetry is related to 
the photon-nucleon spin asymmetry by $A_{l N}^{c\bar c}
=DA^{c\bar c}_{\gamma N}$, where $D$ is the depolarization 
factor introduced before.

The NLO corrections have recently been 
calculated by Bojak and Stratmann~\cite{bojak99} and
Contogouris et al.~\cite{contogouris00}. The scale
uncertainty is considerably reduced in NLO, but 
the dependence on the precise value of the charm quark 
mass is sizable at fixed target energies. 

Besides the total charm cross section, one
can study the distributions of the cross section
in the transverse momentum or rapidity of the charm
quark. The benefit of doing this is that one can avoid
the region of small $x_G$ where the asymmetry 
is very small~\cite{stratmann97}.

Open charm production can be measured experimentally
by detecting $D^0$ mesons from charm quark fragmentation. 
On average, a charm quark has about 60\% probability
of fragmenting into a $D^0$. The $D^0$ meson can be 
reconstructed through its two-body decay mode
$D^0\rightarrow K^-+\pi^+$; the branching ratio
is about 4\%. Additional background reduction can 
be achieved by tagging $D^{*+} \rightarrow D^0 \pi^+$ through
detection of the additional $\pi^+$.

$J/\psi$ production is, in principle, also sensitive to the
gluon densities. However, because of ambiguities in the production
mechanisms~\cite{jap00}, any information on $\Delta G(x)$ is likely 
to be highly model-dependent.

\subsection{$\Delta G(x)$ from Direct Photon Production
in $p-p$ Collisions \label{subsec:direct} }

\begin{figure}[t]
\begin{center}
\includegraphics*[clip=,height=4cm,angle=0]{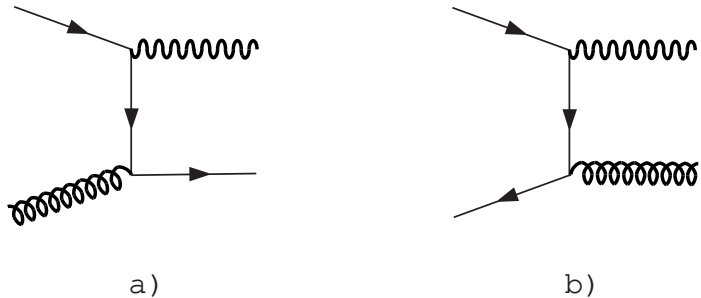}
\end{center}
\caption{Feynman diagrams for Direct photon production.}
\label{directp}
\end{figure}

$\Delta G(x)$ can be measured through direct (prompt)
photon production in proton-proton or 
proton-antiproton scattering~\cite{bergerqiu}.  
At tree level, the direct photon can be 
produced through two underlying 
subprocesses: Compton scattering 
$qg\rightarrow q\gamma$ and quark-antiquark 
annihilation $q\bar q\rightarrow \gamma g$, as shown in 
Fig. \ref{directp}.  
In proton-proton scattering, because the antiquark
distribution is small, direct photon production
is dominated by the Compton process and
hence can be used to extract the gluon distribution
directly.

Consider the
collision of hadron $A$ and $B$ with momenta
$P_A$ and $P_B$, respectively. The invariant
mass of the initial state is $s=(P_A+P_B)^2$.
Assume parton $a$ ($b$) from the hadron $A$
($B$) carries longitudinal momentum $x_aP_A$
($x_aP_B$). The Mandelstam variables for the 
parton subprocess $a+b\rightarrow \gamma+c$ are
\begin{equation}
    \hat s = x_ax_b s, ~~ \hat t = x_a t, ~~ \hat u = x_b u \ , 
\end{equation}
where we have neglected the hadron mass. 
The parton-model cross section for inclusive
direct-photon production is then
\begin{equation}
    E_\gamma {d\sigma_{\rm AB}\over d^3p_\gamma}
   = \sum_{ab} \int dx_a dx_b
   f_A^a(x_a, \mu^2) f_B^b(x_b, \mu^2)
    E_\gamma {d\hat \sigma_{ab}\over d^3 p_\gamma} \ .
\label{direct} 
\end{equation}
For the polarized cross section $\Delta \sigma_{\rm AB}$, 
the parton distributions $f_{A,B}$ are replaced
by polarized distributions $\Delta f_{A,B}$,  and
the parton cross sections $\hat \sigma_{ab}$
are replaced by the spin-dependent cross section
$\Delta \hat \sigma_{ab}$. The tree-level parton
scattering cross section is
\begin{equation}
  E_\gamma { d\hat \sigma \over d^3p_\gamma}
  = \alpha_{\rm em} \alpha_s {1\over \hat s}
         |M|^2 \delta(\hat s + \hat t + \hat u) \ , 
\label{pc}
\end{equation}
where the $\delta$-function reduces the parton
momentum integration into one integration over, say,
$x_a$ with range $[-u/(s+t),1]$ and
\begin{equation}
   |M|^2_{qg\rightarrow \gamma q} = 
  -{1\over 2} {\hat s^2+\hat t^2 \over \hat s \hat t}; ~~~
   |M|^2_{q\bar q\rightarrow \gamma g} = 
  {8\over 9} {\hat u^2+\hat t^2 \over \hat u \hat t} \ . 
\end{equation}
For the polarized case, we have the same expression as in Eq.~(\ref{pc})
but with 
\begin{equation}
   |\Delta M|^2_{qg\rightarrow \gamma q} = 
  -{1\over 2} {\hat s^2-\hat t^2 \over \hat s \hat t}; ~~~
   |\Delta M|^2_{q\bar q\rightarrow \gamma g} = -
  {8\over 9} {\hat u^2+\hat t^2 \over \hat u \hat t} \ . 
\end{equation}

In the energy region where the Compton subprocess 
is dominant, we can write the proton-proton cross
section in terms of the deep-inelastic structure
functions $F_2$ and $g_1$ and the gluon distributions 
$G$ and $\Delta G$~\cite{bergerqiu}, 
\begin{eqnarray}
   E_\gamma {d\sigma_{\rm AB} \over d^3 p_\gamma}
   &=& \int dx_a dx_b \left[
   {F_2(x_a, \mu^2) \over x_a}G(x_b, \mu^2)
         E_\gamma {d\hat \sigma_{qg}\over d^3 p_\gamma}
            + (x_a\rightarrow x_b)\right] \ , \nonumber \\
   E_\gamma {d\Delta\sigma_{\rm AB} \over d^3 p_\gamma}
   &=& \int dx_a dx_b \left[
   {2g_1(x_a, \mu^2) \over x_a}\Delta G(x_b, \mu^2)
         E_\gamma {d\Delta \hat \sigma_{qg}\over d^3 p_\gamma}
            + (x_a\rightarrow x_b)\right] \ . 
\end{eqnarray}
Here the factorization scale $\mu$ is usually taken
as the photon transverse momentum $p_T$. 

Unfortunately, the above simple picture of direct
photon production is complicated by high-order QCD
corrections. Starting at next-to-leading order
the inclusive direct-photon production cross section
is no longer well defined because of the infrared
divergence arising when the photon momentum 
is collinear with one of the final state partons.
To absorb this divergence, an additional term
must be added to Eq. (\ref{direct}) which 
represents the production of jets and 
their subsequent fragmentation into photons. Therefore, the 
total photon production cross section  
depends also on these unknown parton-to-photon 
fragmentation functions. Moreover, the separation into
direct photon and jet-fragmented photon is
scheme-dependent as the parton cross section 
$E_\gamma d\hat \sigma_{ab}/ d^3 p_\gamma$
depends on the methods of infrared subtraction~\cite{fri00}.

To minimize the influence of the fragmentation contribution, 
one can impose an isolation cut on the 
experimental data~\cite{ellis93}. Of course the parton cross section
entering Eq. (\ref{direct}) must be calculated in accordance
with the cut criteria. An isolation cut
has the additional benefit of excluding 
photons from $\pi^0$ or $\eta$ decay. When a 
high-energy $\pi^0$ decays, occasionally
the two photons cannot be resolved in a detector
or one of the photons may escape detection.
These backgrounds usually reside
in the cone of a jet and are largely excluded
when an isolation cut is imposed.

The NLO parton
cross sections in direct photon production 
have been calculated 
for both polarized and unpolarized scattering~\cite{fri00}.
Comparison between the experimental data
and theory for the latter case is still
controversial. While the collider
data at large $p_T$ are described very well
by the NLO QCD calculation~\cite{d0direct}, the fixed-target 
data and collider data at low-$p_T$
are under-predicted by theory. Phenomenologically,
this problem can be solved by introducing 
a broadening of the parton transverse
momentum in the initial state~\cite{owens99}. 
Theoretical ideas attempting to resolve the
discrepancy involve a resummation of threshold
corrections~\cite{lai98} as well as a resummation of double
logarithms involving the parton
transverse momentum~\cite{laenen98,catani98}. 
Recently, it has been shown that a combination of
both effects can reduce the discrepancy
considerably~\cite{laenen00}. 

\subsection{$\Delta G(x)$ from Jet and Hadron 
Production in $p-p$ Collisions \label{subsec:jet} }

Jets are produced copiously in high-energy hadron
colliders. The study of jets is now at a mature stage
as the comparison between experimental data 
from Tevatron and other facilities and 
the NLO QCD calculations are in 
excellent agreement. Therefore, single and/or 
di-jet production in polarized colliders can be
an excellent tool to measure the polarized parton 
distributions, particularly the gluon helicity 
distribution~\cite{soffer}.

There are many underlying subprocesses contributing
to leading-order jet production: $qq'\rightarrow
qq'$, $q\bar q'\rightarrow q\bar q'$, $qq\rightarrow
qq$, $q\bar q\rightarrow q'\bar q'$, $q\bar q\rightarrow
q\bar q$, $q\bar q\rightarrow gg$, $gg\rightarrow
q\bar q$, $qg\rightarrow qg$, $gg\rightarrow gg$. 
Summing over all pairs of initial partons $ab$
and subprocess channels $ab\rightarrow cd$, and folding
in the parton distributions $f_{a/A}(x_a)$, etc., in the
initial hadrons $A$ and $B$, the net two-jet
cross section is
\begin{equation}
  {d\sigma\over d^3p_c}
  = \sum_{abcd}\int dx_a dx_b f_{a/A}(x_a)
   f_{b/B}(x_b) {d\hat \sigma \over  d^3p_c}(ab\rightarrow cd) \ . 
\end{equation}
For jets with large transverse momentum, it is clear
that the valence quarks dominate the production. 
However, for intermediate and small transverse momentum,
jet production is overwhelmed by gluon-initiated 
subprocesses. 

Studies of the NLO corrections are important
in jet production because the QCD structure of the jets
starts at this order. For polarized
scattering, this has
been investigated in a Monte Carlo simulation recently
\cite{florian99}. The main result of the 
study shows that the scale dependence is
greatly reduced. Even though the jet asymmetry is 
small, because of the large abundance of jets, the statistical
error is actually very small.

Besides jets, one can also look for leading
hadron production, just as in  
electroproduction considered previously. This is useful 
particularly when jet construction is difficult 
due to the limited geometrical coverage of the detectors. 
One generally expects that the hadron-production asymmetry has 
the same level of sensitivity to the gluon density 
as the jet asymmetry.

\subsection{Experimental Measurements \label{subsec:exp} }

The first information on $\Delta G$
has come from NLO fits to inclusive 
deep-inelastic scattering data as discussed in 
Sec.~\ref{subsec:nlo}. Also recent semi-inclusive data 
from the HERMES experiment
indicates a positive gluon polarization 
at a moderate $x_G$. Future measurements
from COMPASS at CERN, polarized RHIC, and
polarized HERA promise to provide much
more accurate data. 

\subsubsection{Inclusive DIS Scattering \label{subsubsec:incl} }

As discussed in Sec.~\ref{subsec:nlo}, the spin-dependent
structure function $g_1(x, Q^2)$ is sensitive to the
gluon distribution at NLO. However, to extract the gluon 
distribution, which appears as an additive
term, one relies on the different
$Q^2$-dependence of the quark and gluon
contributions. 

The biggest uncertainty in the procedure of the NLO fits
is the parametric form of the gluon distribution at $Q_0^2$.
It is known that by taking different 
parameterizations, one can get quite different
results.

\subsubsection{HERMES Semi-inclusive Scattering \label{subsubsec:semi} }

The HERMES experiment has been described in Sec.~\ref{subsec:desy}. 
In a recent publication~\cite{hermeshipt}, the HERMES
collaboration reported a first measurement of the longitudinal 
spin asymmetry $A_{||}=-0.28\pm 0.12\pm 0.02$ in the photoproduction
of pairs of hadrons with high transverse momentum
$p_T$, which translate into a $\langle \Delta G/G\rangle
=0.41\pm0.18\pm0.03$ at an average  $\langle x_G\rangle
=0.17$. 

\begin{figure}[t]
\begin{center}
\includegraphics*[clip=,height=5cm,angle=0]{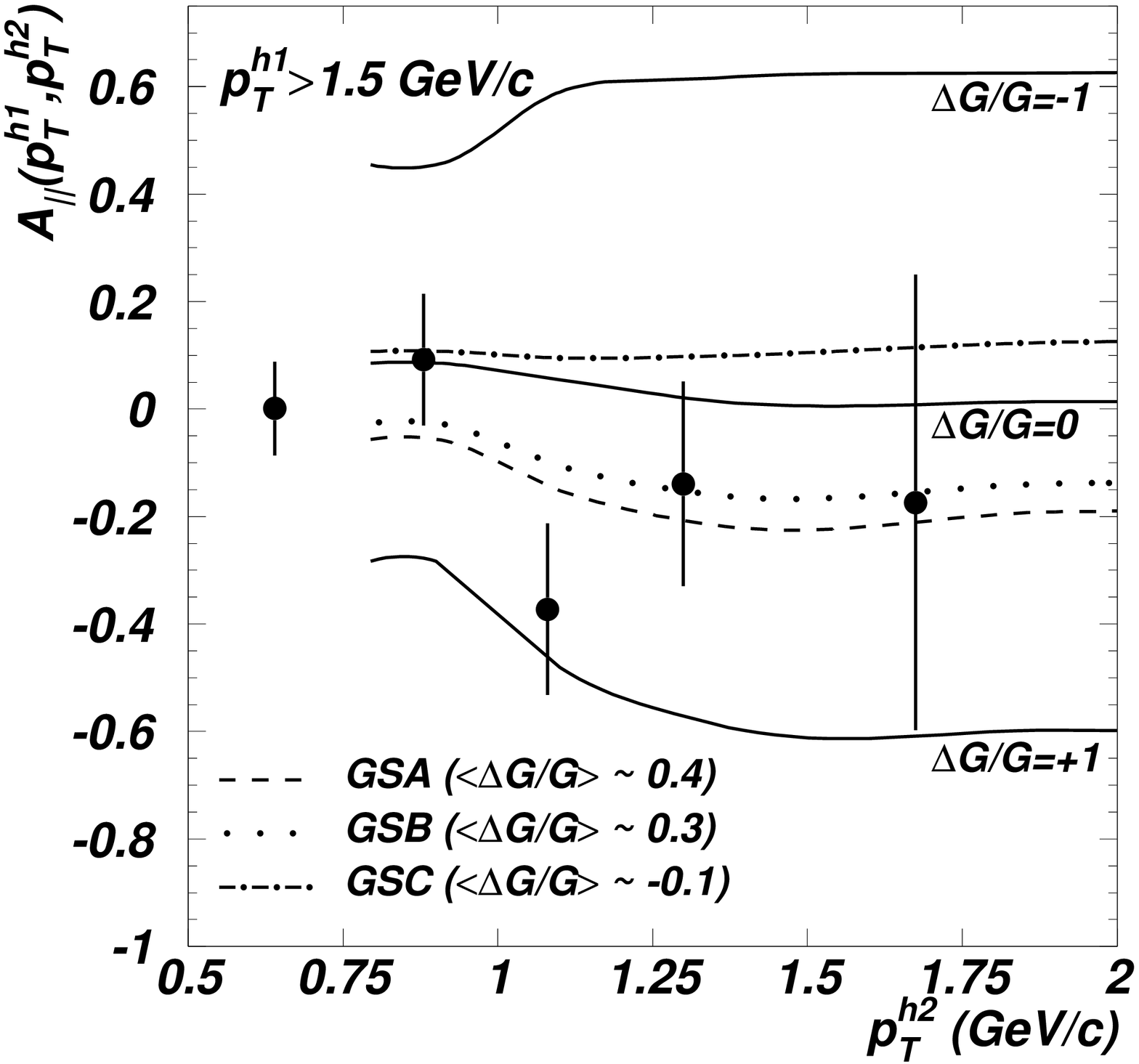}
\hspace{0.3in}
\includegraphics*[clip=,height=5cm,angle=0]{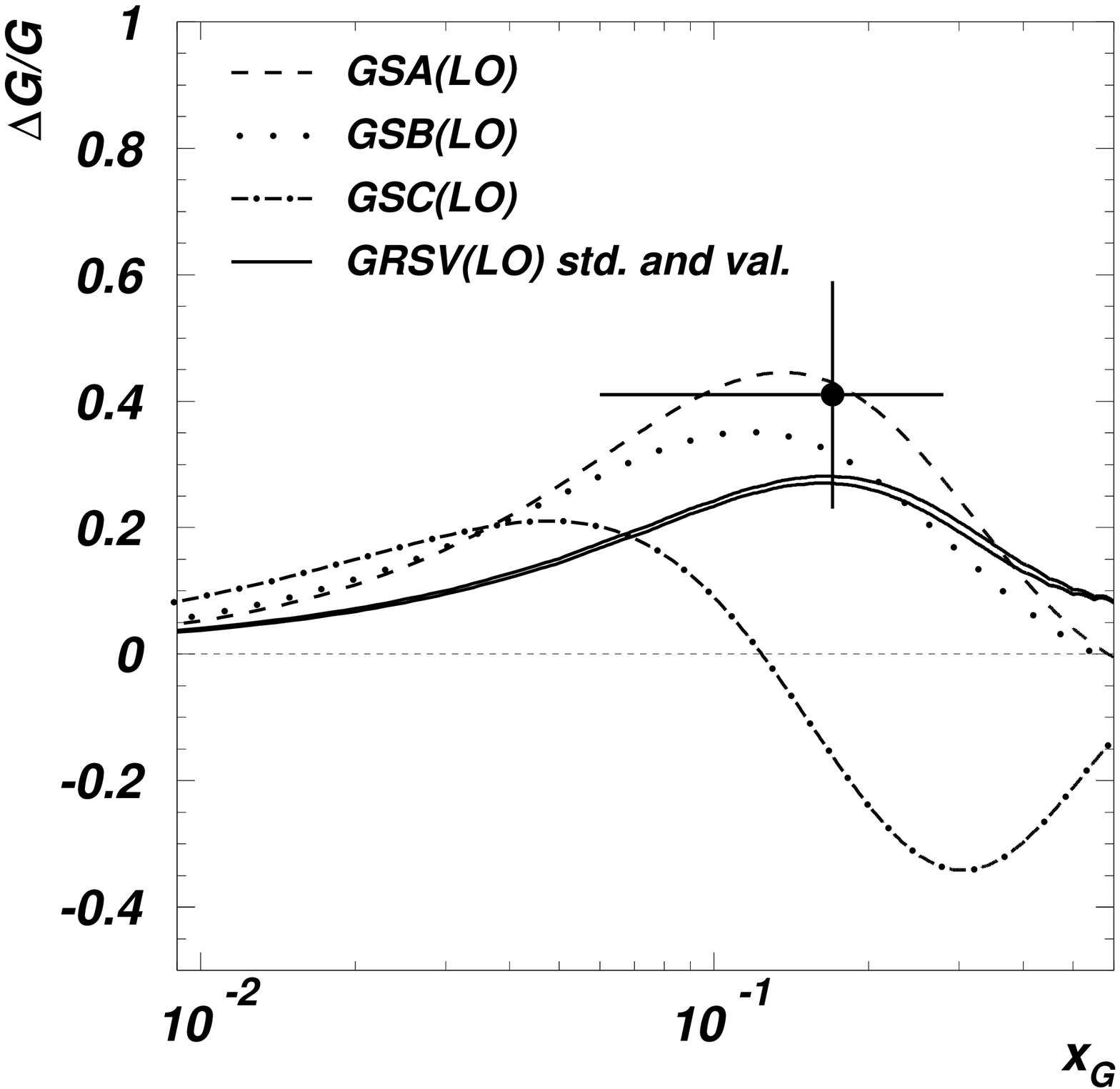}
\end{center}
\caption{Data for high $p_T$ hadrons from the HERMES 
experiment~\cite{hermeshipt} showing
(left) the asymmetry and (right) 
the extracted gluon polarization.}
\label{hermesfig}
\end{figure}

Following the proposal of Ref.~\cite{bravar98}, 
the data sample contains hadron pairs with opposite 
electric charge. The momentum of the hadron is
required to be above 4.5 GeV/c with a transverse 
component above 0.5 GeV/c. The minimum value of
the invariant mass of the two hadrons, in the case
of two pions, is 1.0 GeV/c$^2$. A nonzero asymmetry
is observed if the pairs with $p_T^{h_1}>1.5$ GeV/c
and $p_T^{h_2}>1.0$ GeV/c are selected. The measured
asymmetry is shown in Fig.~\ref{hermesfig} with an average $Q^2$ of
0.06(GeV/c)$^2$. If $p_T^{h_1}>1.5$
GeV/c is not enforced the asymmetry is consistent 
with zero. 

The measured asymmetry was interpreted in terms of the
following processes: lowest-order deep-inelastic
scattering, vector-dominance of the photon, resolved
photon, and hard QCD processes -- Photon Gluon Fusion
and QCD Compton effects. The PYTHIA~\cite{pythia} Monte Carlo generator
was used to provide a model for the data. In the
region of phase space where a negative asymmetry
is observed, the simulated cross section is dominated
by photon gluon fusion. The sensitivity of the measured
asymmetry to the polarized gluon distribution
is also shown Fig.~\ref{hermesfig}. Note that the analysis does not
include NLO contributions which could be important.  
The HERMES collaboration will have more data on this
process in the near future.

\subsubsection{COMPASS Experiment \label{subsubsec:compass} }

The COMPASS expriment at CERN will use a 
high-energy (up to 200 GeV) muon beam
to perform deep-inelastic scattering 
on nucleon targets, detecting final
state hadron production~\cite{compass}. The main goal of the
experiment is to measure the cross section
asymmetry for open charm production to
extract the gluon polarization $\Delta G$. 

For the charm production process, COMPASS estimates a 
charm production cross section 
of 200 to 350 nb. With a luminosity 
of $4.3\times 10^{37}$ cm$^{-2}$day$^{-1}$, 
they predict about 82,000 charm events in this kinematic
region per day. Taking into account branching
ratios, the geometrical acceptance and target
rescattering, etc., 900 of these events can be
reconstructed per day. The number of background events
is on the order of 3000 per day. Therefore the
total statistical error on the spin asymmetry
will be about $\delta A_{\gamma N}^{c\bar c} 
= 0.076$. 

\begin{figure}[t]
\begin{center}
\includegraphics*[clip=,height=4cm,angle=0]{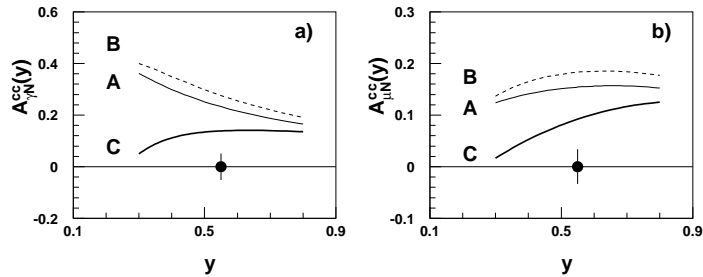}
\end{center}
\caption{Predictions of the open charm asymmetry for the COMPASS 
experiment. The curves are predictions for 
three representative gluon spin
distributions~\cite{gs}. 
Also shown are typical error bars expected from the 
measurement.}
\label{compassasy}
\end{figure}  

Shown in Fig.~\ref{compassasy} are the 
predicted asymmetries $A_{\gamma N}^{c\bar c}$
and $A_{\mu N}^{c\bar c}$ for open charm production as a
function of $y$. The curves correspond
to three different models for $\Delta G$. 
From the results at different $y$, one hopes
to get some information about the variation
of $\Delta G$ as a function of $x$. Measurements with 
high $p_T$ hadrons will also be used to complement the information
from charm production. 

\subsubsection{$\Delta G(x)$ from RHIC Spin Experiments 
\label{subsubsec:rhic} }

One of 
the primary goals of the RHIC spin experiments is to determine
the polarized gluon distribution. This can be 
done with direct photon, jet, and heavy quark production. 

\begin{figure}[t]
\begin{center}
\includegraphics*[clip=,height=5cm,angle=0]{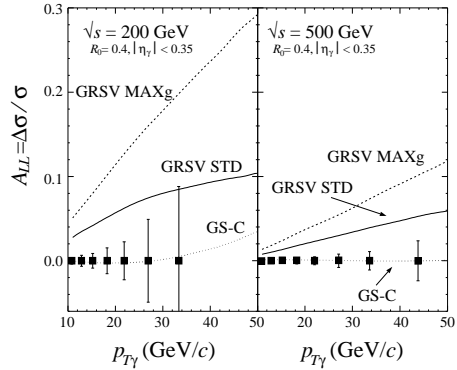}
\end{center}
\caption{Projected sensitivity~\cite{fri00} to the gluon polarization 
from direct photon production at RHIC with the PHENIX detector.
The curves represent different assumptions for the gluon 
spin distribution~\cite{gs,grsv}.}
\label{rhicfig}
\end{figure}  

Direct photon production is unique at RHIC. 
This can either be done on inclusive direct photon events 
(PHENIX) or photon-plus-jet events (STAR). Estimates of 
the background
from $q\bar q$ annihilation show a small effect. 
Shown in Fig. \ref{rhicfig} is the sensitivity of STAR
measurements of $\Delta G(x)$ in the channel
$\vec{p}\vec{p}\rightarrow \gamma+{\rm jet}+X$. The solid line
is the input distribution and the data points represent
the reconstructed $\Delta G(x)$. For inclusive direct photon
events, simulations show very different spin asymmetries from 
different spin-dependent gluon densities.

Jet and heavy flavor productions are also favorable channels
to measure polarized gluons at RHIC. The interested reader
can consult the recent review in Ref.~\cite{rhicspin}.

\subsubsection{$\Delta G(x)$ from Polarized HERA 
\label{subsubsec:phera} }

The idea of a polarized HERA collider ($\vec e^- - \vec p$)
has been described in Sec 2.4. 
Here we highlight a few experiments which can provide a good 
measurement of the polarized gluon distribution~\cite{phera}.

First of all, polarized HERA will provide access to 
very large $Q^2$ and low $x$ regions compared with fixed-target 
experiments. At large $x$, $Q^2$ can be as 
large as $10^4$ GeV$^2$. Thus, one can probe 
the gluon distribution through the $Q^2$ variation
of the $g_1$ structure function.  An estimate from an NLO
pQCD analysis shows that the polarized HERA
data on $g_1$ can reduce
the uncertainty on the total gluon helicity to 
$\pm 0.2$(exp)$\pm0.3$(theory). 

\begin{figure}[t]
\begin{center}
\vspace{0.3in}
\includegraphics*[clip=,height=5cm,angle=0]{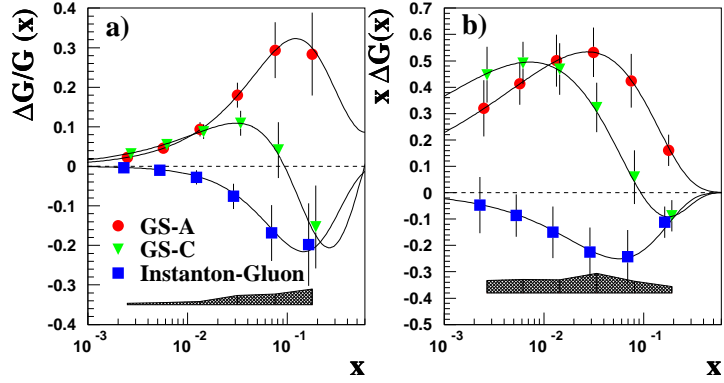}
\end{center}
\caption{Projected statistical and systematic error bars for the 
polarized gluon distribution
using two-jet production with polarized $e-p$ collisions at 
HERA. The curves 
represent different assumptions about the gluon spin 
distribution~\cite{gs}.}
\label{herafig}
\end{figure}  

Polarized HERA can also measure the polarized 
gluon distribution through di-jet production. Assuming
a luminosity 500 $pb^{-1}$ and with the event selection
criteria $5<Q^2<100$ GeV$^2$, $0.3<y<0.85$ and
$p_T^{\rm jet}> 5$ GeV, the expected error bars on the extracted
$\Delta G(x)$ are shown in Fig.~\ref{herafig}. 
The measured $x$ region covers
$0.002<x_G<0.2$. $\Delta G(x)$ can also be measured 
at polarized HERA through high-$p_T$
hadrons and jet production with real photons. 

\section{Transverse Spin Physics \label{sec:trans} }

\subsection{The $g_2(x, Q^2)$ Structure Function of the Nucleon 
\label{subsec:g2} }

As discussed in Sec.~\ref{subsec:extract}, 
the structure function $g_2(x, Q^2)$
can be measured with a longitudinally polarized lepton beam 
incident on a transversely polarized nucleon target. For many years, 
theorists have searched for a physical interpretation of
$g_2(x, Q^2)$ in terms of a generalization of 
Feynman's parton model~\cite{g2,g2parton}, 
as most of the known high-energy processes 
can be understood in terms of {\it incoherent} 
scattering of massless, on-shell and collinear partons~\cite{g2}. 
It turns out, however, that $g_2$ is an example of 
a higher-twist structure function.

Higher-twist processes cannot be understood in 
terms of the simple parton model~\cite{efp}. Instead, one 
has to consider parton correlations initially present in 
the participating hadrons.  Higher-twist processes can be described
in terms of {\it coherent} parton scattering 
in the sense that more than one parton from a particular hadron 
takes part in the scattering. Higher-twist observables
are interesting because they represent 
the quark and gluon correlations in the nucleon 
which cannot otherwise be studied.
 
Why does $g_2(x, Q^2)$ contain information
about quark and gluon correlations? 
According to the optical theorem, $g_2(x, Q^2)$ is the imaginary 
part of the spin-dependent Compton amplitude for the
process 
\begin{equation}
      \gamma^*(+1) + N(1/2) \rightarrow \gamma^*(0) + N(-1/2)
\end{equation}
where $\gamma^*$ and $N$ represent the virtual photon and 
nucleon, respectively, and the labels in the 
brackets are helicities. Thus Compton scattering 
involves a $t$-channel helicity exchange. 
When the process is factorized in terms of parton subprocesses, 
the intermediate partons must carry this helicity 
exchange. Because of the vector coupling, massless quarks 
cannot undergo helicity flip in perturbative processes.
Nonetheless, the required helicity exchange is fulfilled
in two ways in QCD: first, through single quark 
scattering in which the quark carries one unit of 
orbital angular momentum through its transverse momentum; second, 
through quark scattering with an additional transversely-polarized 
gluon from the nucleon target. These two mechanisms are 
combined in such a way to yield a gauge-invariant result. 

To leading order in $\alpha_s$, $g_2(x_B, Q^2)$ can be expressed
in terms of a simple parton distribution 
$\Delta q_T(x)$~\cite{hey}, 
\begin{equation}
   g_2(x_B, Q^2) = {1\over 2} \sum_i e_i^2 (\Delta q_{iT}(x_B, Q^2) + 
   \Delta \bar q_{iT}(x_B, Q^2)) \ , 
\end{equation}
where 
\begin{equation}
    \Delta q_T(x) = {1\over 2M} \int {d\lambda\over 2\pi}
   e^{i\lambda x} \langle PS_\perp |\bar q(0)
      \gamma_\perp\gamma_5q(\lambda n) 
   |PS_\perp\rangle \ ,
\end{equation} 
and $S_\perp$ is the transverse polarization vector and $\gamma_\perp$
is the component along the same direction.
Although it allows a simple estimate 
of $g_2$ in the models~\cite{jaffeji91}, the
above expression is 
deceptive in its physical content. It has led to 
incorrect identifications of twist-three operators
\cite{ahmed,sasaki} and incorrect 
next-to-leading order coefficient functions~\cite{kodaira79}.
When the leading-logarithmic corrections 
were studied, it was found that $\Delta q_T(x,Q^2)$ mixes with 
other distributions under scale 
evolution~\cite{sv}. 
In fact, $\Delta q_T(x,Q^2)$ is a special moment of  
more general parton distributions involving two 
light-cone variables  
\begin{equation}
    \Delta q_T(x) = 
{2\over x}\int^1_{-1} dy\left(K_1(x, y) + K_2(x, y)\right)\ , 
\end{equation}
where the $K_i(x, y)$ are defined as
\begin{eqnarray}
  && \int {d\lambda\over 2\pi}{d\mu\over 2\pi}
      e^{ix\lambda+i\mu(y-x)}
  \langle PS|\bar \psi(0)iD^\alpha(\mu n) \psi(\lambda n)
  |PS\rangle  \nonumber \\
 & =& S^\alpha\gamma_5\not\!p K_1(x,y) + 
    iT^\alpha \not\! p K_2(x, y) + ...
\end{eqnarray}
where $T^\alpha = \epsilon^{\alpha\beta\gamma\delta} 
S_\perp p_\gamma n_\delta$. 
Under a scale transformation, the general
distributions $K_i(x, y)$ evolve autonomously while the $\Delta q_{iT}(x)$
do not~\cite{bkl,et}. The first result for the
leading logarithmic evolution of the twist-three 
distributions (and operators)~\cite{bkl} has now been 
confirmed by many studies~\cite{ratcliffe86,othertw3}.

Thus an all-order $g_2$ factorization formula is much more 
subtle than is indicated by the leading-order result. It
involves the generalized two-variable distributions,
$K_i(x,y)$,  
\begin{equation}
    g_T(x_B, Q^2) =\sum_{ia} \int^1_{-1}
         {dxdy\over xy} \left(C_{a}\left({x_B\over x},
       {x_B\over y}, \alpha_s\right)K_{i}(x,y)  +
     (x_B\rightarrow -x_B) \right) \  ,
\label{g2fac}
\end{equation}
where $C_{a}$ are the coefficient functions with $a$ summing
over different quark flavors and over gluons. Accordingly, 
a perturbative calculation of $g_2$ in terms of quark 
and gluon external states must be interpreted carefully~\cite{quarkonly}. 
Recently, the complete one-loop radiative 
corrections to the singlet and non-singlet $g_2$ have been 
published~\cite{jilu00}.  The result is represented
as the order-$\alpha_s$ term in $C_i$ and is one of the necessary
ingredients for a NLO analysis of $g_2$ data. 
Note that the Burkhardt-Cottingham sum rule, $\int^1_0 
g_2(x,Q^2)dx=0$, survives
the radiative corrections provided the order 
of integrations can be exchanged~\cite{bc}.

As an example of the interesting physics associated with $g_2$, 
we consider its second moment in $x$
\begin{equation}
   \int^1_0 dx x g_2(x, Q^2) 
= {1\over 3}\left(-a_2(Q^2) + d_2(Q^2)\right) \ , 
\end{equation}
where $a_2(Q^2)$ is the second moment of the $g_1(x)$ structure function. 
Here $d_2(Q^2)$ is the 
matrix element of a twist-three operator, 
\begin{equation}
\langle PS|{1\over 4} \bar\psi g\tilde 
F^{\sigma(\mu}\gamma^{\nu)} \psi |PS\rangle
  = 2d_2 S^{[\sigma} P^{(\mu]} P^{\nu)} \ , 
\end{equation}
where $\tilde F^{\mu\nu} = (1/2)\epsilon^{\mu\nu\alpha\beta}
F_{\alpha\beta}$, and the different brackets --
$(\cdots)$ and $[\cdots]$ -- denote
symmetrization and antisymmetrization of indices, respectively.
The structure of this twist-three operator suggests that it
measures a quark {\it and} a gluon amplitude 
in the initial nucleon wave function. 

To better understand the significance of $d_2(Q^2)$, 
we consider a polarized nucleon 
in its rest frame and consider how the gluon field inside of the 
nucleon responds to the polarization. Intuitively, 
because of parity conservation, the color magnetic 
field $\vec{B}$ can be induced along the 
nucleon polarization and the color electric field $\vec{E}$
in the plane perpendicular to the polarization. 
Introducing the color-singlet operators $\hat O_B=\psi^\dagger
g\vec{B}\psi$ and $\hat O_E=\psi^\dagger \vec{\alpha}
\times g\vec{E}\psi$, we define the gluon-field 
polarizabilities $\chi_B$ and $\chi_E$ in the rest frame of the
nucleon, 
\begin{equation}
    \langle PS|\hat O_{B,E}|PS\rangle = \chi_{B,E} 2M^2\vec{S} \ . 
\end{equation}
Then it is easy to show 
\begin{equation}
     d_2 = (2\chi_B + \chi_E)/3 \ . 
\end{equation}
Thus $d_2$ measures the response of the color electric
and magnetic fields to the polarization of the nucleon.

\begin{figure}[!hbp]
\begin{center}
\includegraphics*[clip=,height=9cm,angle=0]{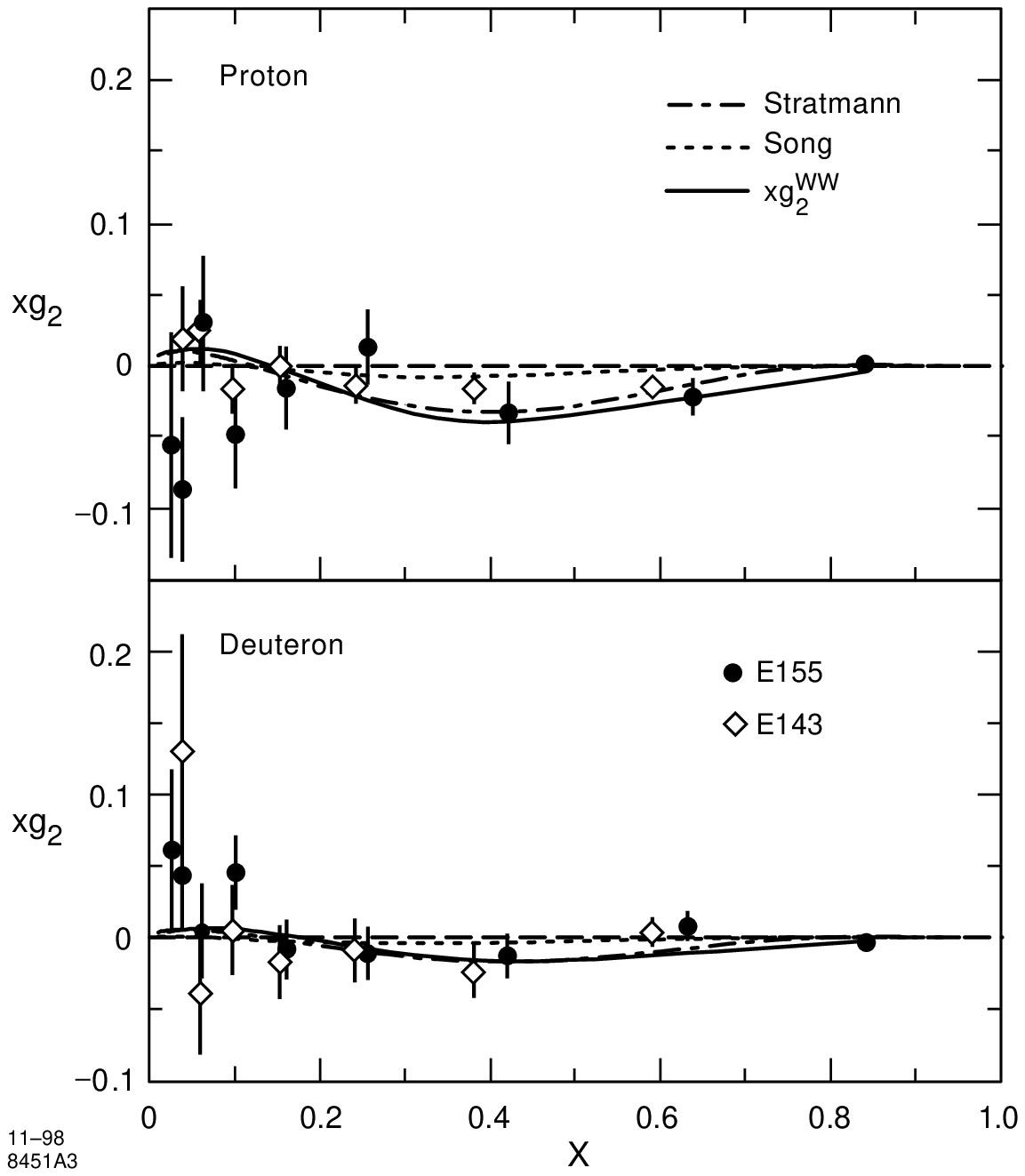}
\hspace{0.5in}
\includegraphics*[clip=,height=8cm,angle=0]{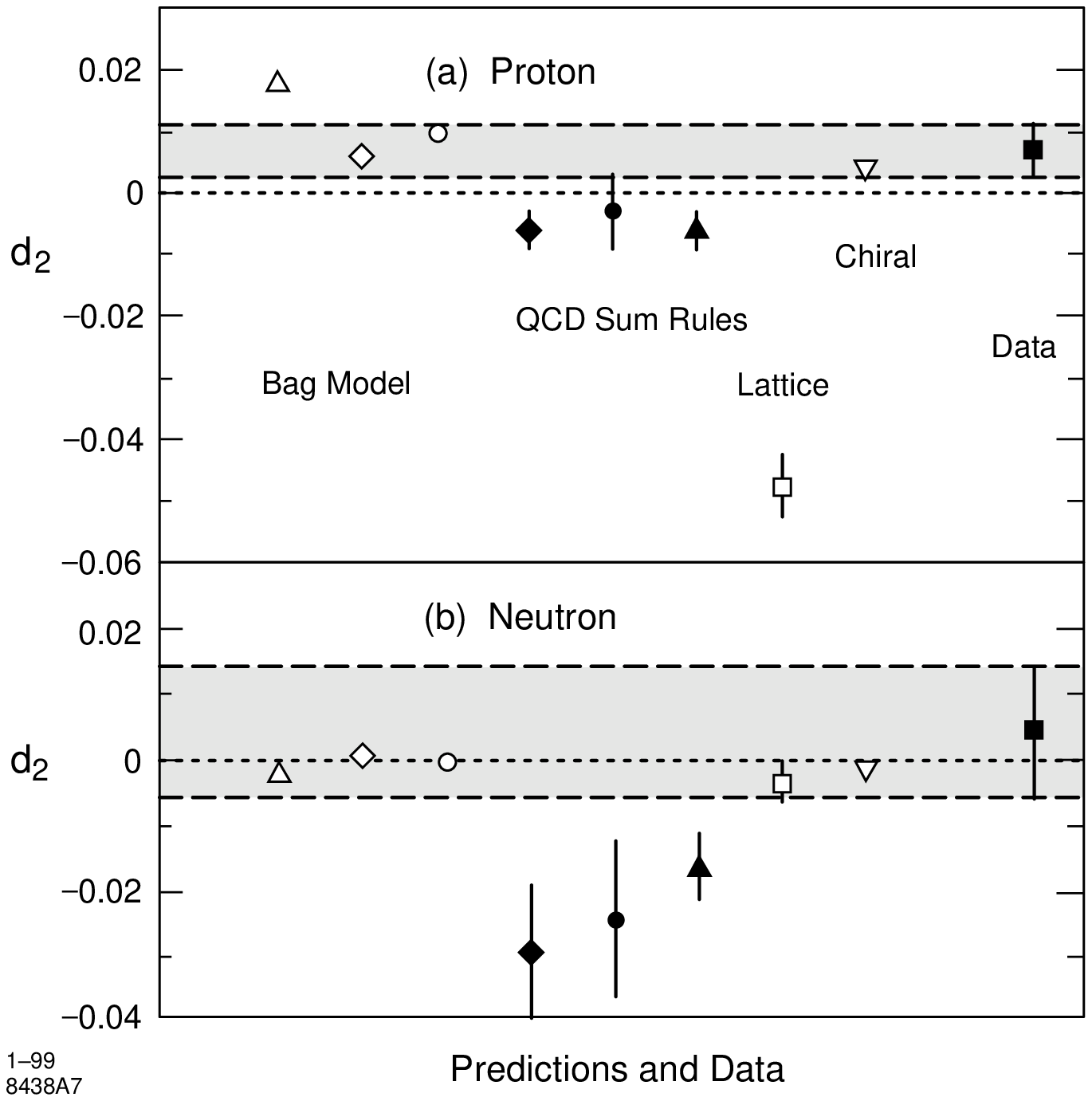}
\end{center}
\caption{(Upper) Data for $g_2$ for the proton and deuteron 
from E143~\cite{g2e143} and E155~\cite{g2e155} collaborations. 
(Lower) Comparison between data
and calculations for $d_2$, the second moment of $g_2$.}
\label{g2data}
\end{figure}

The experimental measurements of the $g_2$ 
structure function started with the SMC~\cite{g2smc} 
and E142~\cite{e142n1,e142fin} collaborations. Subsequently,
the E143~\cite{g2e143}, E154~\cite{g2e154}, 
and E155~\cite{g2e155,g2e155x} collaborations
have also measured and published their data. 
The combined E143 and E155 data for proton and 
deuteron are shown in Fig.~\ref{g2data}. The solid line
shows the twist-two contribution to $g_2$
only~\cite{wilczek}. The dashed and dash-dotted lines
are the bag model calculations by Song~\cite{g2song} 
and Stratmann~\cite{g2stratmann}. 

Neglecting the contributions from $x<0.02$ and $x>0.8$
and the $Q^2$ dependence, the E155 collaboration~\cite{g2e155,g2e155x} 
has integrated their data
to get $\int dx g_2^p = -0.022\pm 0.071$ and
$\int dx g_2^d = 0.023\pm 0.044$. The results
are consistent with the Burkardt-Cottingham sum rules
within the relatively large errors. The second moments allow
an extraction of the $d_2$ matrix elements. E155 
found $d_2^p=0.005\pm 0.008$ and $d_2^d=0.008\pm 0.005$
at an average $Q^2$ of 5 GeV$^2$. 
A combined analysis of the E142, E143, E154, and E155 data yields
$d_2^p= 0.007\pm 0.004$ and $d_2^n = 0.004\pm 0.010$. 
These numbers are generally consistent with bag 
model~\cite{g2song,g2stratmann,jiunrau,jm97} and chiral quark 
model~\cite{weigel} estimates, and are 1 to $2\sigma$ away from 
QCD sum rule calculations~\cite{stein,bal90,ehrnsperger}. 
The error bars on the present lattice calculation are
still relatively large~\cite{gockeler}. 

According to the simple
quark model, the $d_2$ matrix element in the neutron 
should be much smaller than that in the proton because of 
SU(6) spin-flavor symmetry. While the proton
$d_2$ has been constrained with reasonable precision, 
the neutron $d_2$ has a much larger
error bar. In the near future, JLab experiments 
with a polarized $^3He$ target~\cite{jlabg2} can 
improve the present error on the neutron $d_2$
and hence test the quark model predictions. 

\subsection{Tranversity Distribution \label{subsec:trans} }

Along with the unpolarized and polarized quark distributions -
$q_i(x,Q^2)$ and $\Delta q_i(x,Q^2)$ - discussed above, a 
third quark distribution exists at the same order (twist two) 
as the other two distributions. Note that no corresponding 
transverse spin distribution exists for gluons (due to helicity
conservation). 

This transversity distribution,
$\delta q_i(x,Q^2)$, can be described in the Quark-Parton
Model as the difference in the distribution of quarks with spin
aligned along the nucleon spin vs. anti-aligned for a nucleon
polarized transverse to its momentum. The 
structure function related to transversity is given by 
\be
h_1(x,Q^2) = {1\over 2}\sum_i e_i^2 \delta q_i(x,Q^2) \ .
\ee

The first moment of the transversity distributions also leads to 
an interesting observable - the nucleon's tensor charge $a^t_i$:
\be
a^t_i = \int_0^1 \left[\delta q_i(x) - \delta\bar q_i(x)\right]dx \ .
\ee
In terms of nucleon matrix elements, this 
tensor charge is defined~\cite{heji95} as:
\be 
2 a^t_i \left( S^\mu P^\nu - S^\nu P^\mu\right) = 
\langle PS | \bar\psi_i \sigma^{\mu\nu} i \gamma_5 \psi_i | PS \rangle \ .
\ee
Recent calculations have made estimates of the tensor charges
using QCD Sum Rules~\cite{heji96,jin97}, Lattice QCD~\cite{ao97}, 
and within the Chiral Quark Model~\cite{kim96}.

In a non-relativistic model the transversity is equal to the
longitudinal spin distribution ($\delta q_i = \Delta q_i$) because
the distribution would be invariant under the combination of a rotation
and a Lorentz boost. Relativistically, this is not the case 
and $\delta q_i$ could be
significantly different from $\Delta q_i$. The challenge to gaining
experimental information on $\delta q_i$ lies in its chiral structure. 
In the helicity basis~\cite{jj93,jj93_2}
$\delta q_i$ represents a quark helicity flip,
which cannot occur in any hard process for massless quarks
within QED or QCD. This chiral-odd property of transversity 
makes it unobservable in inclusive DIS. In order to observe 
$h_1(x,Q^2)$ a second non-perturbative process that is also 
chiral-odd must take place. This was first discussed by Ralston and 
Soffer~\cite{ralston} in connection with Drell-Yan production
of di-muons in polarized $p-p$ collisions. Here the transversity 
distribution of both protons results in a chiral-even interaction. 

Several calculations have suggested that the transversity 
distribution may be accessible in semi-inclusive lepton-nucleon
scattering~\cite{collins,jaffe,kot95,mt96,ans95}. In this process 
a chiral-odd fragmentation function, leading to a lepto-produced
hadron, offsets the chiral-odd transversity distribution. Many of these
calculations take advantage of an inequality 
\be
|\delta q_i(x)| \leq {q_i(x) + \Delta q_i(x)\over 2}
\ee
discovered by Soffer~\cite{sof95} to limit the possible
magnitude of $h_1(x)$. 

Calculations~\cite{kot95,mt96,km97} have also detailed a set of 
spin distribution and fragmentation functions that are accessible
from leading and next-to-leading twist processes. In fact in some
cases the next-to-leading twist processes can dominate, especially
at low $Q^2$ and with longitudinally polarized targets. 

\begin{figure}[t]
\begin{center}
\includegraphics*[clip=,height=5.5cm,angle=0]{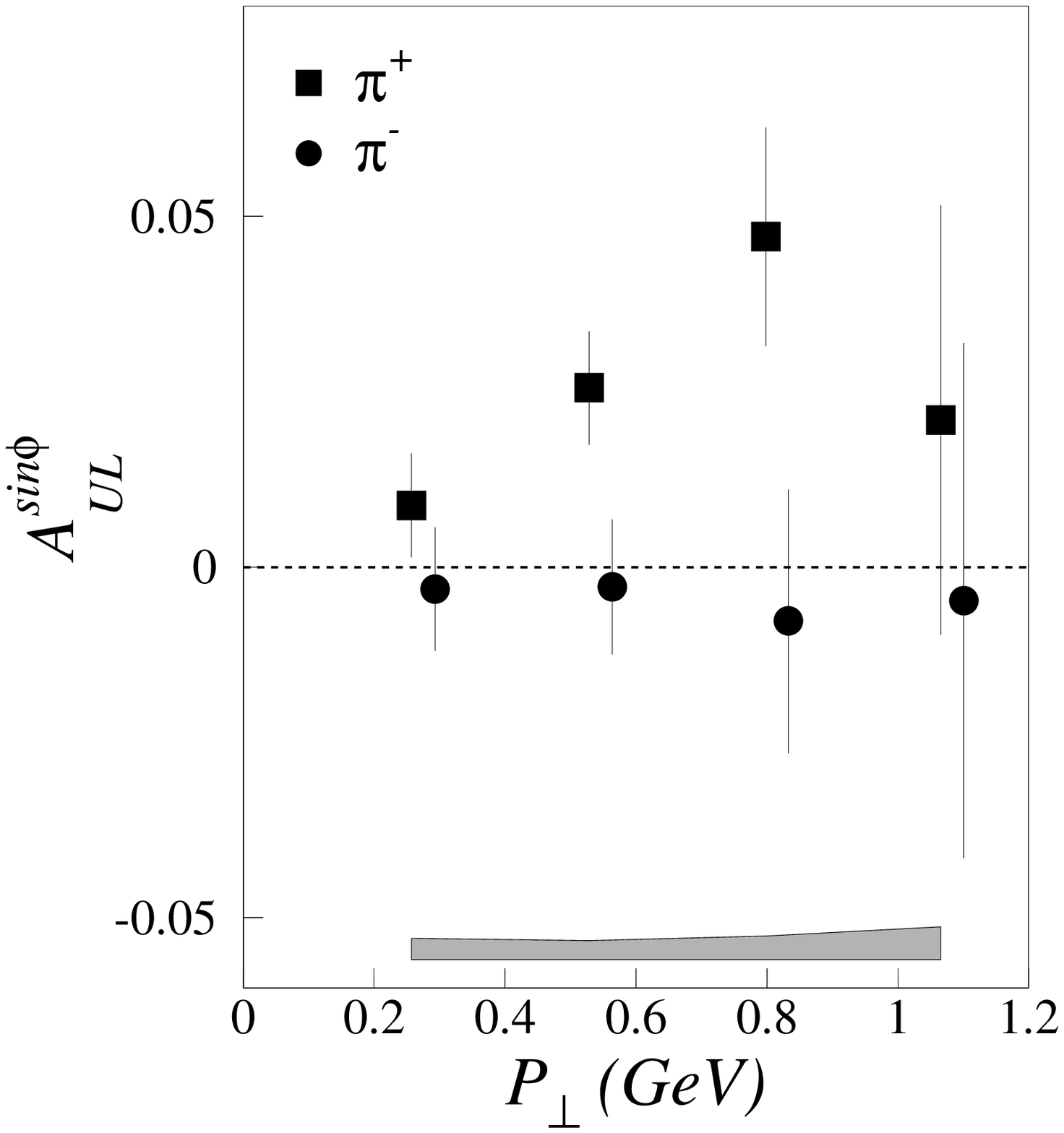}
\hspace{0.3in}
\includegraphics*[clip=,height=5.5cm,angle=0]{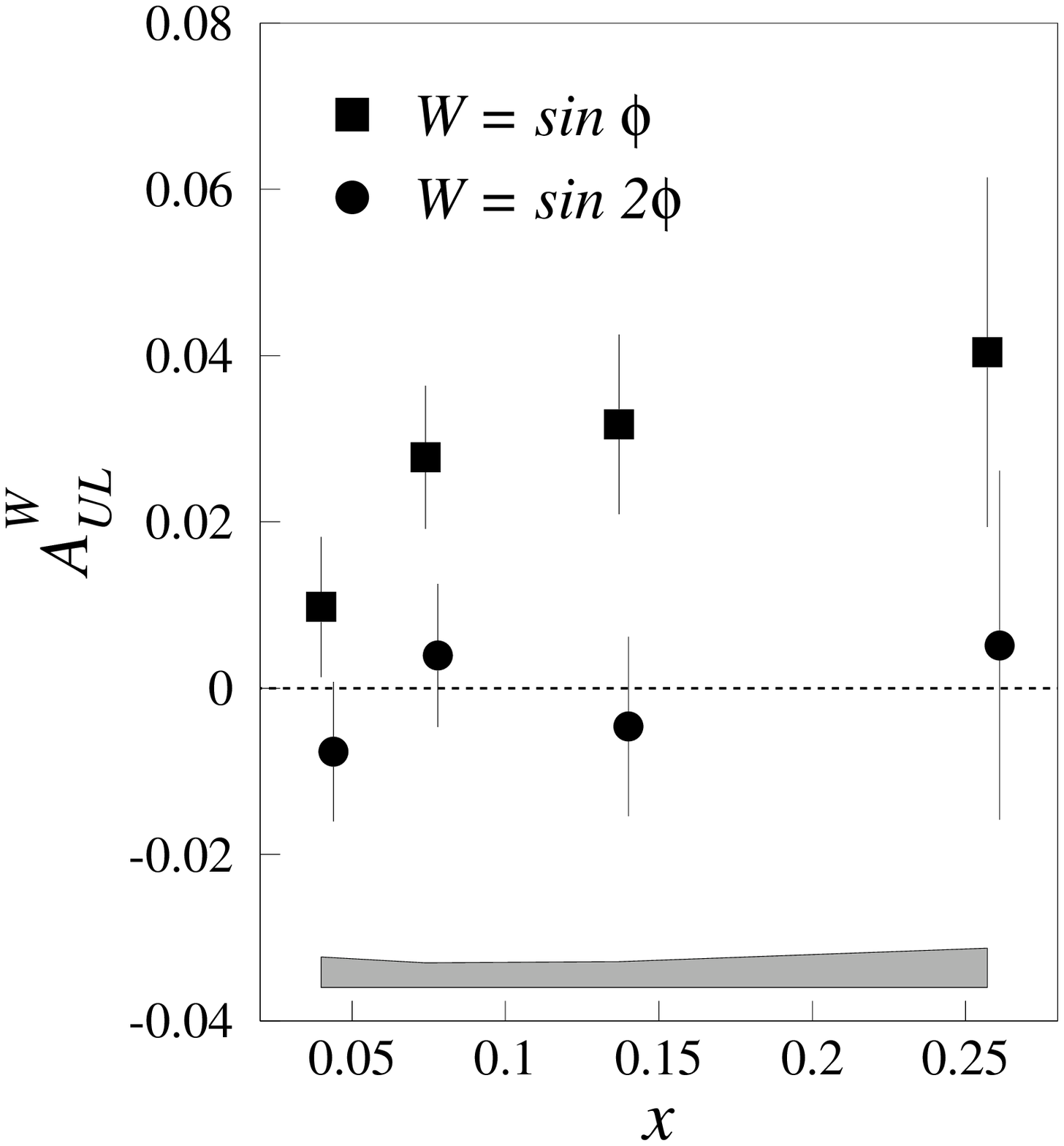}
\end{center}
\caption{Single spin asymmetry for pion electroproduction from the HERMES 
experiment~\cite{airapetian00} vs. $p_T$ (left) and $x$ (right).}
\label{hermesssa}
\end{figure}  

Potentially relevant experimental information has recently come from the
HERMES collaboration and SMC collaboration. HERMES has
measured the single-spin azimuthal
asymmetry for pions produced in deep-inelastic
scattering of unpolarized positrons from a 
longitudinally polarized hydrogen target~\cite{airapetian00} 
(see Fig. \ref{hermesssa}). A related measurement has been reported 
by SMC~\cite{smcssa} using a transversely polarized target.
The HERMES asymmetry 
is consistent with a $\sin \phi$ distribution, where $\phi$ is the
angle between the lepton scattering plane and the plane
formed by the virtual photon and pion momenta. While
the $x$ dependence of the asymmetry is relatively weak
except for the smallest $x=0.04$ point, the $p_T$ dependence
shows a rapid rise up to 0.8 GeV$^{-1}$. The average
$\pi^+$ asymmetry averaged over the full acceptance is
$0.022 \pm 0.005 \pm 0.003$ while the asymmetry for 
$\pi^-$ production is consistent with zero. 
Some models~\cite{ssamod}
suggest that the product of the transversity distribution
times the chiral-odd fragmentation function can account for the
observed asymmetry. 

Transversity may also play a role in observed single-spin 
asymmetries in $p - p$ collisions. These possibilities are
discussed in the next section.

\subsection{Single-Spin Asymmetries From Strong Interactions 
\label{subsec:single} }

As we have discussed in Sec.~\ref{subsec:high}, single-spin asymmetries can 
arise from processes involving parity-violating interactions, 
such as $W$ and $Z$ boson production in $\vec{p}-p$ collisions
at RHIC. In this subsection, we discuss a different
class of single-spin asymmetries which are generated 
entirely from strong interaction effects. While we do not have
enough space here to make a thorough examination of the subject, 
we briefly discuss the phenomena, a few
leading theoretical ideas, 
and some other related topics. A recent review 
of the subject can be found in Ref.~\cite{liang00}. 
 
For more than two decades, it has been known that 
in hadron-hadron scattering with one beam
transversely polarized, the single-particle inclusive
yield at nonzero $p_T$ has an azimuthal 
dependence in a coordinate system where $z$ is chosen 
to lie along 
the direction of the polarized beam, and $x$ along the  
beam polarization~\cite{ssadata}. It is easy to see
that the angular dependence is allowed by strong interaction
dynamics. If the momentum of the polarized beam is $\vec{p_b}$ 
and that of the observed particle $\vec{p_o}$, the angular
distribution reflects the existence of a 
triple correlation,
\begin{equation}
           \vec{p}_b \times \vec{p}_o \cdot \vec{S} \ , 
\end{equation}
where $\vec{S}$ is the beam polarization. The correlation
conserves parity and hence is not forbidden in strong 
interactions. Although it is nominally time-reversal 
odd, the minus sign can be canceled, under the time-reversal 
transformation, by a factor of 
$i$ from an interference of two amplitudes with different
phase factors. 

\begin{figure}[t]
\begin{center}
\includegraphics*[clip=,height=5.3cm,angle=0]{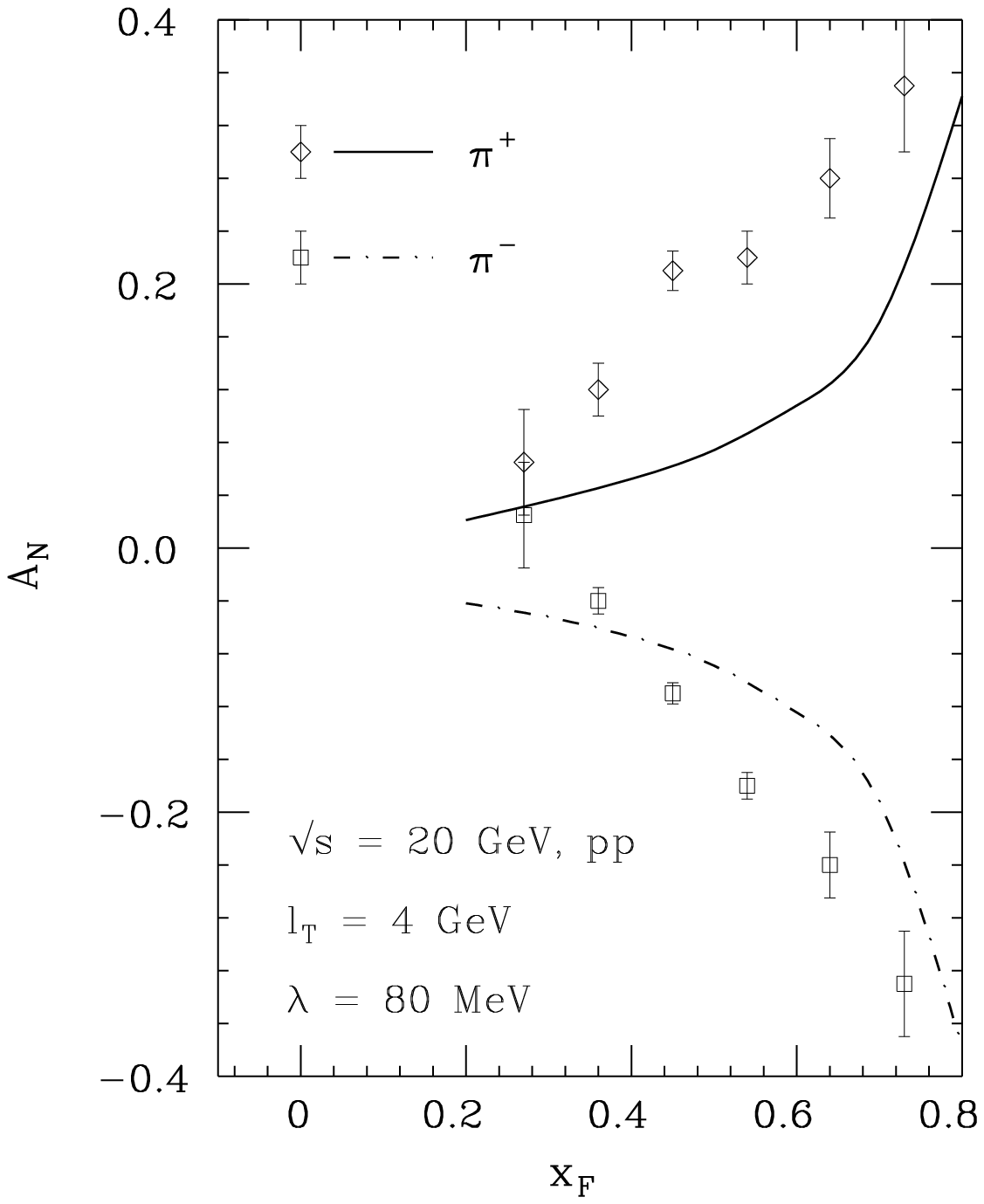}
\hspace{0.3in}
\includegraphics*[clip=,height=5.3cm,angle=0]{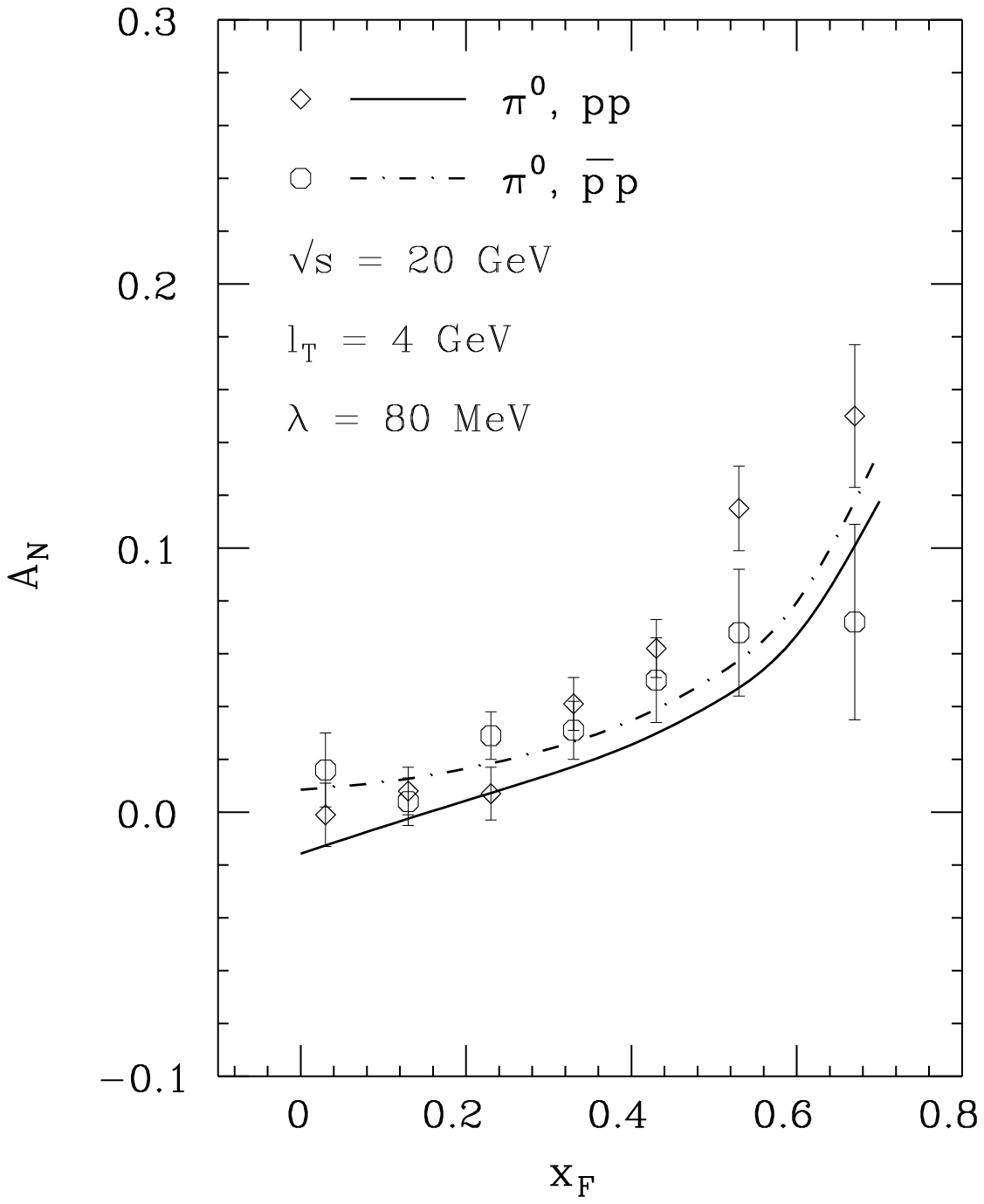}
\end{center}
\caption{Measured single-spin asymmetries from $\pi^{\pm}$ production in
$\vec{p}-p$ scattering (left) and $\pi^0$ production in $\vec{p}-p$ and 
$\vec{p}-\bar p$ scattering (right) at Fermilab~\cite{ssadata}. 
The curves are theoretical fits from Ref.~\cite{qiu99}.}
\label{ssafig}
\end{figure}

The angular correlation is usually characterized by
the spin asymmetry
\begin{equation}
     A_N(x_F, p_T) = {d\sigma^\uparrow - d\sigma^\downarrow
        \over d\sigma^\uparrow + d\sigma^\downarrow} \ , 
\end{equation}
where $d\sigma^{\uparrow,\downarrow}$ are the cross sections with
reversed polarizations, and $p_T$ is 
the transverse momentum of the produced particle.
$x_F$ is the Feynman $x$ variable,
$x_F = p_L/p_L^{\rm max}$, where $p_L$ is the longitudinal
momentum of the produced hadron and $p_L^{\rm max}$ is the maximum
allowed longitudinal momentum. An example of 
a single spin asymmetry for $\pi$ production is shown
in Fig.~\ref{ssafig}. After examining the existing
data, one finds the following interesting systematic effects
\cite{liang00}:
\begin{itemize}
\item{$A_N$ is significant only in the fragmentation region
of the polarized beam. It increases almost linearly 
with $x_F$ when the target is unpolarized.}
\item{$A_N$ is large only for moderate transverse 
momentum $p_T$.}
\item{$A_N$ and its sign show a strong dependence on the 
type of polarized beam ($p, \bar p$) and produced particles ($\pi^\pm,
\pi_0$).}
\end{itemize}
That $A_N$ is strikingly large is the most
impressive aspect of the phenomenon.

The simplest theory explaining $A_N$ is one that assumes 
an underlying parton process: partons from the parent 
hadrons scatter and fragment to produce the observed particle. 
To get the single spin asymmetry, one requires,
for instance, that quarks change their helicity during
hard scattering. However, chiral symmetry then
dictates that the asymmetry is proportional 
to the quark mass $m_q$ which is vanishingly small 
for light quarks. Thus the simple parton model for $A_N$
cannot yield the magnitude of the observed symmetry
\cite{michigan}.

For the moment, the leading theoretical ideas in the
literature are still based on the parton degrees of 
freedom. However, the spin-flip is introduced through
more complicated mechanisms: Either the initial 
and final partons are assumed to have 
novel nonpertrubative distribution and fragmentation
functions, respectively, or the parton hard scattering 
involves coherent processes. 

In the latter case, the asymmetry can arise 
from the coupling of chiral even (odd) twist-two (twist-three) 
parton correlations in the polarized nucleon 
and chiral even (odd) twist-three (twist-two) fragmentation
functions of the scattered partons~\cite{efremov,qiu91,ji93,qiu99}.
The required phase difference is generated from the 
interference of the hard scattering
amplitudes in which one of the hard propagators 
is on-shell. The predicted asymmetry is 
of order $\Lambda_{\rm QCD}/p_T$ in the large 
$p_T$ limit, which is a characteristic twist-three 
effect. For moderate $p_T$, $A_N$ can be a slowly decreasing
function of $p_T$~\cite{qiu99}. The comparison between
the experimental data and the phenomenological prediction
seems to yield good agreement~\cite{qiu99}. 
It is not clear, however, that the available fixed target
data can be fully described by perturbative parton 
scattering. One needs more data at higher energy to test 
the scaling property inherent in a 
perturbative description. 

The alternative is to consider nonperturbative
mechanisms to generate the phase difference. 
This can be done by introducing transverse-momentum 
dependent parton distributions~\cite{sivers} and fragmentation
functions~\cite{collins}. In a transversely polarized 
nucleon, the transverse momentum distribution may not be 
rotationally invariant. It may depend on the relative
orientation of the spin and momentum vectors. Likewise,
when a transversely polarized quark fragments, the 
amplitude for hadron production can depend on the relative 
orientation between the hadron momentum and the quark spin. 
Both mechanisms have been shown to produce large single
spin asymmetries~\cite{ans95,boer98,boglione99,
anselmino99}. Here again the applicability of the model for
the existing data is not clear. In particular, the fitted
fragmentation functions and parton distributions must be
tested in different kinematic regions. Moreover, the new distributions 
do not possess color gauge invariance. 

A phenomenological model for parton scattering
with formation of large-$p_T$ hadrons was proposed
by Boros, Liang and Meng~\cite{liang}. Although
not derived from field theory, the model
has a very intuitive physical picture and  
successfully describes the data. It would be 
interesting to test the predictive power of the model 
in future experiments. The RHIC spin facility can
test many of these theoretical ideas with a variety of
experimental probes including~\cite{rhic} polarized
Drell-Yan, dimeson production, etc.

A subject closely related to the single-spin asymmetry
is the polarization of hyperons, such as $\Lambda$, 
produced in unpolarized hadron collisions~\cite{heller}. 
The observed polarization is perpendicular to the plane 
formed by the beam and hyperon momenta. 
Many theoretical models have been invented to 
explain the polarization~\cite{lptheory}. Most models
are closely related to those devised to explain the
single spin asymmetry.

\section{Off-Forward Parton Distributions \label{sec:off} }

In this section, we discuss some of the recent
theoretical developments on generalized (off-forward)
parton distributions (OFPD) and their relation to the angular
momentum distributions in the nucleon. We will also
consider possible experimental processes, such as deeply
virtual Compton scattering (DVCS) and meson production,
to measure these novel distributions. 

OFPD's were first introduced in Ref.~\cite{dittes88} and discovered
independently in Ref.~\cite{jiprl97} in studying the spin structure
of the nucleon. Radyushkin and others have introduced slightly 
different versions of the distributions, but the physical content 
is the same~\cite{radmeson,radyushkin97,rad99,cfs}. 
The other names for these functions
range from off-diagonal, non-forward and skewed to generalized 
parton distributions. Here we follow the discussion in
Ref.~\cite{jijpg}. 

One of the most important sources of information
about the nucleon structure is the form factors
of the electroweak currents. It is well known that the vector
current yields two form factors
\begin{equation}
\langle P'|\overline \psi \gamma^\mu \psi|P\rangle
 = F_1(Q^2) \overline U\gamma^\mu U
    + F_2(Q^2) \overline U{i\sigma^{\mu\nu}q_\nu\over 2M}U \, , 
\end{equation}
where $q_\nu = P^\prime - P$ and $F_1$ and $F_2$
are the Dirac and Pauli form factors, respectively.
$F_2$ gives the anomalous magnetic moment of the nucleon,
     $\kappa = F_2(0)$. 
The charge radius
of the nucleon is defined by 
\begin{equation}
     \langle r^2\rangle = -6 \left.{dG_E(Q^2)\over  
  d Q^2}\right|_{Q^2=0} \, , 
\end{equation}
with $G_E = F_1 - Q^2/(4M^2) F_2$.
The axial vector current also defines two form factors,
\begin{equation}
\langle P'|\overline \psi \gamma^\mu \gamma_5\psi|P\rangle
 = G_A(Q^2) \overline U\gamma^\mu\gamma_5 U
    + G_P(Q^2) \overline U{\gamma_5 q^\mu\over 2M}U(P) \, . 
\end{equation}
The axial form factor $G_A$ is related to 
the fraction of the nucleon spin carried by the spin of the quarks, 
$\Delta \Sigma$, and 
can be measured from polarized deep inelastic
scattering as discussed in previous sections.
The pseudoscalar charge, 
$G_P(0)$, can be measured in muon
capture. 

A generalization of the electroweak currents can be made through
the following sets of twist-two operators,
\begin{eqnarray}
     O_q^{\mu_1\cdots\mu_n} 
   &= &\overline \psi_q \gamma^{(\mu_1} iD^{\mu_2} \cdots iD^{\mu_n)} \psi \,
   \nonumber \\
  \tilde O_q^{\mu_1\cdots\mu_n} 
   &= &\overline \psi_q \gamma^{(\mu_1}\gamma_5 iD^{\mu_2} \cdots iD^{\mu_n)}
\psi \,  ,
\end{eqnarray}
where all indices $\mu_1\cdots \mu_n$ are symmetric and traceless as 
indicated by (...) in the superscripts. 
These operators form the totally symmetric representation
of the Lorentz group. One can also introduce gluon currents through the operators:
\begin{eqnarray}
      O_g^{\mu_1\cdots\mu_n} 
     &= &F^{(\mu_1\alpha} i D^\mu_2 \cdots iD^{\mu_{n-1}}
F_\alpha^{~\mu_n)} \nonumber \\
     \tilde O_g^{\mu_1\cdots\mu_n} 
     &= &F^{(\mu_1\alpha} i D^\mu_2 \cdots iD^{\mu_{n-1}} 
     \tilde F_\alpha^{~\mu_n)}  \, . 
\end{eqnarray}
For $n>1$, the above operators are not conserved currents
from any global symmetry. 
Consequently, their matrix elements 
depend on the momentum-transfer scale $\mu$ at which they are probed. 
For the same reason, there is no low-energy probe 
that couples to these currents. 

One can then define the generalized charges $a_n(\mu^2)$
from the forward matrix elements of these currents
\begin{equation}
   \langle P|O^{\mu_1\cdots\mu_2}|P\rangle
   = 2 a_n(\mu^2)P^{(\mu_1} P^{\mu_2}\cdots P^{\mu_n)} \, . 
\end{equation}
The moments of the Feynman parton distribution $q(x,\mu^2)$ 
are related to these charges
\begin{equation}
   \int^1_{-1} dx x^{n-1} \tilde q(x,\mu^2) = 
\int^1_{0} dx x^{n-1} \left[q(x,\mu^2) + (-1)^n \bar q(x,\mu^2)\right] 
= a_n(\mu^2) \ , 
\end{equation}
where $\tilde q(x,\mu^2)$ is defined in the range $-1<x<1$. 
For $x>0$, $\tilde q(x, \mu^2)$ is just the
density of quarks which carry the fraction $x$ of
the parent nucleon momentum. The density of antiquarks
is customarily denoted as $\bar q(x,\mu^2)$, which in
the above notation is $-\tilde q(-x,\mu^2)$ for $x<0$. 

One can also define the form factors 
($A_{qn,m}(t), B_{qn,m}(t)$, and $C_{qn}(t)$) 
of these currents using constraints from 
charge conjugation, parity, time-reversal and
Lorentz symmetries 
\begin{eqnarray}
&&  \langle P'| O^{\mu_1\cdots \mu_n}_q |P\rangle  \nonumber \\
&=& {\overline U}(P') \gamma^{(\mu_1} U(P) \sum_{i=0}^{[{n-1\over 2}]}
       A_{qn,2i}(t) \Delta^{\mu_2}\cdots \Delta^{\mu_{2i+1}}
      \overline{P}^{\mu_{2i+2}}\cdots
      \overline{P}^{\mu_n)}  \nonumber \\
&+& \overline {U}(P'){\sigma^{(\mu_1\alpha}
     i\Delta_\alpha \over 2M}U(P)   \sum_{i=0}^{[{n-1\over 2}]}
     B_{qn,2i}(t) \Delta^{\mu_2}\cdots \Delta^{\mu_{2i+1}}
      \overline{P}^{\mu_{2i+2}}\cdots
    \overline{P}^{\mu_n)} \nonumber \\
&+&  C_{qn}(t) {\rm Mod}(n+1,2)~{1\over M} \overline {U}(P') U(P)
    \Delta^{(\mu_1} \cdots \Delta^{\mu_n)} \, ,
\label{form} 
\end{eqnarray}
where $\overline {U}(P')$ and $U(P)$ are Dirac spinors, 
$\Delta^2 = (P' - P)^2 = t$, $\overline {P} = (P' + P)/2$ and
${\rm Mod}(n+1,2)$ is 1 when $n$ is even and 0 when $n$ is odd. Thus
$C_{qn}$ is present only when n is even. We suppress the 
renormalization scale dependence for simplicity. 
In high energy experiments, it is difficult
to isolate the individual form factors. Instead
it is useful to consolidate them into generalized 
distributions --- the off-forward parton distributions
(OFPD's). To accomplish this a light-light
vector $n^\mu$ ($n^2=0$) is chosen such that
\begin{equation}
   n\cdot \overline P = 1 \ , \;\; \xi = -n\cdot\Delta/2 \, .
\end{equation}
Then,
\begin{equation}
   n_{\mu_1}\cdot n_{\mu_n}
    \langle P'|O^{\mu_1\cdots\mu_n} |P\rangle
  = H_n(\xi,t) \overline U \not\! n U 
   + E_n(\xi,t) \overline U {i\sigma^{\mu\alpha}n_\mu 
   \Delta_\alpha \over 2M} U \ , 
\end{equation}
where $H_n(\xi,t)$ and $E_n(\xi,t)$ are polynomials in 
$\xi^2$ of degree $n/2$ ($n$ even) or $n-1/2$ ($n$ odd).
The coefficients of the polynomials are form factors. 
The OFPD  $E(x,\xi,t)$ and $H(x,\xi,t)$ are then defined as:
\begin{eqnarray}
   \int^1_{-1} dx x^{n-1} E(x, \xi, t) &= &E_n(\xi, t) \nonumber \\
   \int^1_{-1} dx x^{n-1} H(x, \xi, t) &= &H_n(\xi, t) \, .
\end{eqnarray}
Since all form factors are real, the new distributions
are also real. Moreover, because of time-reversal
and hermiticity, they are {\it even}
functions of $\xi$.

The OFPD's are more complicated than the
Feynman parton distributions because of their
dependence on the momentum transfer $\Delta$. As such, they 
contain two more scalar variables besides
the $x$ variable. The variable $t$ is the usual $t$-channel
invariant which is always present in a form factor.
The $\xi$ variable is a natural product of marrying the concepts
of the parton distribution and the form factor: The former requires
the presence of a prefered momentum $p^\mu$ along which the partons are
predominantly moving, and the latter requires a four-momentum
transfer $\Delta$; $\xi$ is just a scalar product of these two
momenta. 

\subsection{Properties of the Off-Forward Parton Distributions 
\label{subsec:prop} }

The physical interpretation of parton distributions
is transparent only in light-cone
coordinates and light-cone gauge. To see this,
we sum up all the local twist-two operators into
a light-cone bilocal operator and express
the parton distributions in terms of the latter,
\begin{eqnarray}
   F_q(x, \xi, t) \hspace*{-.05in}&= {1\over 2}\int {d\lambda\over 2\pi} 
e^{i\lambda x}  
 \left\langle P'\left|\overline \psi_q \left(-{\lambda \over
2}n\right)
      \not \! n {\cal P}e^{-ig\int^{-\lambda/2}_{\lambda/ 2}
       d\alpha ~n\cdot A(\alpha n)}
    \psi_q\left({\lambda \over 2}n\right) \right| P\right\rangle
    \nonumber \\
  \hspace*{-.1in}&= H_q(x, \xi, t)~ {1\over 2}\overline U(P')\not\! n U(P)
    + E_q(x, \xi, t){1\over 2}\overline U(P') {i\sigma^{\mu\nu}
  n_\mu \Delta_\nu \over 2M} U(P) \, .
\nonumber \\* \hspace*{-.2in}
\label{string}
\end{eqnarray}
The light-cone bilocal operator (or light-ray
operator) arises frequently in hard scattering processes
in which partons propagate along the light-cone.
In the light-cone gauge $n\cdot A=0$, the gauge link
between the quark fields can be ignored. 
Using the light-cone coordinate system
\begin{equation}
     x^{\pm} = {1\over \sqrt{2}}(x^0 \pm x^3); ~~
     x_\perp = (x^1, x^2) \ , 
\end{equation}
we can expand the Dirac field
\begin{eqnarray}
  \psi_+(x^-,x_\perp) &=& \int {dk^+ d^2\vec{k}_\perp \over
2k^+(2\pi)^3}
     \theta(k^+) \sum_{\lambda = \pm}
      \left(b_\lambda(k^+,\vec{k}_\perp) u_\lambda(k)      
       e^{-i(x^-k^+-\vec{x}_\perp
    \cdot \vec{k}_\perp)} \right. \nonumber \\
  && \left. +~ d_\lambda^\dagger(k^+,\vec{k}_\perp) v_\lambda(k)
   e^{i(x^-k^+-\vec{x}_\perp
    \cdot \vec{k}_\perp)} \right) \, ,
\label{ff}
\end{eqnarray}
where $\psi_+=P_+\psi$ and $P_\pm = {1\over 2} \gamma^{\mp}
\gamma^{\pm}$. The quark (antiquark)
creation and annihilation operators, $b_{\lambda k}^\dagger$
($d_{\lambda k}^\dagger)$ and $b_{\lambda k}$ ($d_{\lambda k})$,
obey the usual commutation relation.
Substituting the above into Eq. (\ref{string}),
we have~\cite{jijpg}
\begin{eqnarray}
\hspace*{-.2in} && F_q(x,\xi)  = {1\over 2p^+V}
          \int {d^2k_\perp \over 2\sqrt{|x^2-\xi^2|} (2\pi)^3}
         \sum_\lambda  \nonumber \\
\hspace*{-.2in} && \times\left\{ \begin{array}{ll}
         \left\langle P'\left|b^\dagger_\lambda\left((x-\xi)p^+,
         \vec{k}_\perp+\vec{\Delta}_\perp \right)
        b_\lambda\left((x+\xi)p^+, \vec{k}_\perp\right)\right|
        P\right\rangle,\\ 
~~~~~~~~~~~~~~~~~~~~~~~~~~~~~~~~~~~~~~~~~~~~~~~~~~~~~~~~~~~~~~~~~~~~~~~~ {\rm for }~ x > \xi  \\ 
         \left \langle P'\left|d_\lambda\left((-x+\xi)p^+,
         -\vec{k}_\perp-\vec{\Delta}_\perp \right)
        b_{-\lambda}\left((x+\xi)p^+, \vec{k}_\perp\right)\right|P
       \right\rangle,  \\
~~~~~~~~~~~~~~~~~~~~~~~~~~~~~~~~~~~~~~~~~~~~~~~~~~~~~~~~~~~~~~~~~~~~~~~~ {\rm for }~ \xi >  x > -\xi \\
         -\left\langle P'\left| d^\dagger_\lambda\left((-x-\xi)p^+,
         \vec{k}_\perp+\vec{\Delta}_\perp \right)
        d_\lambda\left((-x+\xi)p^+, \vec{k}_\perp\right)\right|
         P\right\rangle, \\
~~~~~~~~~~~~~~~~~~~~~~~~~~~~~~~~~~~~~~~~~~~~~~~~~~~~~~~~~~~~~~~~~~~~~~~~ {\rm for }~ x <-\xi \end{array} \right.
\nonumber \\* \hspace*{-.2in}
\label{fock}
\end{eqnarray}
where $V$ is a volume factor.  The distribution has different physical
interpretations in the three different regions. In the region $x>\xi$,
it is the amplitude for taking a quark of momentum $k$ out of the
nucleon, changing its momentum to $k+\Delta$, and inserting it back to
form a recoiled nucleon.  In the region $\xi> x> -\xi$, it is the
amplitude for taking out a quark and antiquark pair with momentum
$-\Delta$. Finally, in the region $x<-\xi$, we have the same
situation as in the first, except the quark is replaced by an
antiquark. The first and third regions are similar to those present in
ordinary parton distributions, while
the middle region is similar to that in a meson amplitude.

By recalling the definition of $J_{q,g}(\mu)$ in terms of the QCD 
energy-momentum tensor $T^{\mu\nu}_{q,g}$
\begin{equation}
      J_{q,g}(\mu) = \left\langle P{1\over 2} \left|
         \int d^3x (\vec{x}\times \vec{T}_{q,g})_z
 \right|P{1\over 2}\right\rangle \, ,
\label{matrix}
\end{equation}
it is clear that they can be extracted from the
form factors of the quark and gluon parts of
the $T^{\mu\nu}_{q,g}$.
Specializing Eq. (\ref{form}) to ($n=2$),
\begin{eqnarray}
  \langle P'| T_{q,g}^{\mu\nu} |P\rangle
   &=& \overline U(P') \Big[ A_{q,g}(t) \gamma^{(\mu} \overline P^{\nu)} +
  B_{q,g}(t) \overline P^{(\mu} i\sigma^{\nu)\alpha}\Delta_\alpha/2M \nonumber \\
 &+& C_{q,g}(t)\Delta^{(\mu} \Delta^{\nu)}/M \Big] U(P) \, .
\end{eqnarray}
Taking the forward limit of the $\mu=0$ component and integrating
over three-space, one finds that the $A_{q,g}(0)$ give   
the momentum fractions of the nucleon carried by
quarks and gluons ($A_q(0)+A_g(0)= 1$).
On the other hand, substituting
the above into the nucleon matrix element of Eq. (\ref{matrix}),
one finds~\cite{jiprl97}
\begin{eqnarray}
      J_{q, g} = {1\over 2} \left[A_{q,g}(0) + B_{q,g}(0)\right] \, .
\end{eqnarray}
Therefore, the matrix elements of the energy-momentum
tensor provide the fractions of
the nucleon spin carried by quarks and gluons.
There is an analogy for this. If one knows the Dirac and Pauli     
form factors of the electromagnetic current,
$F_1(Q^2)$ and $F_2(Q^2)$,
the magnetic moment of the nucleon, defined as
the matrix element of (1/2)$\int d^3x (\vec{x} \times \vec{j})^z$,
is $F_1(0) +F_2(0)$.

Since the quark and gluon energy-momentum tensors
are just the twist-two, spin-two, parton helicity-independent
operators, we immediately have the following
sum rule from the off-forward distributions;
\begin{eqnarray}
     \int^1_{-1} dx x [H_q(x, \xi, t) +
       E_q(x, \xi, t) ]
     = A_q(t) + B_q(t) \, ,
\end{eqnarray}
where the $\xi$ dependence, or $C_q(t)$
contamination, drops out. Extrapolating the sum rule
to $t=0$, the total quark 
contribution to the nucleon spin is obtained.
When combined with measurements of the quark spin contribution
via polarized DIS measurements, the quark orbital contribution to
the nucleon spin can be extracted. 
A similar sum rule exists for gluons.    
Thus a deep understanding of the spin structure of the
nucleon may be achieved by measuring OFPD's
in high energy experiments.                      

A few rigorous results about OFPD's
are known. First of all, in the limit 
$\xi\rightarrow 0$ and $t \rightarrow 0$,
they reduce to the ordinary parton
distributions. For instance,
\begin{eqnarray}
     H_q(x, 0, 0) &=& q(x) \, ,  \nonumber \\
     \tilde H_q(x, 0, 0) &=& \Delta q(x) \, ,
\end{eqnarray}
where $q(x)$ and $\Delta q(x)$ are the unpolarized and
polarized quark densities.  Similar equations hold
for gluon distributions. For practical purposes,
in the kinematic region where
\begin{equation}
       \sqrt{|t|}<\!\!< M_N~~~ {\rm and} ~~~ \xi<\!\!<x
\end{equation}
an off-forward distribution may be approximated
by the corresponding forward one. The first condition,
$\sqrt{|t|}<\!\!< M_N$, is crucial---otherwise       
there is a significant form-factor
suppression which cannot be neglected at any $x$ and $\xi$.
For a given $t$, $\xi$ is restricted to
\begin{equation}
    |\xi| < \sqrt{-t/(M^2-t/4)} \, .
\label{gs}
\end{equation}
Therefore, when $\sqrt{|t|}$ is small, $\xi$ is
automatically limited and there is in fact
a large region of $x$ where the forward approximation holds.

The first moments of the off-forward distributions
are constrained by the form factors of
the electromagnetic and axial currents. Indeed,
by integrating over $x$, we have~\cite{jiprl97}
\begin{eqnarray}
     \int^1_{-1} dx H_q(x, \xi, t) &=& F_1^q(t) \ , 
     \int^1_{-1} dx E_q(x, \xi, t) = F_2^q(t) \ ,  \nonumber  \\
     \int^1_{-1} dx \tilde H_q(x, \xi, t) &=& G_A^q(t) \ , 
      \int^1_{-1} dx \tilde E_q(x, \xi, t) = G_P^q(t) \ ,
\end{eqnarray}
where $F_1$, $F_2$, $G_A$ and $G_P$ are the Dirac, Pauli,
axial, and pseudo-scalar elastic form factors, respectively.
The $t$ dependence of the form factors are
characterized by hadron mass scales. Therefore,
it is reasonable to speculate that similar
mass scales control the $t$ dependence of
the off-forward distributions. 

The first calculation of the OFPD
has been done in the MIT bag model~\cite{jms}.
The parameters are adjusted so that the electromagnetic
form factors and the Feynman parton distributions
are well reproduced. The shapes of the distributions 
as a function of $x$ are rather similar 
at different $t$ and $\xi$. The $t$ dependence of the 
energy-momentum form factors is controlled by 
a mass parameter between 0.5 and 1 GeV$^2$.
The same distributions were also studied
in the chiral quark-soliton model by Petrov et al. 
\cite{petrov98}. 
In contrast to the bag model results, the chiral soliton
model yields a rather strong $\xi$ dependence. 
The model also predicts qualitatively different behaviors 
in the regions $|x|>\xi$ and $|x|<\xi$, in line with the physical
interpretation of the distributions. In the case
of the $\tilde E(\xi,t)$ distribution, the pion pole contribution
is important~\cite{fpps}. The OFPD's have also been modeled
directly without a theory of the structure of the nucleon. 
In Ref.~\cite{vg}, the distributions 
are assumed to be a product of the usual parton
distributions and some $t$-dependent form factors, 
independent of the variable $\xi$. In Ref.~\cite{rad99,rmodel}, 
the so-called double distributions are modeled in 
a similar ansatz from which a strong $\xi$ dependence
is generated. 

Scale evolution of the OFPD's has received wide
attention and is now completely solved up to two 
loops. In the operator form, the evolution has been 
studied at the leading 
logarithmic approximation long before~\cite{opev}.
In terms of the actual distributions, the evolution equations
at the leading-log can
be found in Refs.~\cite{dittes88,jidvcs,radmeson,
radyushkin97,rad99,disev}
in different cases and forms. 

In a series of interesting papers, Belitsky and M\"uller
have calculated the evolution of the off-forward distributions
at two loops~\cite{twoloop}. The key observation
is that perturbative QCD is approximately conformally
invariant. The breaking of the conformal symmetry can be
studied through conformal Ward identities, which allows
one to obtain the two-loop anomalous dimension by 
calculating just the one-loop conformal anomaly. 

\subsection{Deeply Virtual Exclusive Scattering \label{subsec:deep} }

Of course the eventual utility of the OFPD's depends on
whether they can actually be measured in any
experiment. The simplest, and possibly
the most promising, type of experiments
is deep-inelastic exclusive production of photons, 
mesons, and perhaps even lepton pairs. Here we 
consider two experiments that have been studied 
extensively in the literature: deeply virtual
Compton scattering (DVCS) in which a real photon
is produced, and diffractive
meson production. There are practical advantages
and disadvantages from both processes. Real
photon production is, in a sense, cleaner but the
cross section is reduced by an additional power
of $\alpha_{\rm em}$. The Bethe-Heitler contribution
can be important but can actually be used to extract
the DVCS amplitude.
Meson production may be easier to detect, however,
it has a twist suppression of $1/Q^2$.
In addition, the theoretical cross section
depends on the unknown light-cone meson wave
function. 

Deeply virtual Compton scattering was first proposed
in Ref.~\cite{jiprl97,jidvcs} as a practical way to measure
the off-forward distributions. 
Consider virtual photon scattering in which 
the momenta of the incoming (outgoing)    
photon and nucleon are $q(q')$ and $P(P')$,
respectively. The Compton amplitude is defined
as
\begin{equation}
    T^{\mu\nu} = i\int d^4z e^{\bar q\cdot z}
    \left\langle P'\left|{\rm T}J^\mu\left(-{z\over 2}\right)
    J^\nu\left({z\over 2}\right)\right|P\right\rangle
\end{equation}
where $\overline q = (q+q')/2$. In the Bjorken
limit, $-q^2 $ and $ P\cdot q\rightarrow
\infty$ and their ratio remains finite,
the scattering is dominated by the single quark
process in which a quark absorbs    
the virtual photon, immediately radiates a real one,
and falls back to form the recoiling nucleon.
In the process, the initial and final photon
helicities remain the same. The leading-order
Compton amplitude is then
\begin{eqnarray}
     T^{\mu\nu} & =& g^{\mu\nu}_\perp \int^1_{-1}
       dx \left({1\over x-\xi+ i\epsilon}
       + {1\over x+\xi-i\epsilon}\right) \sum_q
       e_q^2 F_q(x, \xi, t, Q^2)
    \nonumber \\
 &+& i \epsilon^{\mu\nu\alpha\beta} p_\alpha n_\beta
      \int^1_{-1}
      dx  \left({1\over x-\xi + i\epsilon}
       - {1\over x+\xi-i\epsilon}\right)\sum_q e_q^2
      \tilde F_q(x, \xi, t, Q^2) 
\nonumber \\* \hspace*{-.2in}
\label{dvcsamp}
\end{eqnarray}
where $n$ and $p$ are the conjugate light-cone vectors
defined according to the collinear direction of
$\overline q$ and $\overline P$, and $g^{\mu\nu}_\perp$
is the metric tensor in transverse space. $\xi$ is related   
to the Bjorken variable $x_B = - q^2 /(2 P\cdot q)$
by $x_B=2\xi/(1+\xi)$.   

Much theoretical work has been devoted to DVCS 
in the last few years. The one-loop corrections to DVCS have
been studied by Ji and Osborne~\cite{jiosborne}. 
An all-order proof of the
DVCS factorization has been given 
in Ref.~\cite{radmeson,radyushkin97,rad99,jiosborne,collinsf}.  
Suggestions have also been made to test the 
DVCS scattering mechanism~\cite{diehl97}. Asymmetries for polarized
DVCS have been considered in~\cite{jidvcs} and reconsidered
in~\cite{fs,bmns00}. DVCS with double photon
helicity flips have been investigated 
in Ref.~\cite{hoodbhoyji,belitskym}. The estimates for 
cross sections have been made in Ref.~\cite{vg,guichon00}. 

Development on the experimental front is also
promising. Recently, both ZEUS and H1 collaborations 
have announced the first evidence for a DVCS signature 
\cite{zeush1}, and the HERMES 
collaboration has made a first measurement of the DVCS single-spin 
asymmetry~\cite{hermesdvcs}. More experiments
are planned for COMPASS, JLAB and other future facilities
\cite{hamburg00}. 
                                                                 
Heavy quarkonium production was first studied
by Ryskin as a way to measure the unpolarized gluon
distribution at small $x$~\cite{rys1}.
In the leading-order diagram, the virtual photon
fluctuates into a $c\bar c$ pair which subsequently
scatters off the nucleon target through two-gluon
exchange. In the process, the pair transfers
a certain amount of its longitudinal momentum and
reduces its invariant mass to that of a $J/\Psi$.
The cross section is
\begin{equation}
    {d\sigma \over dt}(\gamma^* +P\rightarrow
       J/\Psi+P') = {16\pi^3M\Gamma_{e^+e^-} \over 3\alpha_{\rm em}Q^6}
       \alpha_s^2(\overline Q^2)[\xi g(\xi, \overline Q^2)]^2   
\end{equation}
where $\overline Q^2 = (Q^2+M^2)/4$, $M$ is the
$J/\psi$ mass, and $\Gamma_{e^+e^-}$ is the decay
width into the lepton pair. The equation was derived
in the kinematic limit $s\gg Q^2\gg M^2\gg t$
and the Fermi motion of the quarks in the
meson was neglected. Two other important
approximations were used in the derivation.
First, the contribution from the real part
of the amplitude is neglected, which may be   
justifiable at small $x$. Second, the
off-forward distributions are identified with the
forward ones. 
                     
The above result was extended to the case
of light vector-meson production by
Brodsky et al., who considered the effects
of meson structure in perturbative QCD~\cite{bro1}.
They found a similar cross section,
\begin{equation}
  \left. {d\sigma \over dt}\right|_{t=0}
     (\gamma^*N\rightarrow VN)
   = {4\pi^3\Gamma_V m_V \alpha_s^2(Q)
   \eta_V^2\left(xg(x, Q^2)\right)^2 \over
    3\alpha_{\rm em} Q^6} \, ,
\end{equation}
where the dependence on the meson structure
is in the parameter
\begin{equation}
\eta_V = {1\over 2} \int {dz\over z(1-z)} \phi^V(z)
     \left( \int dz \phi^V(z) \right)^{-1} \, ,
\end{equation}
and $\phi^V(z)$ is the leading-twist light-cone
wave function. Evidently, the above formula
reduces to Ryskin's in the heavy-quark limit 
$(\phi^V(x) = \delta(x-1/2)).$

The amplitude
for hard diffractive electroproduction 
can be calculated in terms
of off-forward gluon distributions~\cite{radmeson}. With
the virtual photon and vector meson both polarized longitudinally
(i.e. determined using a Rosenbluth separation, with the vector
meson polarization measured via its decay products),
one finds
\begin{eqnarray}
 {d\sigma_{LL} \over dt}  &&      
     (\gamma^*N\rightarrow VN)
    =  {4\pi\Gamma_V m_V \alpha_s^2(Q)
   \eta_V^2 \over  3\alpha_{\rm em} Q^6} \nonumber\\
   && ~~ \times  \left|2x_B\int^1_{-1}dx
      \left({1\over x-\xi+i\epsilon} +
        {1\over x+\xi-i\epsilon}\right)F_g(x,\xi,t)\right|^2 \, ,
\end{eqnarray}
where again $x_B = 2\xi/(1+\xi)$. The above formula
is valid for any $x_B$ and $t$ smaller than typical
hadron mass scales. Hoodbhoy has also studied the effects of
the off-forward distributions in the case of
$J/\psi$ production~\cite{hoo1}. He found
that Ryskin's result needs to be modified
in a similar way once the off-forward effects
become important.        

More detailed theoretical studies of meson production 
have been done in Refs.~\cite{mankiewicz,martin,fpps}. 
Longitudinal $\rho^0$ production
data has been collected by the E665 and the HERMES 
collaborations~\cite{mesondata} and the comparison 
with model calculations is encouraging~\cite{vg,fks}.

\section{Related Topics in Spin Structure \label{sec:relat} }

In this section, we review two interesting topics
related to the nucleon spin.
First, we consider the Drell-Hearn-Gerasimov sum rule 
and its generalization to finite $Q^2$. Then we briefly
review polarized $\Lambda$ production from fragmentation of
polarized partons where the $\Lambda$ polarization can be measured 
through its weak non-leptonic decay. 

\subsection{The Drell-Hearn Gerasimov Sum Rule and Its
Generalizations \label{subsec:dhg} }

The Drell-Hearn-Gerasimov (DHG) sum rule~\cite{dhg} 
involves the spin-dependent photo-nucleon production
cross section. Consider a polarized 
real photon of energy $\nu$, scattering
from a longitudinally 
polarized nucleon and producing arbitrary hadronic
final states. The total cross sections 
are denoted as $\sigma_{{3\over 2},{1\over 2}}(\nu)$, where the 
subscripts 3/2
and 1/2 correspond to the helicity of the photon 
being parallel or antiparallel to the spin of 
the nucleon. The sum rule 
relates the $1/\nu$-weighted integral of the 
spin-dependent cross section from the inelastic 
threshold to infinity to the anomalous magnetic moment
of the nucleon $\kappa$, 
\begin{equation}
   \int^\infty_{\nu_{\rm in}}
   {d\nu\over \nu}(\sigma_{3\over 2}(\nu) - \sigma_{1\over 2}(\nu))
  = {2\pi^2\alpha_{\rm em} \kappa^2\over M^2} \ . 
\end{equation}
For the proton and neutron, the sum rule is $204.5\mu b$
and $232.8\mu b$, respectively. 

\begin{figure}[t]
\begin{center}
\includegraphics*[clip=,height=5cm,angle=0]{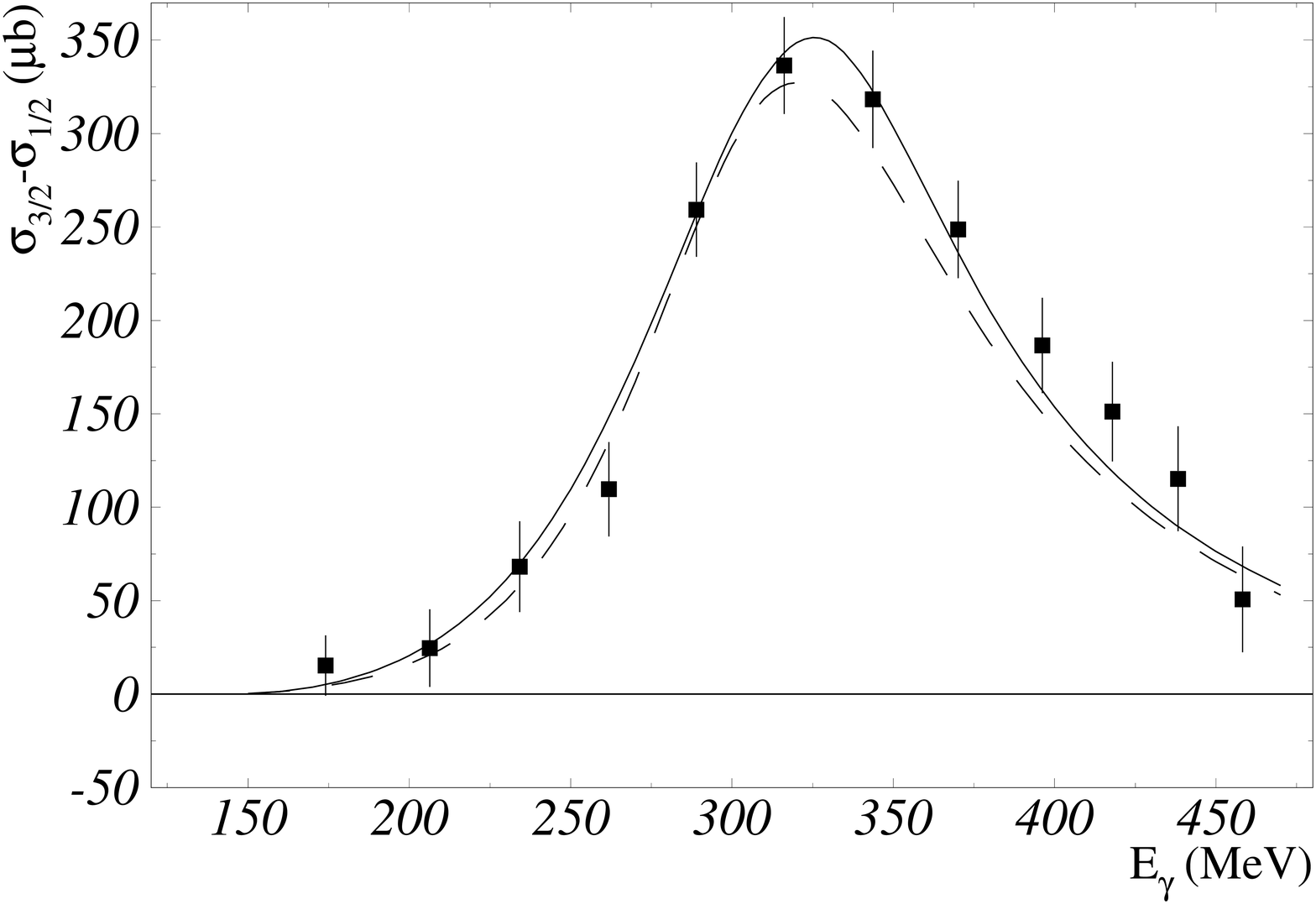}
\hspace{0.3in}
\includegraphics*[clip=,height=5cm,angle=0]{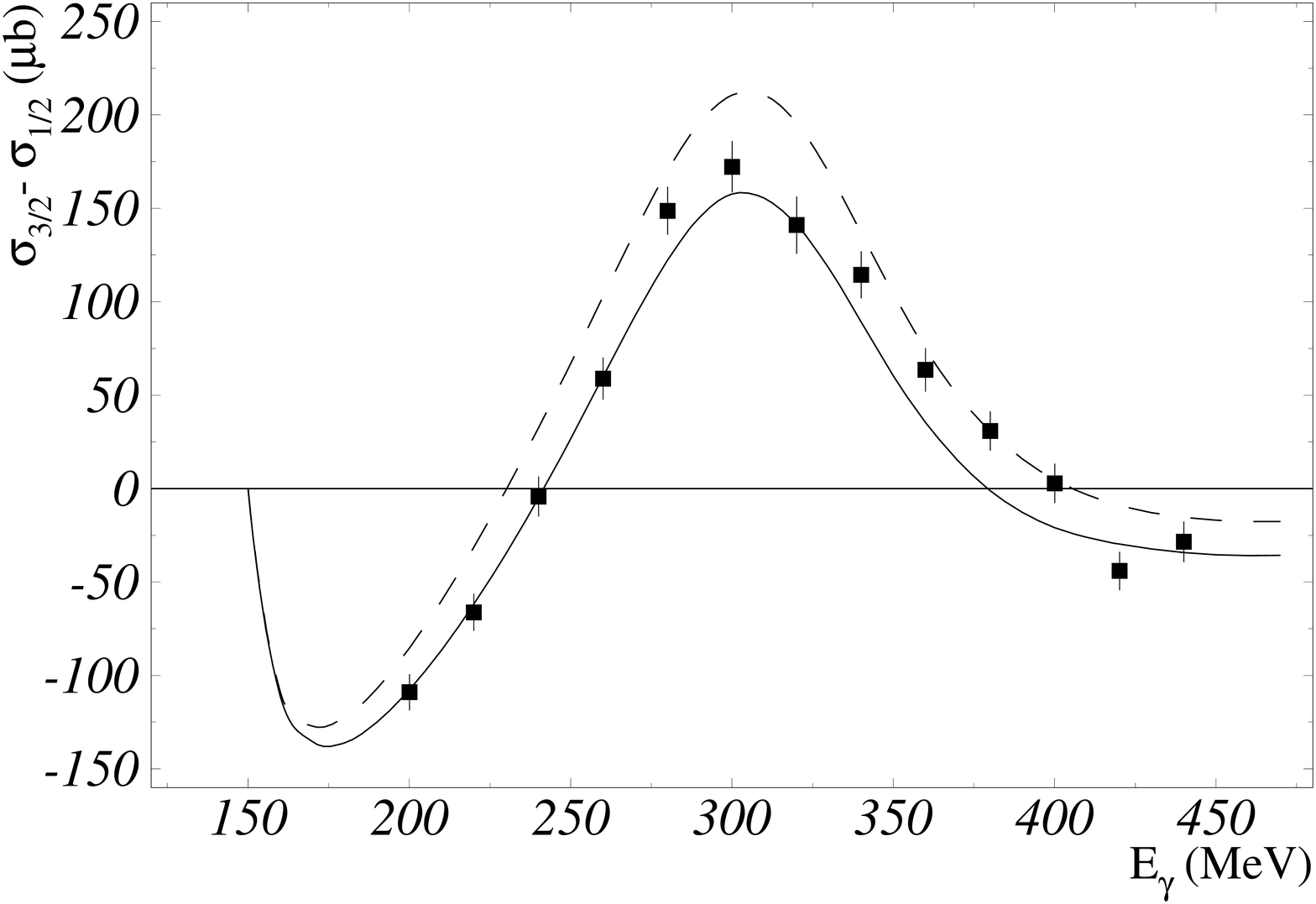}
\end{center}
\caption{Spin-dependent cross section for 
$\vec{\gamma}\vec{p}\rightarrow p\pi^0$ (upper) and $n\pi^+$ 
(lower) as a function
of the laboratory photon energy~\cite{ahrens}.}
\label{dhgdata}
\end{figure}

There has been much interest in recent years in testing 
the above sum rule by determining the integral on the left-hand 
side. Direct experimental
data on the spin-dependent photoproduction cross section
has become available recently~\cite{ahrens} 
(see Fig.~\ref{dhgdata}), and more data at higher energy 
are coming soon~\cite{moredhgdata}. However,
many of the published ``tests" in the literature
rely on theoretical models for the photoproduction 
helicity amplitudes which are only partially 
constrained by unpolarized photoproduction data 
\cite{vpisaid,hdt}. Because of the $1/\nu$ weighting, 
the low energy amplitudes play a dominant role 
in the DHG integral~\cite{drechselkrein}. In fact, one can show
that in the large $N_c$ limit, the integral is
entirely dominated by the $\Delta$ resonance 
contribution~\cite{cohenji}.

We will not discuss in detail how the 
phenomenological estimates of the DHG integral are done in 
the literature~\cite{karliner,workman,burkertli,sandorfi}. 
The interested reader can 
consult a recent review on the subject~\cite{drechsel95}. The main 
conclusion from these calculations is that
the isoscalar part of the sum rule ($219\mu b$) 
is approximately satisfied, whereas a large discrepancy
remains for the isovector part ($-14\mu$ b). 
Typically, the proton integral is estimated to be
in the range of $260\mu b$ to $290\mu b$. A more up-to-date 
analysis~\cite{drechsel00} including the recent data from MAMI
and the extrapolation of DIS data gives a result of
$202\pm 10 \mu b$ for the proton, but disagrees with the expected
neutron sum by $\sim 60 \mu b$. 

What do we learn about nucleon spin physics
by testing the sum rule? Moreover, the DHG sum rule is the analogue of 
the Bjorken sum rule at $Q^2=0$~\cite{anselmino89} (here, we 
discuss the Bjorken sum rule in the generalized
sense that the first moment of $g_1(x, Q^2)$ is related to 
nucleon axial charges in the asymptotic limit). 
If both sum rules are important to study, how do we extend
these sum rules away from the kinematic limits
($Q^2=0$ and $Q^2=\infty$)? Finally, how is 
the DHG sum rule evolved to the Bjorken sum rule
and what can we learn from the $Q^2$ evolution? 
In recent years, there has been much discussion 
in the literature about the generalized DHG integrals 
and their $Q^2$ dependence
\cite{burkertioffe,burkertli,bernard93,soffer93,dkht,scholten99}. 
A summary of different definitions of the generalized 
DHG integrals can be found in Ref.~\cite{pantforder,drechsel00}. 
As pointed out in~\cite{jiosborne99},
the key to addressing the above questions
is the dispersion relation for the spin-dependent
Compton amplitude $S_1(\nu, Q^2)$.

The virtual-photon forward scattering tensor 
defines the spin-dependent amplitude $S_{1}(\nu,
\omega)$, 
\begin{eqnarray}
   T^{\alpha\beta}(P, q)
  &=& i\int e^{iq\cdot\xi}
  d^4\xi \langle PS|{\rm T} J^\alpha(\xi)J^\beta(0)|PS\rangle \ , 
\nonumber \\
  & =& -i\epsilon^{\mu\nu\alpha\beta}
    q_\alpha S_\beta S_1(\nu, Q^2)/M^2 
  + ... \ , 
\end{eqnarray}
where $J^\mu$ is the electromagnetic current. 
From general principles, such as causality and unitarity as well 
as assumptions about the large-$\nu$ behavior of $S_1(Q^2,\nu)$, 
one can write down a dispersion relation 
\begin{equation}
   S_1(\nu, Q^2) 
  = 4 \int^\infty_{Q^2/2M}
   {\nu'd\nu' G_1(\nu', Q^2) \over \nu'^2-\nu^2}\ , 
\label{dispersion}
\end{equation}
where $G_1(\nu', Q^2)$ is the spin-dependent 
structure function discussed in Sec.~\ref{subsubsec:quark}
Whenever $S_1$ is known, in theory or experiment,
the above relation yields a dispersive sum rule. For instance, 
the Bjorken and DHG sum rules are obtained from theoretical predictions 
for $S_1(0, Q^2)$ at $Q^2=\infty, 0$, respectively
\cite{bjsum,low}.

What do we learn by testing these dispersive sum rules? 
First, we learn about the assumptions required for 
the derivation of the relation; in particular,
the high-energy behavior of the Compton amplitude 
\cite{regge}. Second, we learn about the scattering 
mechanisms in the virtual-Compton process. For
the Bjorken sum rule, it is perturbative QCD and
asymptotic freedom; for the DHG sum rule, it is 
nucleon-pole dominance and gauge symmetry~\cite{low}. 
Finally, if the sum rules are reliable, 
we have a new way to measure nucleon observables. In earlier
sections, we discussed how to extract $g_A$
and $\Delta \Sigma$ (the fraction of the nucleon spin carried
by quark spin) from polarized DIS data. Assuming
the validity of the DHG sum rule, we obtain the magnetic moment of the
nucleon from inclusive photoproduction. 

How do we extend these sum rules to other kinematic
regions? According to Eq.~\ref{dispersion},
the virtual Compton amplitude is the key. 
As discussed in Sec.~\ref{subsec:theory}, at large but finite $Q^2$, 
perturbative QCD introduces two types of corrections. 
The first are the radiative corrections: gluons are radiated 
and absorbed by active quarks, etc. The second are the higher 
twist corrections in which more than one parton from 
the target participates in the scattering. With these corrections, 
we can extend Bjorken's result for the Compton amplitude from 
$Q^2=\infty$ to finite $Q^2$~\cite{jiunrau,jm97}. 
Since the scale that controls the twist expansion is on 
the order of $0.1-0.2$ GeV$^2$, the perturbative
QCD prediction for  $S_1(0, Q^2)$ is valid down to $Q^2\sim 0.5$
GeV$^2$. Combined with Eq.~\ref{dispersion}, 
it yields a {\it generalized} Bjorken sum rule.
It is the generalized Bjorken sum rule that
is commonly tested experimentally. 

At small but finite $Q^2$, chiral perturbation theory
provides a sound theoretical method to calculate corrections 
to the low-energy theorem~\cite{bernard93,jiosborne99}. 
Recently, a fourth-order chiral perturbation theory
calculation for the inelastic part of $S_1(0,Q^2)$ yielded~\cite{koj}
\begin{equation}
   \bar S_1(0, Q^2)  
  =  -{\kappa^2}
  + {g_A^2M\over 12(4\pi f_\pi)^2m_\pi}
    (1+3\kappa_V+2(1+3\kappa_S)\tau^3)Q^2 + \cdots  
\end{equation}
The result shows a rapid $Q^2$-dependence near $Q^2\sim 0$, which
is qualitatively, though not quantitatively,  
consistent with a recent phenomenological 
analysis~\cite{drechsel00}. For a
quantitative test, one needs polarized electron scattering
data soon available from JLab~\cite{gdhgdata}.

How does the DHG sum rule at $Q^2=0$ evolve 
to the Bjorken sum rule at $Q^2=\infty$? The physically most
interesting quantity which connects both sum rules
is
\begin{equation}
   \Gamma(Q^2) \equiv {Q^2\over 8M^2} S_1(0, Q^2) 
          = {Q^2\over 8M^2} \bar S_1(0, Q^2) 
     + {1\over 2} F_1(Q^2)(F_1(Q^2)+F_2(Q^2)) \ , 
\end{equation}
where $F_1(Q^2)$ and $F_2(Q^2)$ are the elastic nucleon form 
factors. It is the elastic contribution
which dominates at low $Q^2$~\cite{jiosborne99}. 
$\Gamma(Q^2)$ starts at $1(0)+\kappa/2$ from the
proton (neutron) at $Q^2=0$
and rapidly decreases to about 0.2 at 
$Q^2=0.7$ GeV$^2$ and remains essentially 
flat as $Q^2\rightarrow \infty$. 
The interpretation for the $Q^2$ variation is 
as follows~\cite{jiosborne99}. The forward Compton amplitude
is an amplitude for the photon to scatter from a nucleon target 
and remain in the forward direction. This is very much
like a diffraction process and $\Gamma(Q^2)$ is the 
``brightness'' of the diffraction center. 
For low $Q^2$ photons, scattering from the different parts
of the proton is coherent, and the scattered photons
produce a large diffraction peak at the center.
As $Q^2$ becomes larger, the photon sees some large scale 
fluctuations in the nucleon; the scattering becomes
less coherent. The large scale fluctuations can largely be 
understood in terms of the dissociation of the 
nucleon into virtual hadrons. When $Q^2 > 0.5$ GeV$^2$, 
the photons see parton fluctuations at the scale
of $1/Q$. As $Q^2\rightarrow \infty$, the photons
see individual quarks inside the nucleon and the
scattering is completely incoherent. The diffraction 
peak is just the sum of diffractions from individual
quarks. In short, the $Q^2$ variation of the sum rules
reflects the change of the diffraction intensity
of the virtual photon as its mass is varied.

A clear theoretical understanding of the 
virtual photon diffraction at $Q^2\sim 0.1-0.5$ GeV$^2$ is 
not yet available, but there are
two distinct possibilities. First, there is a gap in which
neither parton nor hadron language describes the 
scattering well. In this case, an interesting 
theoretical question is how the 
transition from low to high $ Q^2$ happens.
Second, some extensions of the twist expansion 
and chiral perturbation theory may
overlap in the intermediate region. If so, we have
parton-hadron duality at a new level. In any case,
a lattice calculation of $S_1(0, Q^2)$ may shed 
important light on this~\cite{jijung}. 

\subsection{Spin-Dependent $\Lambda$ Fragmentation \label{subsec:frag} }

In the constituent quark model, the spin structure of 
the $\Lambda$ baryon is simple: the $ud$ 
quark pair couples to give zero angular momentum 
and isospin, and the spin of the $\Lambda$
is entirely carried by the spin of the remaining $s$ quark. 
From our present knowledge of the spin structure of 
the nucleon, we expect that this naive picture will fail to 
explain the actual spin structure of the $\Lambda$. 
In fact, if $SU(3)$ flavor 
symmetry is valid, we can deduce 
from the beta decay data
and polarized deep-inelastic scattering on the
nucleon that (in leading order) 
$\Delta u_\Lambda= \Delta d_\Lambda \sim -0.23$ 
and $\Delta s_\Lambda \sim 0.58$~\cite{burkardt93}.

Unfortunately the spin structure of the $\Lambda$ cannot 
be measured because of the lack of a stable target. 
However, the spin-dependent fragmentation of 
partons to the $\Lambda$ baryon can be studied experimentally 
because the $\Lambda$ polarization
can be measured through the self-analyzing decay 
$\Lambda\rightarrow p\pi^-$. The fragmentation 
functions are difficult to calculate in QCD, even in principle. 
We have little experience in modeling the 
fragmentation functions compared with 
the internal structure of the nucleon. Nevetherless, one
hopes that the spin physics in the fragmentation process
corroborates what we learn about the spin structure. 
Moreover, if a $\Lambda$ or $\bar \Lambda$ is exclusively produced 
from the fragmentation of a strange or
antistrange quark, respectively, the measurement of the $\Lambda$ 
polarization is a way to access the strange quark 
polarization in the nucleon. 

A relatively simple process from which the spin-dependent
fragmentations to $\Lambda$ can be studied is 
$e^+e^-$ annihilation with one of the beams
(say, electron) polarized. Considering only the intermediate 
photon state, the asymmetry in polarized $\Lambda$
production is
\begin{equation}
    {d^2\Delta \sigma(\vec{e^-}e^+\rightarrow \vec{\Lambda} X)
      \over d\Omega dz} 
   = {\alpha_{\rm em}^2\over  2s}
   \cos\theta \sum_q e_q^2 \left(\Delta \hat D_q(z) + \Delta \hat
   {\bar D_q}(z)\right) \ , 
\end{equation}
where $\Delta \hat D_q(z) = \hat D_q^+(z) -\hat D_q^-(z)$ is
a spin-dependent fragmentation function and $D_q^\pm(z)$
are the fragmentations of the quarks with helicities $\pm 1/2$
to a $\Lambda$ of helicity $+1/2$.
At the $Z^0$ peak, the parity violating
coupling induces polarizations in the quark-antiquark
pairs produced. Hence even without beam polarization,
the $\Lambda$ particles produced through fragmentations
are polarized~\cite{burkardt93}. 
Recently, several collaborations at LEP have extracted
the $\Lambda$ polarization from quark fragmentation
at the $Z_0$ peak~\cite{leplamdata}. A number of models for
spin-dependent quark fragmentation functions have
been proposed to explain the results~\cite{kotzinian98,
florian98,boros98,ma00}, and data are consistent
with very different scenarios about the flavor
structure of fragmentation. 

The polarized fragmentation functions can also be measured
in deep-inelastic scattering, in which the polarized
beam produces a polarized quark from an unpolarized target,
which then fragments~\cite{jaffe96}.
Within the QPM, the measured $\Lambda$ polarization from 
a lepton beam with polarization $P_l$ is,
\begin{equation}
    P_{\rm exp} = P_b D(y) 
   {\sum_a e_a^2 q_a(x, Q^2) \Delta \hat q_a(z, Q^2)
   \over \sum_a e_a^2 q_a(x, Q^2) \hat q_a(z, Q^2)}
\end{equation}
where $D(y)$ is the depolarization factor. A 
process-independent $\Lambda$ polarization can be defined
from $P_\Lambda = P_{\rm exp}/(P_b D(y))$. 

The $\Lambda$ polarization from DIS scattering was first
measured by the E665 Collaboration with a 470 GeV/c$^2$ polarized
muon beam ($P_\mu=-0.7\pm 0.1$)~\cite{adams99}. 
The data sample was taken at $10^{-4}<x_B<10^{-1}$ with
$\langle x_B\rangle = 5\cdot 10^{-3}$, $0.25<Q^2<2.5$ GeV$^2$
with $\langle Q^2\rangle=1.3$ GeV$^2$, and $\langle \nu\rangle
=150$ GeV. The $\Lambda$ polarization was
found to be $-1.2\pm 0.5$ at $0<x_F<0.3$ and $-0.32\pm 0.7$
at $0.3<x_F<1.0$. The $\bar \Lambda$ polarization
was $0.26\pm 0.6$ and $1.1\pm 0.8$ for the two bins, 
respectively. The comparisons with different
fragmentation models can be found in Refs.~\cite{lipkin,ma}. 

Recently, HERMES has also reported a measurement of
the $\Lambda$ polarization from polarized
deep-inelastic positron scattering from an unpolarized proton
target. The result is $P_\Lambda = 0.11\pm 0.17\pm 0.03$ at 
an average $z=0.45$~\cite{airapetian99}. The result
seems to be consistent with the assumption of the naive quark
model that the $\Lambda$ polarization is entirely 
carried by the valence $s$ quark~\cite{florian98}. 

In Ref.~\cite{florian981}, predictions for $\Lambda$ production from 
$\vec{p} - p$ collisions at RHIC and HERA-${\vec N}$ with a single
beam polarization was studied.
Spin asymmetry measurements as a function of 
the rapidity provide a way to 
discriminate various models of the 
spin-dependent fragmentation. The main theoretical 
uncertainties, such as the NLO corrections and
the unknown polarized parton distributions, 
have no major impact on the asymmetry. In Ref.~\cite{boros00},
it is argued that the hyper-fine interaction responsible
for the N-$\Delta$ mass splitting induces a sizable
fragmentation of polarized up and down quarks into a $\Lambda$,  
which leads to large positive $\Lambda$ polarizations at 
large rapidity. 

\section{Conclusions}

Since the EMC publication of the measurement on 
the fraction of the nucleon spin carried by quarks,
understanding the spin structure of the nucleon has become 
an important subfield in hadron physics. In this
review, we have tried to highlight some of the important
developments over the last ten years and discuss some of the future 
prospects in this field. 

\section{Acknowledgements}

The authors would like to thank J. W. Martin for a careful reading of the 
manuscript.


\begin{thebibliography}{99}

\bibitem{d0direct}(D0) Abbott et al., Phys. Rev. Lett. 84, 2786 (2000). 

\bibitem{e143p1}(E143) K. Abe et al., \Journal{\PRL}{74}{346}{1995}.

\bibitem{e143d}(E143) K. Abe et al., \Journal{\PRL}{75}{25}{1995}.

\bibitem{e143q2}(E143) K. Abe et al., \Journal{\PLB}{364}{61}{1995}.

\bibitem{g2e143}(E143) K. Abe et al., \Journal{\PRL}{76}{587}{1996}.

\bibitem{e154n1}(E154) K. Abe et al., \Journal{\PRL}{79}{26}{1997}

\bibitem{g2e154}(E154) K. Abe et al., \Journal{\PLB}{404}{377}{1997}.

\bibitem{e154nlo}(E154) K. Abe et al., \Journal{\PLB}{405}{180}{1997}.

\bibitem{e143fin}(E143) K. Abe et al., \Journal{\PRD}{58}{112003}{1998}. 

\bibitem{hermesn}(HERMES) K. Ackerstaff et al., \Journal{\PLB}{404}{383}{1997}.

\bibitem{hermspec}(HERMES) K. Ackerstaff, et al., \Journal{\NIMA}{417}{230}{1998}.

\bibitem{hermessemi}(HERMES)K. Ackerstaff et al., \Journal{\PLB}{464}{123}{1999}.

\bibitem{smcp1}(SMC) D. Adams et al., \Journal{\PLB}{329}{399}{1994}.

\bibitem{g2smc}(SMC) D. Adams et al., \Journal{\PLB}{336}{125}{1994}.

\bibitem{smcd2}(SMC) D. Adams et al., \Journal{\PLB}{357}{248}{1995}.

\bibitem{smcp2big}(SMC) D. Adams et al., \Journal{\PRD}{56}{5330}{1997}.       

\bibitem{smcd3}(SMC) D. Adams et al., \Journal{\PLB}{396}{338}{1997}.

\bibitem{adams99}(E665) M. R. Adams, hep-ex/9911004. 

\bibitem{smcd1}(SMC) B. Adeva et al., \Journal{\PLB}{302}{533}{1993}.

\bibitem{smcsemi1}(SMC) B. Adeva et al., \Journal{\PLB}{369}{93}{1996}.

\bibitem{smcp2}(SMC) B. Adeva et al., \Journal{\PLB}{412}{414}{1997}.      

\bibitem{smcfin}(SMC) B. Adeva et al., \Journal{\PRD}{58}{112001}{1998}.

\bibitem{smcsemi2}(SMC) B. Adeva et al., \Journal{\PLB}{420}{180}{1998}.      

\bibitem{smcnlo}(SMC) B. Adeva et al., \Journal{\PLB}{58}{112002}{1998}.       

\bibitem{smcloq}B. Adeva et al., \Journal{\PRD}{60}{072004}{1999}.

\bibitem{ahmed}M. A. Ahmed and G. G. Ross, \Journal{\NPB}{111}{441}{1976}.

\bibitem{ahrens}(GDH/A2) J. Ahrens et al., \Journal{\PRL}{84}{5950}{2000}.

\bibitem{moredhgdata}
For example, J. Ahrens et al., Mainz MAMI Proposal A2/2-95; G. Anton
et. al., Bonn ELSA Proposal, 1992. 

\bibitem{h195}(H1) S. Aid et al., \Journal{\NPB}{445}{3}{1995};
contributed paper to ICHEP97, Jerusalem (1997). 

\bibitem{hermesp}(HERMES) A. Airapetian et al., \Journal{\PLB}{442}{484}{1998}.

\bibitem{airapetian99}(HERMES) A. Airapetian et al., hep-ex/9911017. 

\bibitem{hermeshipt}(HERMES) A. Airapetian et al., \Journal{\PRL}{84}{2584}{2000}. 

\bibitem{airapetian00}(HERMES) A. Airapetian et. al., \Journal{\PRL}{84}{4047}{2000}.

\bibitem{radcor} I.V.~Akushevich and N.M.~Shumeiko, \Journal{\JPG}{20}{513}{1994}.

\bibitem{e80}(E80) M. J. Alguard et al., \Journal{\PRL}{37}{1261}{1976},\Journal{\PRL}{
41}{70}{1976}.

\bibitem{abh}
A. Ali, V. M. Braun and G. Hiller, \Journal{\PLB}{266}{117}{1991}.

\bibitem{altarelli}
G. Altarelli and G. G. Ross, 
\Journal{\PLB}{212}{391}{1988}; 
R. D. Carlitz, J. C. Collins, and A. H. Mueller, 
\Journal{\PLB}{214}{229}{1988}.  

\bibitem{qcd5}G. Altarelli, R.D. Ball, S. Forte and G. Ridolfi
\Journal{\APP}{B29}{1145}{1998}.

\bibitem{afr98}G. Altarelli, S. Forte and G. Ridolfi, \Journal{\NPB}{534}{277}{1998}.

\bibitem{hermesdvcs}
M. Amarian (for HERMES Collaboration), 
talk given at the DESY workshop on Skewed Parton Distributions and 
Lepton-Nucleon
Scattering, Sept., 2000, Hamburg, Germany.

\bibitem{ans95}M. Anselmino, M. Boglione and F. Murgia, 
\Journal{\PLB}{362}{164}{1995}.

\bibitem{anselmino99}
M. Anselmino, M. Boglione and F. Murgia,
\Journal{\PRD}{60}{054027}{1999}.

\bibitem{ael95}M. Anselmino, A. Efremov, and E. Leader, 
\Journal{\PR}{261}{1}{1995}; Erratum:\Journal{\PR}{281}{399}{1997}. 

\bibitem{anselmino89}
M. Anselmino, B. L. Ioffe, and E. Leader, 
\Journal{\SJN}{49}{136}{1989}.

\bibitem{e142n1}(E142) P. L. Anthony et al., \Journal{\PRL}{71}{959}{1993}.

\bibitem{e142fin}(E142) P.L. Anthony et al., \Journal{\PRD}{54}{6620}{1996}. 

\bibitem{g2e155}(E155) P. L. Anthony {\it et al.}, 
\Journal{\PLB}{458}{530}{1999}.

\bibitem{e155d1}(E155) P. L. Anthony et al., \Journal{\PLB}{463}{339}{1999}.

\bibitem{e155pd}(E155) P. L. Anthony et al., hep-ph/0007248.

\bibitem{ao97}S. Aoki et al., \Journal{\PRD}{56}{433}{1997}.

\bibitem{owens99}L. Apanasevich et al., \Journal{\PRD}{59}{074007}{1999}.

\bibitem{vpisaid}
R. A. Arndt, I. I. Strakovsky, and R. 
Workman, \Journal{\PRC}{53}{430}{1996}.

\bibitem{EMCfrag}(EMC) A. Arneodo et al,. \Journal{\NPB}{321}{541}{1989}.

\bibitem{mesondata}(NMC) M. Arneodo et al., \Journal{\NPB}{429}{503}{1994};
(E665) M. R. Adams et al., \Journal{\ZPC}{74}{237}{1997};
(ZEUS) Collaboration, M. Derrick et al., \Journal{\PLB}{356}{601}{1995};
J. Breitweg et al., \Journal{\EJC}{6}{603}{1999};
(HERMES) A. Airapetian et al., hep-ex/0004023.  

\bibitem{am90}X. Artru and M. Mekhfi, \Journal{\ZPC}{45}{669}{1990}.

\bibitem{lipkin}
D. Ashery and H. J. Lipkin, \Journal{\PLB}{469}{263}{1999}.

\bibitem{emc1}(EMC) J. Ashman et al., \Journal{\PLB}{206}{364}{1988}.

\bibitem{emc2}(EMC) J. Ashman et al., \Journal{\NPB}{328}{1}{1989}.

\bibitem{jlabg2}T. Averett and W. Korsch, et al., JLAB Proposal E97-103,
(www.jlab.org).

\bibitem{othertw3}
I. I. Balitsky and V. M. Braun, \Journal{\NPB}{311}{541}{1989}; 
X. Ji and C. Chou, \Journal{\PRD}{42}{3637}{1990};
B. Geyer, D. M\"uller and D. Robaschik, hep-ph/9611452;
J. Kodaira, Y. Yasui, K. Tanaka and T. Uematsu, \Journal{\PLB}{387}{855}{1996}.

\bibitem{bal1}
I. I. Balitsky and X. Ji, \Journal{\PRL}{79}{1225}{1997}.

\bibitem{bal90}I. I. Balitsky, V. M. Braun and A. V. Kolesnichenko, 
\Journal{\PLB}{242}{245}{1990};\Journal{\PLB}{318}{648}{1993}.

\bibitem{bfr96}R. D. Ball, S. Forte and G. Ridolfi, \Journal{\PLB}{378}{255}{1996}.

\bibitem{hermtp}D. P. Barber, et al., \Journal{\NIMA}{329}{79}{1993}.

\bibitem{barone}
V. Barone, T. Calarco, A. Drago, 
\Journal{\PLB}{431}{405}{1998}. 

\bibitem{e130}(E130) G. Baum et al.,\Journal{\PRL}{51}{1135}{1983},\Journal{\PRL}{45}{2
000}{1980}.

\bibitem{belitskym}
A. V. Belitsky and D. Muller, \Journal{\PLB}{486}{369}{2000}. 

\bibitem{bmns00}
A. Belitsky, D. Muller, L. Niedermeier, and A. Schafer,
hep-ph/0004059. 

\bibitem{twoloop}
A. V. Belitsky, D. M\"uller, \Journal{\NPB}{527}{207}{1998};
\Journal{\NPB}{537}{397}{1999}; 
\Journal{\NPB}{546}{279}{1999}; 
\Journal{\PLB}{450}{126}{1999}; 
\Journal{\PLB}{464}{249}{1999}; 
\Journal{\PLB}{486}{369}{2000};
A. V. Belitsky, D. M\"uller, L. Niedermeier, 
A. Schafer, \Journal{\PLB}{437}{160}{1998}; 
A. V. Belitsky, A. Freund, and D. M\"uller,
\Journal{\PLB}{461}{270}{1999}. 
\Journal{\NPB}{574}{347}{2000}; hep-ph/0008005. 
 
\bibitem{bergerqiu}
E. Berger and J. W. Qiu, \Journal{\PRD}{44}{2002}{1991}. 

\bibitem{bernard93}
V. Bernard, N. Kaiser, and Ulf-G. Meissner, 
\Journal{\PRD}{48}{3062}{1993}. 

\bibitem{spol}
V. Bernard, N. Kaiser, and U. -G. Meissner, \Journal{\IJE}{4}{193}{1995};
X. Ji, C. Kao, and J. Osborne, \Journal{\PRD}{61}{074003}{2000}; 
K. B. Vijaya Kumar, J. A. McGovern, and M. C. Birse, hep-ph/9909442; 
\Journal{\PLB}{479}{167}{2000}; 
G. C. Gellas, T. R. Hemmert, U.-G. Meissner, \Journal{\PRL}{85}{12}{2000}. 

\bibitem{nlofrag}J. Binnewies, B.A. Kniehl, G. Kramer, 
\Journal{\ZPC}{65}{471}{1995}; J. Binnewies,  hep-ph/9707269.

\bibitem{bjsum}J. D. Bjorken, \Journal{\PR}{148}{1467}{1966}; 
\Journal{\PRD}{1}{1376}{1970}.

\bibitem{blank} R. Blankleider and R. M. Woloshyn,\Journal{\PRC}{29}{538}{1984}.

\bibitem{boer98}
D. Boer and P. Mulders, \Journal{\PRD}{57}{5780}{1998};
D. Boer, P. J. Mulders, and O. Teryaev, \Journal{\PRD}{57}{3057}{1998}. 

\bibitem{bof93}S. Boffi, C. Giusti and F. D. Pacati, \Journal{\PR}{226}{1}{1993}.

\bibitem{boglione99}
M. Boglione and P. Mulders, \Journal{\PRD}{60}{054007}{1999}. 

\bibitem{bojak99}
I. Bojak and M. Stratmann, \Journal{\NPB}{540}{345}{1999}. 

\bibitem{boros98}
C. Boros and Z. Liang, \Journal{\PRD}{57}{4491}{1998}. 

\bibitem{liang}
C. Boros, Z. Liang, and T. Meng, 
\Journal{\PRL}{70}{1751}{1993}; \Journal{\PRD}{49}{3759}{1994}. 

\bibitem{boros00}
C. Boros, J. T. Londergan, and A. W. Thomas, \Journal{\PRD}{61}{014007}{2000};
\Journal{\PRD}{62}{014021}{2000}. 

\bibitem{g2e155x}(E155x) P. Bosted (for E155x Collaboration),  
\Journal{\NPA}{663}{297}{2000}.

\bibitem{soffer}
C. Bourrely, J. Ph Guillet, and J. Soffer,
\Journal{\NPB}{361}{72}{1991}; 
P. Chiappetta, P. Colangelo, J. Ph Guillet, and G. Nardulli, 
\Journal{\ZPC}{59}{629}{1993}; 
J. Soffer and J. M. Virey, \Journal{\NPB}{509}{297}{1998}. 

\bibitem{bs95} C. Bourrely and J. Soffer, \Journal{\NPB}{445}{341}{1995}.

\bibitem{bou98} C. Bourrely, F. Buccella, O. Pisanti, P. Santorelli, and J. 
Soffer, \Journal{\PTP}{99}{1017}{1998}.

\bibitem{bravar98}A. Bravar, D. von Harrach, 
and A. Kotzinian, \Journal{\PLB}{421}{349}{1998}.

\bibitem{smcssa}(SMC) A. Bravar, \Journal{\NPB}{79}{520c}{1999}. 

\bibitem{bro1}
S. J. Brodsky, L. L. Frankfurt, J. F. Gunion, A. H. Mueller,
and M. Strikman, \Journal{\PRD}{50}{3134}{1994}.  

\bibitem{bkl}
A. P. Bukhvostov, E. A. Kuraev, and L. N. Lipatov, \Journal{\SJN}{38}{263}{1983}; 
\Journal{\JETPL}{37}{484}{1983}; 
\Journal{\SJN}{39}{121}{1983};  
\Journal{JETP}{60}{22}{1984}.

\bibitem{rhicspin}
G. Bunce, N. Saito, J. Soffer, and W. Vogelsang, hep-ph/0007218, 
to be published {\em Ann. Rev. Nucl. Part. Sci.}.

\bibitem{burkardt93}
M. Burkardt and R. L. Jaffe, \Journal{\PRL}{70}{2537}{1993}. 

\bibitem{burkertioffe}
V. D. Burkert and B. L. Ioffe, \Journal{\PLB}{296}{223}{1992}; 

\bibitem{burkertli}
V. Burkert and Z. Li, \Journal{\PRD}{47}{46}{1993}. 

\bibitem{gdhgdata}
V. D. Burkert et al., JLab proposal 91-23 (1991);
S. Kuhn et al., JLab proposal 93-09 (1993); 
Z. E. Meziani et al., Jlab proposal 94-10 (1994); 
J. P. Chen et al., Jlab proposal 97-110 (1997);
(www.jlab.org). 

\bibitem{bc}
H. Burkhardt and W. N. Cottingham, \Journal{\APNY}{56}{453}{1970}.

\bibitem{leplamdata}
(ALEPH) D. Buskulic et al., \Journal{\PLB}{374}{319}{1996}; 
(DELPHI) Report No. DELPHI 95-86
PHYS 521, CERN-PPE-95-172;
(OPAL) K. Ackerstaff, \Journal{\EJC}{2}{49}{1998}. 

\bibitem{zeush1}(ZEUS/H1)
P. Bussy (for ZEUS Collaboration) and E. Lobodzinska
(for H1 Collaboration), talks given
at the DESY workshop on Skewed Parton Distributions and Lepton-Nucleon
Scattering, Sept., 2000, Hamburg, Germany.

\bibitem{butterworthxx}
J. M. Butterworth, N. Goodman, M. Stratmann, W. Vogelsang, 
{\em Proceedings of the 1997 Workshop on Physics with Polarized Protons
at HERA}, Hamburg, Germany, Sept. 1997.

\bibitem{ck}R. Carlitz and J. Kaur, \Journal{\PRL}{38}{673}{1976}.

\bibitem{ccm88}R. D. Carlitz, J. C. Collins and A. H. Mueller, \Journal{\PLB}{214}{229}{1988}.

\bibitem{catani98}
S. Catani, M. L. Mangano, and P. Nason, JHEP 9807, 024 (1998). 

\bibitem{disev}
Z. Chen, \Journal{\NPB}{525}{369}{1998};
L. L. Frankfurt et al., \Journal{\PLB}{418}{345}{1998}. 
J. Bl\"ulein, B. Geyer, D. Robaschik, \Journal{\PLB}{413}{114}{1997}; 
I. I. Balitsky and A. V. Radyushkin, \Journal{\PLB}{413}{114}{1997}.
A. V. Belitsky and D. M\"uller, \Journal{\PLB}{417}{129}{1998}.  

\bibitem{che98}H-Y Cheng, \Journal{\PLB}{427}{371}{1998}.

\bibitem{che00}H-Y Cheng, \Journal{\CJP}{38}{753}{2000}.

\bibitem{nuclcorr} C.~Ciofi degli Atti et al.,\Journal{\PRC}{48}{968}{1993}.

\bibitem{cohenji}
T. D. Cohen and X. Ji, \Journal{\PLB}{474}{251}{2000}. 

\bibitem{collins}
J. Collins, \Journal{\NPB}{394}{169}{1993}. 

\bibitem{collinsf}
J. C. Collins and A. Freund, \Journal{\PRD}{59}{074009}{1999}. 

\bibitem{cfs}
J. C. Collins, L. Frankfurt, and M. Strikman, \Journal{\PRD}{56}{2982}{1997}.

\bibitem{compass}(COMPASS) CERN/SPSLC 96-14 (1996); http://wwwcompass.cern.ch/.

\bibitem{contogouris00}
A. P. Contogouris, Z. Merebashvili, and G. Grispos, 
\Journal{\PLB}{482}{1}{2000}. 

\bibitem{florian99}
D. de Florian, S. Frixione, A. Signer, and W. Vogelsang, 
\Journal{\NPB}{539}{455}{1999}. 

\bibitem{seminlo}D. de Florian, C. A. Garcia Canal and R. Sassot
\Journal{\NPB}{470}{195}{1996}; 
D. de Florian, et al., \Journal{\PLB}{389}{358}{1996}; 
E. Christova and E. Leader, hep-ph/0007303.

\bibitem{semifits}D. de Florian, O. A. Sampayo and R. Sassot, 
\Journal{\PRD}{57}{5803}{1998}. D. de Florian and R. Sassot, 
\Journal{\PRD}{62}{094025}{2000}.

\bibitem{florian98}
D. de Florian et al., \Journal{\PRD}{57}{5811}{1998}. 

\bibitem{florian981}
D. de Florian, M. Stratmann, and W. Vogelsang, \Journal{\PRL}{81}{530}{1998}; 
D. de Florian, J. Soffer, M. Stratmann, W. Vogelsang, 
\Journal{\PLB}{439}{176}{1998}. 

\bibitem{deroeck96}
A. De Roeck et. al, hep-ph/9610315.

\bibitem{deroeck99}
A. De Roeck et. al., \Journal{\EJC}{6}{121}{1999}. 

\bibitem{phera}
A. Deshpande, talk at the workshop on "Plarized Protons at
High Energies: Accelerator Challenges and Physics Opportunities", 
DESY, Hamburg, May, 1999. 

\bibitem{hamburg00}
DESY workshop on Skewed Parton Distributions and Lepton-Nucleon
Scattering, Sept., 2000, Hamburg, Germany.

\bibitem{diehl97}
M. Diehl, T. Gousset, B. Pire, and J. P. Ralston, 
\Journal{\PLB}{411}{193}{1997}. 

\bibitem{dittes88}
F. M. Dittes et al., \Journal{\PLB}{209}{325}{1988}; 
D. M\"uller et al., \Journal{\FP}{42}{101}{1994}. 

\bibitem{posit}M. G. Doncel and E. de Rafael, \Journal{\NCA}{4A}{363}{1971}.

\bibitem{don95}S. J. Dong, J-F. Lagae, and K-F. Liu, \Journal{\PRL}{75}{2096}{1995}.

\bibitem{dhg}
S. D. Drell and A. C. Hearn, \Journal{\PRL}{16}{908}{1966}; 
S. B. Gerasimov, \Journal{\SJN}{2}{430}{1966}.

\bibitem{drechsel95} 
D. Drechsel, \Journal{\PPNP}{34}{181}{1995}. 

\bibitem{drechselkrein}
D. Drechsel and G. Krein, \Journal{\PRD}{58}{116009}{1998}. 

\bibitem{drechsel00}
D. Drechsel, S. S. Kamalov, and L. Tiator, hep/ph-0008306. 

\bibitem{dkht}
D. Drechsel, S. S. Kamalov, G. Krein, and L. Tiator, 
\Journal{\PRD}{59}{094021}{1999}. 

\bibitem{et}
A. V. Efremov and O. V. Teryaev, \Journal{\SJN}{36}{140}{1982}; 
A. V. Efremov and O. V. Teryaev, \Journal{\YF}{39}{1517}{1984}. 

\bibitem{efremov}
A. V. Efremov and O. V. Teryaev, \Journal{\PLB}{150}{383}{1985}. 

\bibitem{ssamod}A. V. Efremov, et al., \Journal{\PLB}{478}{94}{2000}.

\bibitem{ehrnsperger}
B. Ehrsperger, A. Sch\"afer, \Journal{\PRD}{52}{2709}{1995}. 

\bibitem{ek88} J. Ellis and M. Karliner, \Journal{\PLB}{213}{73}{1988}.

\bibitem{ek95}J. Ellis and M. Karliner, \Journal{\PLB}{341}{397}{1995}.

\bibitem{ejsum}J. Ellis and R. Jaffe, \Journal{\PRD}{9}{1444}{1974};
\Journal{\PRD}{10}{1669E}{1974}.

\bibitem{efp}
R. K. Ellis, W. Furmanski, and Petronzio, 
\Journal{\NPB}{212}{29}{1983}. 

\bibitem{ellis93}
S. D. Ellis and S. Soper, \Journal{\PRD}{48}{3160}{1993}. 

\bibitem{lptheory}
J. Felix, \Journal{\MPLA}{14}{827}{1999}; 
J. Soffer, hep-ph/9911373; 
J. Soffer and N. E. Tornqvist, \Journal{\PRL}{68}{907}{1992}; 
Y. Hama and T. Kodama, \Journal{\PRD}{48}{3116}{1993}; 
R. Barni, G. Preparata and P. Ratcliffe, \Journal{\PLB}{296}{251}{1992}; 
S. M. Troshin and N. E. Tyurin, \Journal{\PRD}{55}{1265}{1997}; 
Z. T. Liang and C. Boros, \Journal{\PRL}{79}{3608}{1997}; 
\Journal{\PRD}{61}{117503}{2000}. 

\bibitem{feltesse96}
J. Feltesse, F. Kunne, E. Mirkes, \Journal{\PLB}{388}{832}{1996}.

\bibitem{g2}
R. F. Feynman, {\it Photon-Hadron Interactions}
(Benjamin, New York, 1972). 

\bibitem{feynman}
R. F. Feynman, Feynman Lecture in Physics, Vol. III. 
(Addison-Wesley, Reading, 1963).

\bibitem{fontannaz81}
M. Fontannaz, D. Schiff, B. Pire, \Journal{\ZPC}{8}{349}{1981}.

\bibitem{fs}
A. Freund and M. Strikman, \Journal{\PRD}{60}{071501}{1999}. 

\bibitem{friar} J.L.~Friar et al., \Journal{\PRC}{42}{2310}{1990}.

\bibitem{fri00}S. Frixione and W. Vogelsang, \Journal{\NPB}{568}{60}{2000}.

\bibitem{fks}
L. Frankfurt, W. Koepf, and M. Strikman, \Journal{\PRD}{54}{3194}{1996}. 

\bibitem{fpps}
L. L. Frankfurt, P. Pobylitsa, M. V. Polyakov, and 
M. Strikman, \Journal{\PRD}{60}{014010}{1999}. 

\bibitem{fuk95}M. Fukugita, Y. Kuramashi, M. Okawa, and A. Ukawa, \Journal{\PRL}{75}{2092}{1995}.

\bibitem{gs}T. Gehrmann and W. J. Stirling, \Journal{\PRD}{53}{6100}{1996}.

\bibitem{gp74} H. Georgi and H. D. Politzer, \Journal{\PRD}{9}{416}{1974}.

\bibitem{gluck88}
M. Gl\"uck and E. Reya, \Journal{\ZPC}{39}{569}{1988}; 
\Journal{\PLB}{83}{98}{1979}. 

\bibitem{gluck91}
M. Gl\"uck, E. Reya, and W. Vogelsang, \Journal{\NPB}{351}{579}{1991}; 
A. D. Watson, \Journal{\ZPC}{12}{123}{1982}. 

\bibitem{gluck92}
M. Gl\"uck and W. Vogelsang, \Journal{\ZPC}{55}{353}{1992}; 
\Journal{\ZPC}{57}{309}{1993}; 
M. Gl\"uck, M. Stratmann, and W. Vogelsang, \Journal{\PLB}{337}{373}{1994}. 

\bibitem{grsv} M. Gluck et al., \Journal{\PRD}{53}{4775}{1996}.

\bibitem{grv}M. Gluck, E. Reya and A. Vogt, \Journal{\ZPC}{67}{433}{1995},
\Journal{\EJC}{5}{461}{1998}.

\bibitem{goc96}M. Gockeler, et al., \Journal{\PRD}{53}{2317}{1996}.

\bibitem{gockeler}
G\"ockeler et al., hep-ph/9909253. 

\bibitem{martin}
K. J. Golec-Biernat, A. D. Martin, M. G. Ryskin, 
\Journal{\PLB}{456}{232}{1999}; 
A. G. Shuvaev, K. J. Golec-Biernat, A. D. Martin, M. G. Ryskin, 
\Journal{\PRD}{60}{014015}{1999};
A. D. Martin, M. G. Ryskin, \Journal{\PRD}{57}{6692}{1998};
\Journal{\PRD}{62}{014002}{2000}. 

\bibitem{got00} Y. Goto, et al., \Journal{\PRD}{62}{034017}{2000}.

\bibitem{gor98} L. E. Gordon, M. Goshtasbpour, and G. P. Ramsey, 
\Journal{\PRD}{58}{094017}{1998}; 
G. P. Ramsey, \Journal{\PPNP}{39}{599}{1997}. 

\bibitem{dglap}V. N. Gribov and L. N. Lipatov, \Journal{\SJN}{15}{138}{1972}, 
Yu. L. Dokahitzer, \Journal{\JETP}{16}{161}{1977},
G. Altarelli and G. Parisi, \Journal{\NPB}{126}{298}{1977}. 

\bibitem{gro}
D. Gross and F. Wilczek, \Journal{\PRD}{9}{980}{1974}. 

\bibitem{gus99}S. Guskin et al., \Journal{\PRD}{59}{114502}{1999}.

\bibitem{hdt}
O. Hanstein et al., \Journal{\NPA}{632}{561}{1999}; 
D. Dreschsel et al., \Journal{\NPA}{645}{145}{1999}. 

\bibitem{star}(STAR) J. W. Harris, et al., \Journal{\NPA}{566}{277c}{1994}.

\bibitem{heji95}H. He and X. Ji, \Journal{\PRD}{52}{2960}{1995}.

\bibitem{heji96}H. He and X. Ji, \Journal{\PRD}{54}{6897}{1996}.

\bibitem{regge}
R. L. Heimann, \Journal{\NPB}{64}{429}{1973};
P. V. Landshoff and O. Nachtmann, \Journal{\ZPC}{35}{405}{1987};
J. Ellis and M. Karliner, \Journal{\PLB}{213}{73}{1988};
F. E. Close and R. G. Roberts, \Journal{\PLB}{336}{257}{1994}.

\bibitem{heller}
For a review of data, see K. Heller, in Proceedings of
Spin 96, D. W. de Jager, T. J. Ketel, and P. Mulders, Eds., 
World Scientific (1997). 

\bibitem{hey}
A. J. G. Hey and J. E. Mandula, \Journal{\PRD}{5}{2610}{1972}. 

\bibitem{hoo1}
P. Hoodbhoy, \Journal{\PRD}{56}{388}{1997}. 

\bibitem{hoodbhoyji}
P. Hoodbhoy and X. Ji, \Journal{\PRD}{58}{054006}{1998}. 

\bibitem{hjw99}
P. Hoodbhoy, X. Ji, and Wei Lu, \Journal{\PRD}{59}{074010}{1999}.  

\bibitem{hv99}E. W. Hughes and R. Voss, \Journal{\ANN}{49}{303}{1999}.

\bibitem{hk83}V. W. Hughes and J. Kuti, \Journal{\ANN}{33}{611}{1983}.

\bibitem{jaffe96}
R. L. Jaffe, \Journal{\PRD}{54}{6581}{1996}. 

\bibitem{jj91}R. L. Jaffe and X. Ji, \Journal{\PRL}{67}{552}{1991}.

\bibitem{jaffeji91}
R. L. Jaffe and X. Ji, \Journal{\PRD}{43}{724}{1991}.  


\bibitem{jj93}R. L. Jaffe and X. Ji, \Journal{\NPB}{375}{527}{1992}.

\bibitem{jj93_2}R. L. Jaffe and X. Ji, \Journal{\PRL}{71}{2547}{1993}.

\bibitem{jaffe}R. L. Jaffe, Xuemin Jin, and Jian Tang, \Journal{\PRL}{80}{1166}{1998}. 

\bibitem{jap00}G. Japaridze, W-D. Nowak and A. Tkabladze, \Journal{\PRD}{62}{034022}{2000}.

\bibitem{ji93}
X. Ji, \Journal{\PLB}{289}{137}{1992}. 

\bibitem{jiprl97}
X. Ji, \Journal{\PRL}{78}{610}{1997}.  

\bibitem{jidvcs}X. Ji, \Journal{\PRD}{55}{7114}{1997}.

\bibitem{jiinvariant}
X. Ji, \Journal{\PRD}{58}{056003}{1998}. 

\bibitem{jijpg}
X. Ji, \Journal{\JPG}{24}{1181}{1998}. 

\bibitem{jijung}
X. Ji and C. Jung, to be published.

\bibitem{koj}
X. Ji, C. Kao, and J. Osborne, \Journal{\PLB}{472}{1}{2000}. 

\bibitem{jilu00}
X. Ji, W. Lu, J. Osborne, and Song, hep-ph/0006121; 
A. Belitsky, X. Ji, W. Lu, and J. Osborne, hep-ph/0007305. 

\bibitem{jm97}X. Ji and W. Melnitchouk, \Journal{\PRD}{56}{R1}{1997}.

\bibitem{jms}
X. Ji, W. Melnitchouk, and X. Song, \Journal{\PRD}{56}{5511}{1997}. 

\bibitem{jiosborne}
X. Ji and J. Osborne, \Journal{\PRD}{57}{R1337}{1998};
\Journal{\PRD}{58}{094018}{1998}.

\bibitem{jinc} 
X. Ji and J. Osborne, \Journal{\EJC}{9}{487}{1999}. 

\bibitem{jiosborne99}
X. Ji and J. Osborne, hep-ph/9905410.

\bibitem{jiunrau}
X. Ji and P. Unrau, \Journal{\PLB}{333}{228}{1994};

\bibitem{jin97}X. Jin and J. Tang, \Journal{\PRD}{56}{5618}{1997}.

\bibitem{michigan}
G. L. Kane, J. Pumplin, and W. Repko, \Journal{\PRL}{41}{1689}{1978}. 

\bibitem{karliner}
I. Karliner, \Journal{\PRD}{7}{2717}{1973}. 

\bibitem{kim96}H-C. Kim, M. V. Polyakov, K. Goeke, 
\Journal{\PLB}{387}{577}{1996}.

\bibitem{ssadata}
Some selective examples of data can be found in: 
R. D. Klem et al., \Journal{\PRL}{36}{929}{1976};
J. Antille et al., \Journal{\PLB}{94}{523}{1980};
B. E. Bonner et al., \Journal{\PRD}{41}{12}{1980}; 
D. L. Adams et al., \Journal{\PLB}{264}{462}{1991}; 
A. Bravar, et al., \Journal{\PRL}{77}{2626}{1996}.

\bibitem{kodaira79}
J. Kodaira, S. Matsuda, K. Sasaki, T. Uematsu, \Journal{\NPB}{159}{99}{1979}. 

\bibitem{kn97}V. A. Korotkov and W-D. Nowak, \Journal{\NPA}{622}{78c}{1997}.

\bibitem{kot95}A. M. Kotzinain, \Journal{\NPB}{441}{234}{1995}.

\bibitem{km97}A. M. Kotzinian and P. J. Mulders, 
\Journal{\PLB}{406}{373}{1997}.

\bibitem{kotzinian98}
A. Kotzinian, A. Bravar, and D. von Harrach, \Journal{\EJC}{2}{329}{1998}. 

\bibitem{eefrag}S. Kretzer, \Journal{\PRD}{62}{054001}{2000}.

\bibitem{g2parton}
J. Kuti and V. Weisskopf, \Journal{\PRD}{4}{3418}{1971};
B. L. Ioffe, V. A. Khoze, and L. N. Lipatov, 
{\it Hard Processes}, Vol. 1 (North-Holland, Amsterdam, 1984);
J. D. Jackson, R. G. Roberts, G. G. Ross, \Journal{\PLB}{226}{159}{1989}.

\bibitem{lac81}M. Lacombe, et al., \Journal{\PLB}{101}{139}{1981}.

\bibitem{laenen98}
E. Laenen, G. Orderda, and G. Sterman, \Journal{\PLB}{438}{173}{1998}. 

\bibitem{laenen00}
E. Laenen, G. Sterman, and W. Vogelsang, \Journal{\PRL}{84}{4296}{2000}. 

\bibitem{cteq}H. L. Lai, et al., \Journal{\PRD}{55}{1280}{1997}.

\bibitem{lai98}
H.-L. Lai and H.-n. Li, \Journal{\PRD}{58}{114020}{1998}.

\bibitem{gr99}B. Lampe and E. Reya, \Journal{\PR}{332}{2}{2000}.

\bibitem{larver} S. A. Larin and J. A. M. Vermaseren, 
\Journal{\PLB}{259}{345}{1991}.

\bibitem{lss98}E. Leader, A. V. Sidorov and D. B. Stamenov, \Journal{\PLB}{445}{232}{1998}.

\bibitem{lss2-98}E. Leader, A. V. Sidorov and D. B. Stamenov, \Journal{\PRD}{58}{114028}{1998}.

\bibitem{lea98}E. Leader, A. V. Sidrov, and D. B. Stamenov, 
\Journal{\PLB}{445}{232}{1998}; 

\bibitem{lss99}E. Leader, A. V. Sidrov, and D. B. Stamenov, 
\Journal{\PLB}{462}{189}{1999}.

\bibitem{liang00}
Z. Liang and C. Boros, \Journal{\IJA}{15}{927}{2000}.  

\bibitem{low} 
F. E. Low, \Journal{\PR}{96}{1428}{1954}; \Journal{\PR}{110}{974}{1958}.

\bibitem{mankiewicz}
L. Mankiewicz, G. Piller, and T. Weigl, \Journal{\EJC}{5}{119}{1998};
\Journal{\PRD}{59}{017501}{1999}; 
L. Mankiewicz, G. Piller, and A. Radyushkin, \Journal{\EJC}{10}{307}{1999}. 

\bibitem{strain}T. Maruyama, et al., \Journal{\PRB}{46}{4261}{1992}.

\bibitem{liuang}
N. Mathur, S. J. Dong, K. F. Liu, L. Mankiewicz, and N. C. 
Mukhopadhyay, hep-ph/9912289.  

\bibitem{opev}
Y. M. Mekeenko, \Journal{\SJN}{33}{440}{1982};
Th. Ohrndoff, \Journal{\NPB}{186}{153}{1981}; 
\Journal{\NPB}{198}{26}{1982}; 
M. K. Chase, \Journal{\NPB}{174}{109}{1980}; 
M. A. Shifman and M. Vysotsky, \Journal{\NPB}{186}{475}{1981};
V. N. Baier and A. G. Grozin, \Journal{\NPB}{192}{476}{1981};
G. Geyer et al., \Journal{\ZPC}{26}{591}{1985}; 
I. Braunschweig et al., \Journal{\ZPC}{33}{175}{1987}. 

\bibitem{quarkonly}
R. Mertig and W. L. van Neervan, \Journal{\ZPC}{60}{489}{1993}; 
G. Altarelli, B. Lampe, P. Nason and G. Ridolfi, 
\Journal{\PLB}{334}{187}{1994}; 
J. Kodaira, S. Matsuda, T. Uematsu, and K. Sasaki, 
\Journal{\PLB}{345}{527}{1995}; 
P. Mathews, V. Ravindran, and K. Sridhar, hep-ph/9607385; 
A. Gabieli, G. Ridolfi, \Journal{\PLB}{417}{369}{1998}.  

\bibitem{ma00}
B. Ma, I. Schmidt and J. Yang, \Journal{\PRD}{61}{034017}{2000}.
 
\bibitem{ma}
B. Ma, I. Schmidt, J. Soffer and J. Yang, hep-ph/0001259.  

\bibitem{mn96}R. Mertig and W. L. van Neerven,  \Journal{\ZPC}{70}{637}{1996}.

\bibitem{splitnlo}R. Mertig and W.L. van Neerven, \Journal{\ZPC}{70}{637}{1996};W. Vogelsang, \Journal{\PRD}{54}{2023}{1996}. 

\bibitem{mirkes96}
E. Mirkes and D. Zeppenfeld, \Journal{\PLB}{380}{205}{1996}. 

\bibitem{mirkes97}
E. Mirkes and S. Willfahrt, hep-ph/9711434. 

\bibitem{phenix}(PHENIX) D. P. Morrison, et al., \Journal{\NPA}{638}{565c}{1998}.

\bibitem{mt96}P. J. Mulders and R. D. Tangerman, \Journal{\NPB}{461}{197}{1996}.

\bibitem{mt97}D. Muller and O. V. Teryaev, \Journal{\PRD}{56}{2607}{1997}.

\bibitem{hermloq}(HERMES) W-D. Nowak, {\em Proceedings of the Eighth
International Workshop on Deep Inelastic Scattering and QCD (DIS2000)},
Liverpool, England, Apr. 25-30, 2000.

\bibitem{Oga98}K. A. Oganessyan, \Journal{\EJC}{5}{681}{1998}.

\bibitem{pantforder}
R. Pantf\"order, hep-ph/9805434.

\bibitem{pdg00}Particle Data Group, \Journal{\EJC}{15}{1}{2000}.

\bibitem{petrov98}
V. Petrov et al., \Journal{\PRD}{57}{4325}{1998}; 
M. V. Polyakov and C. Weiss, \Journal{\PRD}{60}{114017}{1999}.  

\bibitem{qiu91}
J. W. Qiu and G. Sterman, \Journal{\PRL}{67}{2264}{1991}. 

\bibitem{qiu99} 
J. W. Qiu and G. Sterman, \Journal{\PRD}{59}{014004}{1999}. 

\bibitem{radel97}
G. R\"adel, A. De Roeck, and M. Maul, hep-ph/9711373.

\bibitem{radel99}
G. R\"adel and A. De Roeck, hep-ph/9909403.

\bibitem{radmeson}
A. V. Radyushkin, \Journal{\PLB}{385}{333}{1996}.

\bibitem{radyushkin97}
A. V. Radyushkin, \Journal{\PRD}{56}{5524}{1997}.

\bibitem{rad99}A. V. Radyushkin, \Journal{\PRD}{59}{014030}{1999}.

\bibitem{rmodel}
A. V. Radyushkin, \Journal{\PLB}{449}{81}{1999}; 
I. V. Musatov and A. V. Radyushkin, \Journal{\PRD}{61}{074027}{2000}.  

\bibitem{ralston}J. Ralston and D. E. Soper, \Journal{\NPB}{152}{109}{1979}.

\bibitem{ratcliffe86}
P. G. Ratcliffe, \Journal{\NPB}{264}{493}{1986}. 

\bibitem{gdependent}
P. G. Ratcliffe, \Journal{\PLB}{192}{180}{1987}; 
R. L. Jaffe and A. Manohar, \Journal{\NPB}{267}{509}{1990};
X. Ji, J. Tang, and P. Hoodbhoy, \Journal{\PRL}{76}{740}{1996};
P. Hagler and A. Schafer, \Journal{\PLB}{430}{179}{1998}; 
S. V. Bashinsky and R. L. Jaffe, \Journal{\NPB}{536}{303}{1998};
A. Harinderanath and R. Kundu, \Journal{\PRD}{59}{116013}{1999}; 
X. Chen and F. Wang, hep-ph/9802346; 
P. Hoodbhoy, X. Ji, and W. Lu, \Journal{\PRD}{60}{114042}{1999}. 

\bibitem{rhic}(RHIC) See www.rhic.bnl.gov.

\bibitem{zeus96}
J. Respond, proceedings of workshop `Deep Inelastic 
Scattering and Related Phenomena,' Rome 1996, Eds. 
G. D'Agostini and A. Nigro (World Scientific, Singapore, 1997)
439. 

\bibitem{rys1}
M. G. Ryskin, \Journal{\ZPC}{37}{89}{1993}.

\bibitem{sandorfi}
A. M. Sandorfi, C. S. Whisnant, and M. Khandaker, 
\Journal{\PRD}{50}{R6681}{1994}.  

\bibitem{sasaki}
K. Sasaki, \Journal{\PTP}{54}{1816}{1975}. 

\bibitem{scholten99}
O. Scholten and A. Yu. Korchin, \Journal{\EJA}{6}{211}{1999}. 

\bibitem{shulz} R. W. Shulze and P. U. Sauer,\Journal{\PRC}{48}{38}{1993}.

\bibitem{sv}
E. V. Shuryak and A. I. Vainshtein, \Journal{\NPB}{199}{451}{1982}; 
\Journal{\NPB}{201}{141}{1982}.

\bibitem{sivers}
D. Sivers, \Journal{\PRD}{41}{83}{1990}; 
\Journal{\PRD}{43}{261}{1991}. 

\bibitem{pythia}T. Sjostrand, \Journal{CPC}{82}{74}{1994}.

\bibitem{sof95}J. Soffer, \Journal{\PRL}{74}{1292}{1995}.

\bibitem{soffer93}
J. Soffer and O. Teryaev, \Journal{\PRL}{70}{3373}{1993}. 

\bibitem{sok64}A. A. Sokolov and I. M. Ternov, \Journal{\SPD}{8}{1203}{1964}.

\bibitem{g2song}
X. Song, \Journal{\PRD}{54}{1955}{1996}.

\bibitem{stein}
E. Stein, \Journal{\PLB}{343}{369}{1995}. 

\bibitem{g2stratmann}
M. Stratmann, \Journal{\ZPC}{60}{763}{1993}. 

\bibitem{stratmann97}
M. Stratmann and W. Vogelsang, \Journal{\ZPC}{74}{461}{1997}. 

\bibitem{tat00} S. Tatur, J. Bartelski and M. Kurzela, 
\Journal{\APP}{31}{647}{2000}.

\bibitem{hv72}G. t'Hooft and M. Veltman, \Journal{\NPB}{44}{189}{1972}.

\bibitem{phobos}(PHOBOS) A. Trzupek, et al., \Journal{\APP}{B27}{3103c}{1996}.

\bibitem{vg}
M. Vanderhaeghen, P. A. M. Guichon, and M. Guidal, 
\Journal{\PRL}{80}{5064}{1998}; 
P. A. M. Guichon and M. Vanderhaeghen, \Journal{\PPNP}{41}{125}{1998}. 

\bibitem{guichon00}
M. Vanderhaeghen, P. A. M. Guichon, and M. Guidal, 
\Journal{\PRD}{60}{094017}{2000}. 

\bibitem{brahms}(BRAHMS) F. Videbaek, et al., \Journal{\NPA}{566}{299c}{1994}.

\bibitem{vog96}W. Vogelsang, \Journal{\NPB}{475}{47}{1996}.

\bibitem{wilczek}
W. Wandzura and F. Wilczek, \Journal{\PLB}{172}{195}{1977}.

\bibitem{weigel}
H. Weigel, L. Gamberg, H. Reinhart, \Journal{\PRD}{55}{6910}{1997}. 

\bibitem{witten76}
E. Witten, \Journal{\NPB}{104}{445}{1976}; 
L. M. Jones and H. W. Wyld, \Journal{\PRD}{17}{759}{1978}. 

\bibitem{workman}  
R. L. Workman and R. A. Arndt, \Journal{\PRD}{45}{1789}{1992}. 

\end{thebibliography}
\end{document}